\documentclass[fleqn,usenatbib]{mnras}

\usepackage{newtxtext,newtxmath}

\usepackage[T1]{fontenc}
\usepackage{ae,aecompl}
\usepackage{multirow}

\usepackage{graphicx}	
\usepackage{amsmath}	
\usepackage{amssymb}	
\usepackage{pdflscape}

\newcommand{\G}{\textit{Gaia}}
\newcommand{\DR}{\textit{Gaia}~DR2}

\title[GUCDS III: Ultra-cool dwarfs in multiple systems]{The \G~Ultra-Cool
  Dwarf Sample -- III: Seven new multiple systems containing at least one \DR\ ultra-cool dwarf.}

\author[F. Marocco et al.]{F. Marocco$^{1,2}$\thanks{NASA Postdoctoral Program Fellow}\thanks{E-mail: federico.marocco@jpl.nasa.gov},
R. L. Smart$^{3}$, E. E. Mamajek$^{1}$, L. M. Sarro$^{4}$, A. J. Burgasser$^{5}$, \newauthor
J. A. Caballero$^{6}$, J. M. Rees$^{5}$, D. Caselden$^{7}$, K. L. Cruz$^{8,9,10}$, R. Van Linge$^{5}$, \newauthor D. J. Pinfield$^{11}$ \vspace{0.2cm} \\%
$^{1}$Jet Propulsion Laboratory, California Institute of Technology, 4800 Oak Grove Dr., Pasadena, CA 91109, USA \\
$^{2}$IPAC, Mail Code 100-22, California Institute of Technology, 1200 E. California Blvd., Pasadena, CA 91125, USA \\
$^{3}$Istituto Nazionale di Astrofisica, Osservatorio Astrofisico di Torino, Strada Osservatorio 20, I-10025 Pino Torinese, Italy \\
$^{4}$Department of Artificial Intelligence, Universidad Nacional de Educaci\'on a Distancia, c/ Juan del Rosal 16, E-28040 Madrid, Spain \\
$^{5}$Center for Astrophysics and Space Science, University of California San Diego, La Jolla, CA 92093, USA \\
$^{6}$Centro de Astrobiolog\'ia (CSIC-INTA), ESAC, Camino Bajo del Castillo s/n, E-28692 Villanueva de la Ca\~nada, Madrid, Spain \\
$^{7}$Gigamon Applied Threat Research, 619 Western Avenue, Suite 200, Seattle, WA 98104, USA \\
$^{8}$Department of Physics and Astronomy, Hunter College, City University of New York, 695 Park Avenue, New York, NY 10065, USA \\
$^{9}$Physics, Graduate Center of the City University of New York, 365 5th Avenue, New York, NY 10016, USA \\
$^{10}$Department of Astrophysics, American Museum of Natural History, Central Park West at 79th Street, New York, NY 10024, USA \\
$^{11}$Centre for Astrophysics Research, School of Physics, Astronomy and Mathematics, University of Hertfordshire, College Lane, \\ Hatfield AL10 9AB, UK
}

\date{Accepted 2020 April 7. Received 2020 April 6; in original form 2019 September 26}

\pubyear{2020}

\begin{document}
\label{firstpage}
\pagerange{\pageref{firstpage}--\pageref{lastpage}}
\maketitle

\begin{abstract}
We present ten new ultra-cool dwarfs in seven wide binary systems discovered using \DR\ data, identified as part of our \G\ Ultra-Cool Dwarf Sample project. 
The seven systems presented here include an L1 companion to the G5\,IV star HD~164507, an L1: companion to the V478 Lyr AB system, an L2 companion to the metal-poor K5\,V star CD-28~8692, an M9\,V companion to the young variable K0\,V star LT UMa, and three low-mass binaries consisting of late Ms and early Ls. The HD~164507, CD-28~8692, V478 Lyr, and LT UMa systems are particularly important benchmarks, because the primaries are well characterised and offer excellent constraints on the atmospheric parameters and ages of the companions. 
We find that the M8\,V star 2MASS~J23253550+4608163 is $\sim$2.5~mag overluminous compared to M dwarfs of similar spectral type, but at the same time it does not exhibit obvious peculiarities in its near-infrared spectrum. Its overluminosity cannot be explained by unresolved binarity alone.
Finally, we present an L1+L2 system with a projected physical separation of 959\,au, making this the widest L+L binary currently known.
\end{abstract}




\section{Introduction}
Ultra-cool dwarfs (UCDs, spectral type $\geq$M7) in binary systems with main sequence and post-main sequence stars are valuable benchmarks \citep{2006MNRAS.368.1281P}, providing robust tests of ultra-cool atmospheric and evolutionary models. Under the reasonable assumption of common origin, a bright main sequence primary provides constraints on the metallicity and the age of a system, two parameters that are currently difficult to infer for isolated ultra-cool dwarfs.

UCDs are a mixture of the lowest mass hydrogen fusing stars and sub-stellar non-hydrogen-fusing objects. Mass, age, metallicity, and luminosity are degenerate parameters for these objects, and the presence (and evolution) of dust clouds in the photosphere further complicates the interpretation of their spectra \citep{2006ApJ...640.1063B,2008ApJ...689.1327S}. Furthermore sub-stellar UCDs overlap in both mass and temperature with the gaseous giant planets in exosolar systems \citep[e.g.][]{2016ApJS..225...10F}, but can be studied without the additional complication of the planets' vicinity to a bright host star. A full understanding of ultra-cool atmospheres is therefore of vital importance if we wish to understand exoplanets and their formation and evolution.

The recent second data release from the ESA mission \G\  \citep{2016A&A...595A...1G,2018A&A...616A...1G} provides exquisite astrometry for $\sim$1.3 billion objects within our Galaxy \citep{2018A&A...616A...2L}, allowing access to a huge population of wide binaries consisting of an ultra-cool dwarf in a system with a star or white dwarf \citep{2017MNRAS.470.4885M}. In particular, the greatly increased volume probed by \G, and the resulting increased pool of potential primary stars, offers for the first time the possibility to map the full age--temperature--metallicity parameter space, large regions of which are currently undersampled or completely unexplored \citep[see e.g.][]{2011ASPC..448..833D,2014ApJ...792..119D,2017MNRAS.470.4885M}. While the advent of \G\ expands the pool of potential primaries, existing optical and near-infrared surveys, and the astrometric catalogues that spawned from them (e.g. ULAS, \citealt{2014MNRAS.437.3603S}; VIRAC, \citealt{2018MNRAS.474.1826S}; CatWISE, \citealt{2019arXiv190808902E}), grant access to a vast population of ultra-cool dwarfs across spectral types M, L, T, and Y. We have therefore set out to complete the nearby census of these objects, to fully explore and characterise ultra-cool atmospheres.

In this paper we present seven new multiple systems containing at least one \DR\ ultra-cool dwarf component. 

In Section~\ref{sec:cand} we describe our candidate selection; in Section~\ref{sec:obs} we summarise observing and data reduction procedures; in Section~\ref{sec:new} we discuss in more detail the newly discovered systems; in Section~\ref{sec:features} we compare the main features in the spectra of the new UCDs; finally in Section~\ref{sec:conclusions} we summarise our findings and discuss future work.

\section{Candidate selection}
\label{sec:cand}
We identified an initial list of 8013 UCD candidates from the \DR\ catalogue as follows. First, we queried the catalogue for \G\ sources fainter than the maximum brightness that an UCD at the measured parallax could have, as predicted by the BT-Settl models \citep[][]{2012EAS....57....3A,2013MmSAI..84.1053A}. The maximum distance is 373\,pc, the distance at which the brightest, hottest UCD ($T_{\rm eff}\,\sim\,2500$\,K) would be fainter than the \G\ limiting magnitude ($G = 20.7$\,mag). We required the $G-G_{RP}$ colour to be redder than 1.4\,mag \citep[since UCDs are typically redder than that;][]{2017MNRAS.469..401S,2019MNRAS.485.4423S}. To minimise the number of sources with spurious astrometric measurements, we removed candidates within 5 degrees of the Galactic plane and inside an ellipse centred at the Galactic centre with semi-major axis along the Galactic longitude axis of 50$^\circ$, and 8$^\circ$ along the Galactic latitude axis. To retain only reliable astrometric measurements, we required sources to have more than six visibility periods and astrometric excess noise lower than 5\,mas. We computed posterior probability densities of the distance given the parallax measurements and associated uncertainties using an exponentially decreasing constant volume density prior, and selected sources with a posterior probability to be within 373\,pc greater than 0.5. We then fit a principal curve \citep{princurve} in the $M_G$ versus $G - G_{RP}$ plane to the values of the resulting set, and calibrated the curve in effective temperature using the spectral types of sources in the \G\ Ultra-Cool Dwarfs Sample \citep[][]{2017MNRAS.469..401S,2019MNRAS.485.4423S} and the \citet{2009ApJ...702..154S} conversion between spectral types and effective temperatures. Finally, we computed the projections of the UCD candidate positions in the $M_G$ versus $G_{RP}$ plane along the principal curve and assigned effective temperatures accordingly. A cut at $T_{\rm eff} \leq$\,2500\,K resulted in the 8013 candidates mentioned above.

We searched for binaries among these 8013 UCD candidates using the criteria defined in \citet[hereafter GUCDS~II]{2019MNRAS.485.4423S}: 
\begin{equation} \label{eq1}
\begin{split}
\rho         & < 100\,\varpi \\
\Delta\varpi & < {\rm max}\left[3\sigma_{\varpi}, 1\,{\rm mas}\right]\\
\Delta\mu    & < 0.1\mu\\
\Delta\theta & < 15^\circ,
\end{split}
\end{equation}

\noindent where $\rho$ is the separation on the sky in arcseconds, $\Delta\varpi$ is the difference between the candidate UCD and primary parallax, $\varpi$ and $\sigma_{\varpi}$ are the parallax and parallax uncertainty for the UCD (in mas), $\Delta\mu$ is the difference of the total proper motions, and $\Delta\theta$ is the difference of the position angles. The maximum $\rho$ was chosen to correspond to 100,000\,au as a conservative upper limit for the projected physical separation ($s$). This separation meets the binding energy criterion of $ |U_g^*| > 10^{33} J $ as developed by \cite{2009A&A...507..251C} for a system of a 0.1 and a 2 M$_{\odot}$ objects. The parallax criterion is a compromise between a standard 3$\sigma$ criterion, and a more conservative 1.0\,mas difference to allow for parallaxes that had unrealistically low errors. For the proper motion, using a standard 3$\sigma$ criterion would remove nearby objects with significant orbital motion, so we choose a conservative 10\% agreement, which is large enough to accommodate most orbital motions but small enough to reduce false positives. As discussed in GUCDS~II these criteria fail for the nearby binary systems GJ 1048 A/B and G 239-25 A/B (in both cases because the modulus of the proper motions differs by more than 10\%). Therefore, our catalogue of binary candidates should not be regarded as complete. 

\begin{figure}
    \centering
    \includegraphics[trim={2cm 3cm 3cm 4cm}, clip, width=0.48\textwidth]{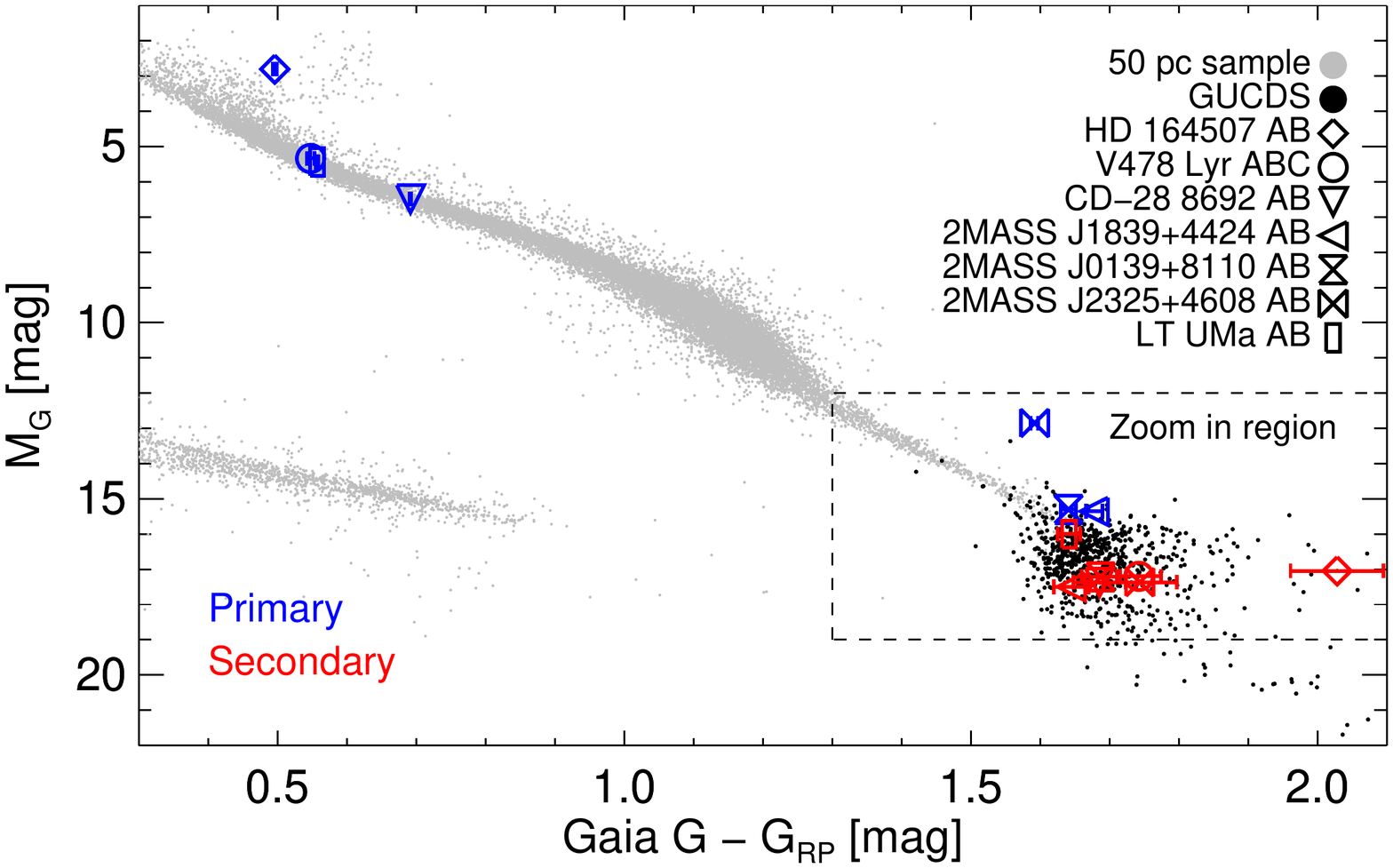}
    \includegraphics[trim={2cm 3cm 3cm 4cm}, clip, width=0.48\textwidth]{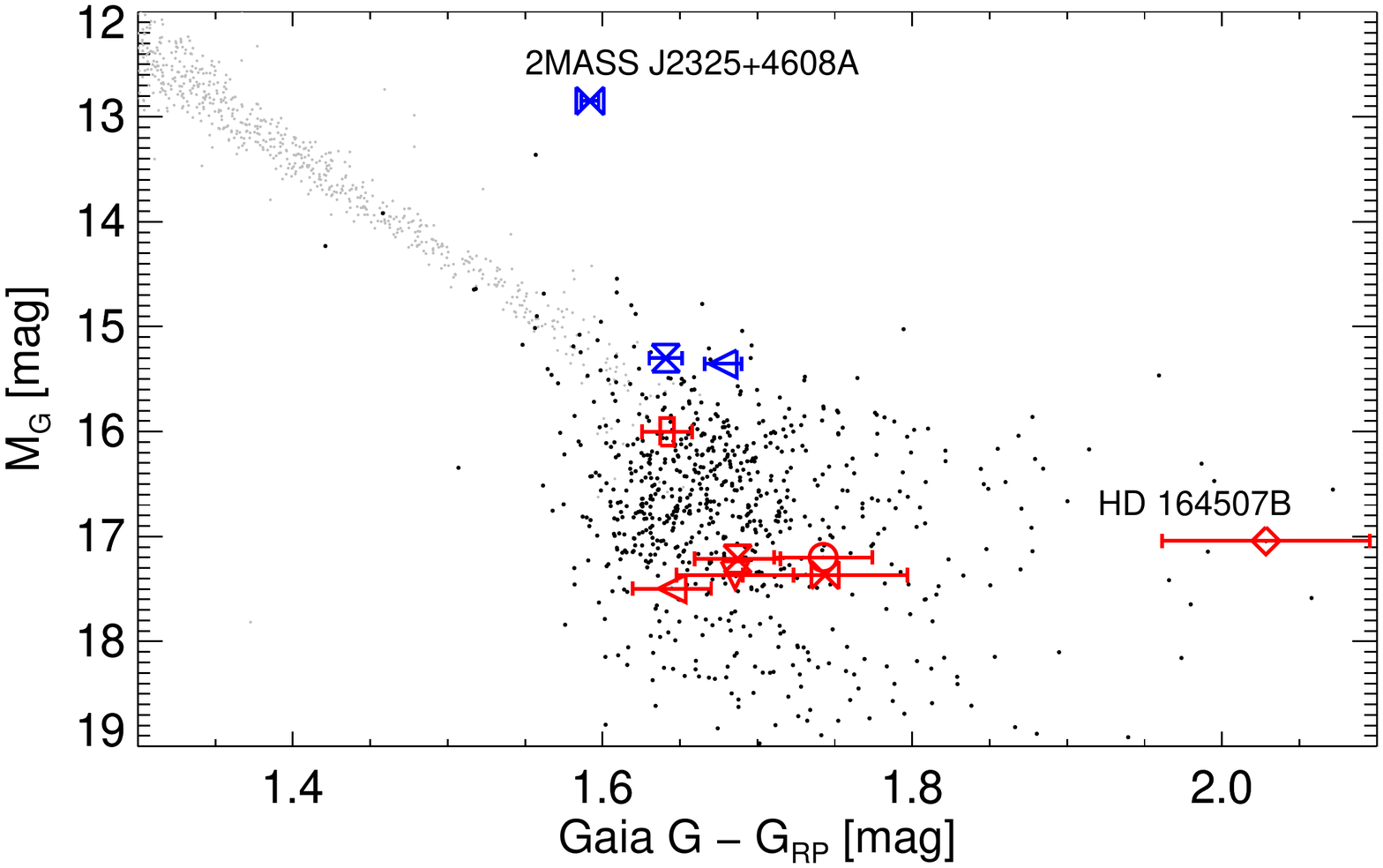}
    \caption{Colour-magnitude diagrams depicting the full stellar sequence (top) and a zoom into the ultra-cool dwarfs region (bottom). The small grey points are stars in \DR\ nominally within 50\,pc, selected using the criteria described in Appendix C of \citet{2018A&A...616A...2L}. Red points are the ultra-cool dwarfs identified in \DR\ by GUCDS~II. The seven systems presented here are plotted with different symbols, with the primary in each system plotted in blue and the companion in green. Vertical error bars are typically smaller than the symbols. Detailed analysis of individual systems can be found in Section~\ref{sec:new}.}
    \label{fig:hr}
\end{figure}

Of the 8013 UCD candidates, 840 have a possible companion according to the criteria above. The seven systems presented here are those that we could observe during our observing nights at the Palomar Observatory. We present their astrometric properties and spectral types in Tables~\ref{tab:systems1} and \ref{tab:systems2}. We collected optical and near-infrared photometry for both components of our newly discovered systems from \DR, 2MASS \citep{2006AJ....131.1163S}, PanSTARRS DR1 \citep{2016arXiv161205560C}, and AllWISE \citep{2013wise.rept....1C}. The photometry is also presented in Tables~\ref{tab:systems1} and \ref{tab:systems2}. In Figure~\ref{fig:hr} we show a colour-magnitude diagram based on \G\ colours and astrometry. The small grey points are objects in \DR\ nominally within 50\,pc, selected using Equation C.1 and C.2 from \citet{2018A&A...616A...2L}. Red points are UCDs identified in \DR\ by GUCDS~II. The position of the seven systems presented here is highlighted with different symbols, with the primary plotted in blue and the companion in green. Two objects stand out at first glance: HD~164507~B, and 2MASS~J23253550+4608163. We will discuss their properties in Sections~\ref{sec:hd164507} and \ref{sec:2m2325}.

\begin{landscape}
\begin{table}
\centering
\caption{Astrometry, photometry, and spectral types for the HD~164507\,AB, V478 Lyr\,ABC, CD-28\,8692\,AB, and 2MASS J1839+4424\,AB systems. \label{tab:systems1}}
\begin{tabular}{l c c c c c c c c}
 \hline 
 & \multicolumn{2}{c}{HD~164507} & \multicolumn{2}{c}{V478 Lyr} & \multicolumn{2}{c}{CD--28~8692} & \multicolumn{2}{c}{2MASS~J1839+4424}\\
 & A & B & AB & C & A & B & A & B \\
 \hline
R.A. (hh:mm:ss.ss) & 18:00:57.22 & 18:00:58.48 & 19:07:32.52 & 19:07:33.23 & 11:10:25.97 & 11:10:29.21 & 18:39:29.22 & 18:39:27.40 \\
Dec. (dd:mm:ss.s)  & +15:05:35.3 & +15:05:18.3 & +30:15:17.8 & +30:15:32.1 & --29:24:51.5 & --29:25:19.8 & +44:24:41.2 & +44:24:51.0 \\
Sep. (arcsec) & \multicolumn{2}{c}{25.01} & \multicolumn{2}{c}{17.05} & \multicolumn{2}{c}{50.91} & \multicolumn{2}{c}{21.89} \\
Sep. (au) & \multicolumn{2}{c}{1136} & \multicolumn{2}{c}{462} & \multicolumn{2}{c}{2026} & \multicolumn{2}{c}{811} \\
P.A. (deg) & \multicolumn{2}{c}{132.58} & \multicolumn{2}{c}{32.63} & \multicolumn{2}{c}{123.84} & \multicolumn{2}{c}{296.62} \\
Sp. type & G5\,IV$^1$ & L1 & G8\,V\,SB$^2$ & L1: & K5\,V$^3$ & L2 & M9\,V$^4$ & L2 \\
$\varpi$ (mas) & $22.009\pm0.037$ & $24.8\pm1.4$ & $36.877\pm0.026$ & $37.02\pm0.46$ & $25.133\pm0.042$ & $24.68\pm0.97$ & $27.00\pm0.41$ & $27.01\pm0.70$ \\
$\mu_\alpha \cos\,\delta$ (mas yr$^{-1}$) & $-53.138\pm0.068$ & $-57.6\pm1.5$ & $110.850\pm0.040$ & $112.39\pm0.62$ & $-9.948\pm0.064$ & $-10.9\pm1.6$ & $20.01\pm0.75$ & $29.3\pm1.5$ \\
$\mu_\delta$ (mas yr$^{-1}$) & $-98.328\pm0.095$ & $-95.2\pm1.4$ & $103.117\pm0.045$ & $103.47\pm0.91$ & $-74.913\pm0.059$ & $-73.2\pm1.4$ & $171.09\pm0.77$ & $173.0\pm2.5$ \\
\textit{Gaia G} (mag) & 6.08969$\pm$0.00046 & 20.070$\pm$0.013 & 7.5035$\pm$0.0016 & 19.3571$\pm$0.0045 & 9.48280$\pm$0.00018 & 20.4050$\pm$0.0077 & 18.1977$\pm$0.0042 & 20.3391$\pm$0.0083 \\
\textit{Gaia G$_{RP}$} (mag) & 5.5936$\pm$0.0028 & 18.041$\pm$0.066 & 6.9557$\pm$0.0054 & 17.615$\pm$0.032 & 8.79104$\pm$0.00078 & 18.719$\pm$0.037 & 16.520$\pm$0.011 & 18.694$\pm$0.024 \\
PS1 $r$ (mag) & \ldots & \ldots & 7.7177$\pm$0.0095 & \ldots & 9.507$\pm$0.064 & 21.14$\pm$0.19 & 20.334$\pm$0.015 & 21.693$\pm$0.044 \\
PS1 $i$ (mag) & \ldots & 18.81$\pm$0.17 & 7.73$\pm$0.22 & 18.474$\pm$0.014 & 9.50$\pm$0.24 & 19.959$\pm$0.024 & 17.6724$\pm$0.0027 & 19.8205$\pm$0.0077 \\
PS1 $z$ (mag) & \ldots & 17.779$\pm$0.029 & 7.4026$\pm$0.0063 & 17.37$\pm$0.14 & 9.848$\pm$0.021 & 18.456$\pm$0.013 & 16.2113$\pm$0.0034 & 18.4309$\pm$0.0076 \\
PS1 $y$ (mag) & \ldots & 16.302$\pm$0.058 & 6.817$\pm$0.022 & 16.1001$\pm$0.0060 & 8.8370$\pm$0.0010 & 17.482$\pm$0.019 & 15.3240$\pm$0.0043 & 17.4922$\pm$0.0083 \\
2MASS $J$ (mag) & 5.19$\pm$0.26 & 15.416$\pm$0.051$^c$ & 6.232$\pm$0.020 & $>13.663$ & 7.922$\pm$0.019 & 15.421$\pm$0.062 & 13.433$\pm$0.029 & 15.432$\pm$0.054 \\
2MASS $H$ (mag) & 4.700$\pm$0.053 & 14.755$\pm$0.082$^c$ & 5.855$\pm$0.016 & $>13.210$ & 7.414$\pm$0.044 & 14.535$\pm$0.046 & 12.792$\pm$0.035 & 14.598$\pm$0.059 \\
2MASS $K_s$ (mag) & 4.551$\pm$0.020 & 14.142$\pm$0.067$^c$ & 5.741$\pm$0.020 & 13.021$\pm$0.042$^c$ & 7.276$\pm$0.018 & 14.085$\pm$0.076 & 12.356$\pm$0.028 & 13.901$\pm$0.047 \\
AllWISE $W1$ (mag) & 4.56$\pm$0.20$^b$ & 11.007$\pm$0.021$^d$ & 5.76$\pm$0.12$^b$ & \ldots & 7.123$\pm$0.035$^b$ & 13.641$\pm$0.024$^a$ & 12.002$\pm$0.023 & 13.540$\pm$0.026 \\
AllWISE $W2$ (mag) & 4.417$\pm$0.056$^b$ & 10.884$\pm$0.021$^d$ & 5.580$\pm$0.045$^b$ & \ldots & 7.298$\pm$0.019 & 13.406$\pm$0.029$^a$ & 11.750$\pm$0.022 & 13.291$\pm$0.028 \\
AllWISE $W3$ (mag) & 4.569$\pm$0.015$^b$ & 10.666$\pm$0.092$^d$ & 5.667$\pm$0.015$^b$ & \ldots & 7.255$\pm$0.017 & $>12.581$ & 11.29$\pm$0.10 & 12.60$\pm$0.36 \\
\hline
\end{tabular} \\
Notes: Coordinates, parallax and proper motion are from \textit{Gaia} DR2. Separation and position angle are computed at the \DR\ epoch (2015.5). Spectral types are assigned using SPLAT (see Section~\ref{sec:obs}), except for HD~164507, V478 Lyr AB, CD--28~8692, and 2MASS~J18392917+4424386, whose spectral types are taken from the literature. References: 1-\citet{1970AJ.....75..165H}; 2-\citet{1988AJ.....95..215F}; 3-\citet{1972AJ.....77..486U}; 4-\citet{2014ApJ...794..143B}. Notes on photometry: $^a$contaminated by bright star halo; $^b$saturated; $^c$contaminated by bright star; $^d$contaminated by diffraction spike.
\end{table}
\end{landscape}

\begin{table*}
\centering
\caption{Astrometry, photometry, and spectral types for the 2MASS J0139+8110\,AB, 2MASS J2325+4608\,AB, and LT UMa\,AB systems presented here. \label{tab:systems2}}
\begin{tabular}{l c c c c c c}
 \hline 
 & \multicolumn{2}{c}{2MASS~J0139+8110} & \multicolumn{2}{c}{2MASS~J2325+4608} & \multicolumn{2}{c}{LT UMa} \\
 & A & B & A & B & A & B \\
 \hline
R.A. (hh:mm:ss.ss) & 01:39:09.00 & 01:38:59.67 & 23:25:35.40 & 23:25:35.09 & 08:44:47.95 & 08:44:50.12 \\
Dec. (dd:mm:ss.s)  & +81:09:59.7 & +81:10:07.9 & +46:08:15.8 & +46:08:09.2 & +55:32:19.7 & +55:32:12.3 \\
Sep. (arcsec) & \multicolumn{2}{c}{23.00} & \multicolumn{2}{c}{7.24} & \multicolumn{2}{c}{19.83} \\
Sep. (au) & \multicolumn{2}{c}{959} & \multicolumn{2}{c}{378} & \multicolumn{2}{c}{879} \\
P.A. (deg) & \multicolumn{2}{c}{290.87} & \multicolumn{2}{c}{205.89} & \multicolumn{2}{c}{111.89} \\
Sp. type & L1 & L2 & M8\,V & L2 & K0\,V & M9\,V \\
$\varpi$ (mas) & $23.98\pm0.23$ & $24.74\pm0.79$ & $19.13\pm0.48$ & $20.3\pm1.4$ & $22.550\pm0.033$ & $22.00\pm0.43$ \\
$\mu_\alpha \cos\,\delta$ (mas yr$^{-1}$) & $-4.25\pm0.51$ & $-5.3\pm1.4$ & $-52.19\pm0.64$ & $-61.2\pm2.6$ & $76.636\pm0.053$ & $77.90\pm0.60$ \\
$\mu_\delta$ (mas yr$^{-1}$) & $-26.61\pm0.33$ & $-26.8\pm1.0$ & $-34.00\pm0.51$ & $-40.4\pm1.8$ & $12.890\pm0.050$ & $14.88\pm0.57$ \\
\textit{Gaia G} & 18.4012$\pm$0.0023 & 20.2468$\pm$0.0070 & 16.4411$\pm$0.0039 & 20.829$\pm$0.012 & 8.6719$\pm$0.0006 & 19.2895$\pm$0.0033 \\
\textit{Gaia G$_{RP}$} & 16.760$\pm$0.010 & 18.560$\pm$0.027 & 14.8492$\pm$0.0038 & 19.086$\pm$0.052 & 8.1151$\pm$0.0020 & 17.648$\pm$0.016 \\
PS1 $r$ & 20.473$\pm$0.046 & $>20.031$ & 17.7361$\pm$0.0088 & $>21.68$ & 7.003$\pm$0.001 & $>$17.26 \\
PS1 $i$ & 17.801$\pm$0.013 & 19.763$\pm$0.066 & 15.6181$\pm$0.0035 & 20.301$\pm$0.027 & 9.149$\pm$0.030 & 18.616$\pm$0.028 \\
PS1 $z$ & 16.4260$\pm$0.0077 & 18.325$\pm$0.020 & 14.6096$\pm$0.0040 & 18.847$\pm$0.016 & \ldots & 17.231$\pm$0.013 \\
PS1 $y$ & 15.6106$\pm$0.0067 & 17.327$\pm$0.017 & 14.0252$\pm$0.0029 & 17.806$\pm$0.019 & 9.475$\pm$0.001 & 16.333$\pm$0.018 \\
2MASS $J$ & 13.891$\pm$0.028 & 15.239$\pm$0.046 & 12.561$\pm$0.020 & 15.868$\pm$0.070 & 7.458$\pm$0.018 & 14.704$\pm$0.035$^c$ \\
2MASS $H$ & 13.233$\pm$0.038 & 14.400$\pm$0.049 & 11.955$\pm$0.021 & 14.783$\pm$0.059 & 7.124$\pm$0.051 & 13.951$\pm$0.043$^c$ \\
2MASS $K_s$ & 12.829$\pm$0.030 & 13.896$\pm$0.053 & 11.573$\pm$0.018 & 14.348$\pm$0.076 & 7.016$\pm$0.026 & 13.491$\pm$0.024$^c$ \\
AllWISE $W1$ & 12.381$\pm$0.022 & 13.419$\pm$0.024 & 11.387$\pm$0.023 & 13.693$\pm$0.080 & 6.927$\pm$0.051$^b$ & \ldots \\
AllWISE $W2$ & 12.125$\pm$0.023 & 13.109$\pm$0.027 & 11.172$\pm$0.021 & 13.493$\pm$0.077 & 7.007$\pm$0.020$^b$ & \ldots \\
AllWISE $W3$ & 11.64$\pm$0.19 & 12.76$\pm$0.46 & 10.90$\pm$0.11 & $>11.946$ & 6.985$\pm$0.017 & \ldots \\
\hline
\end{tabular} \\
Notes: Coordinates, parallax and proper motion are from \textit{Gaia} DR2. Separation and position angle are computed at the \DR\ epoch (2015.5). Spectral types are assigned using SPLAT (see Section~\ref{sec:obs}), except for LT UMa A, whose spectral types is taken from \citet{2000A&AS..142..275S}. Notes on photometry: $^a$contaminated by bright star halo; $^b$saturated; $^c$contaminated by bright star; $^d$contaminated by diffraction spike.
\end{table*}

\section{Observations}
\label{sec:obs} 
We obtained near-infrared spectra for the ultra-cool dwarfs in our newly discovered binary systems using TripleSpec on the 200'' telescope at the Palomar Observatory on 2018 April 27-29, 2018 October 16 and 18, and 2019 April 16 (Proposals 2018A J12, 2018B J08, and 2019A J14; PI: Mamajek; see Appendix~\ref{app}). TripleSpec is a near-infrared echelle spectrograph, that delivers a resolution of 2500--2700 over the wavelength range $1.0-2.4 \mu$m \citep{2008SPIE.7014E..0XH}.

Targets were observed following a standard ABBA nodding pattern with a nod throw of 11\,arcsec. The slit was aligned to the parallactic angle to minimise atmospheric distortion, with the exceptions of HD~164507\,B, V478~Lyr\,C, and 2MASS~J232535.09+460809.3, for which we rotated the slit to avoid the bright primary. We observed an A0\,V star (selected using the Gemini Telluric Standard Search on-line tool\footnote{\url{https://www.gemini.edu/sciops/instruments/nearir-resources/spectroscopic-standards-/telluric-standard-search}}) for telluric correction after each target, matching the airmass of observation as closely as possible. 

\begin{figure}
\includegraphics[width=0.48\textwidth]{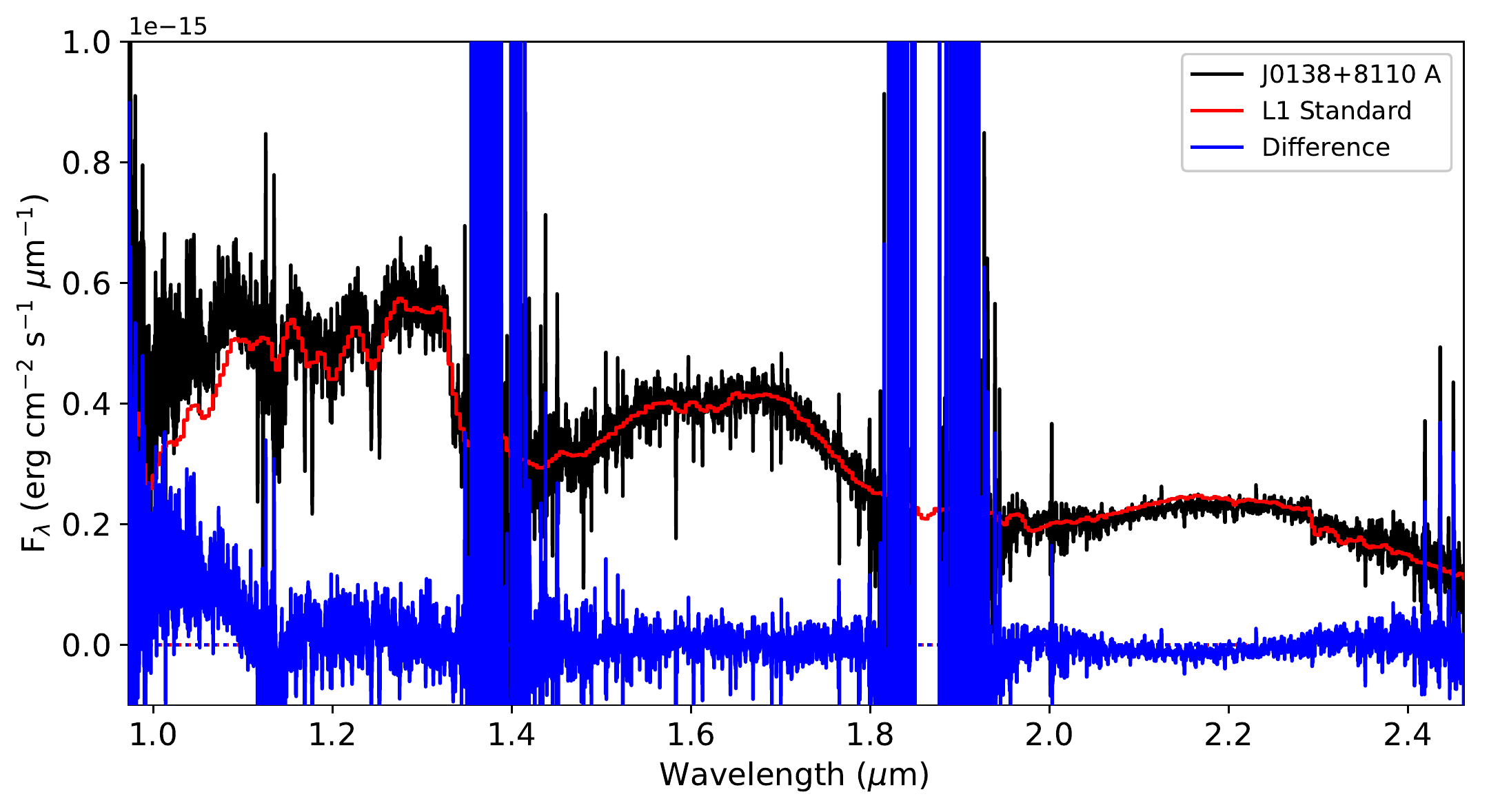}
\includegraphics[width=0.48\textwidth]{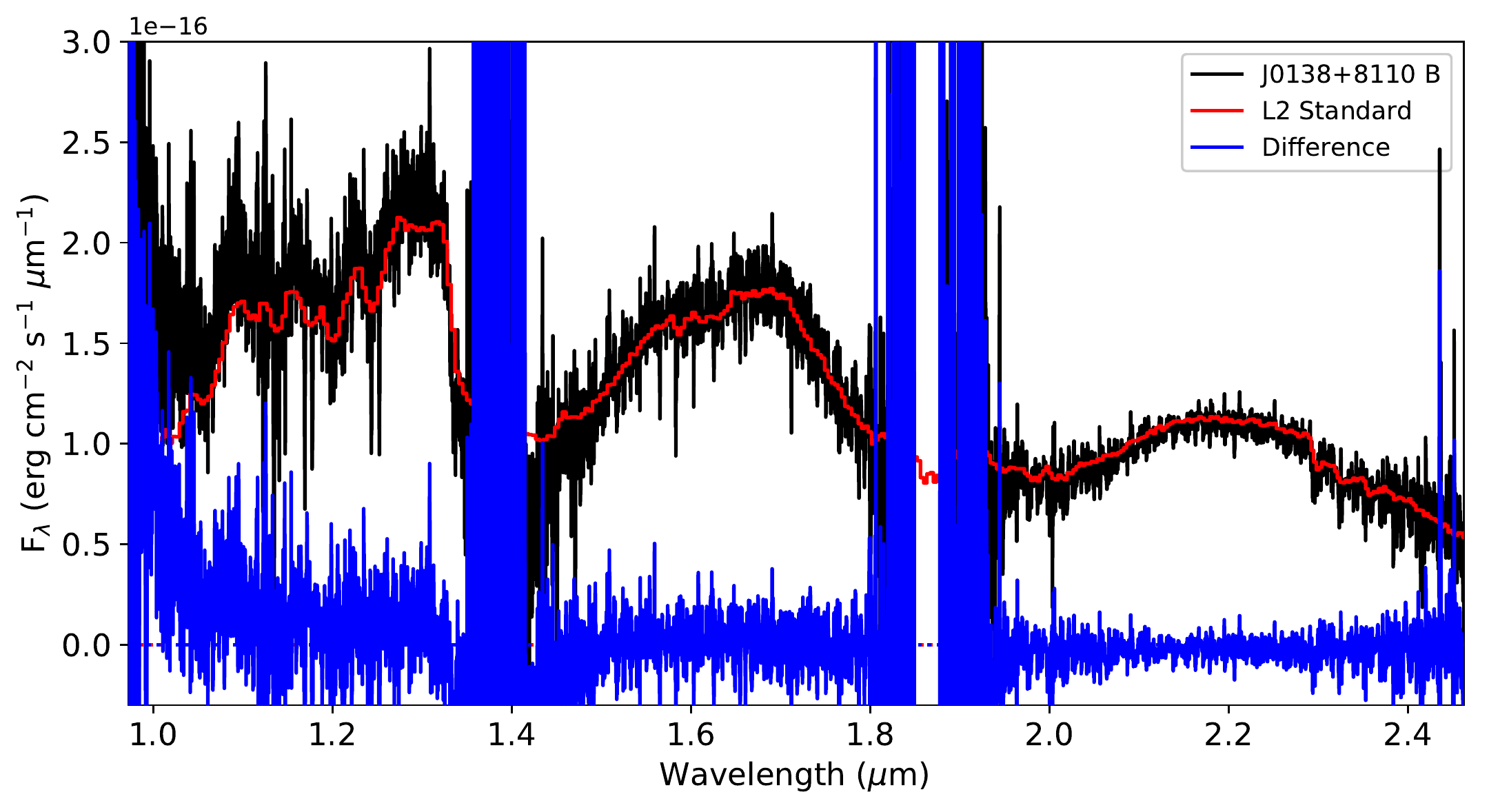}
\includegraphics[width=0.48\textwidth]{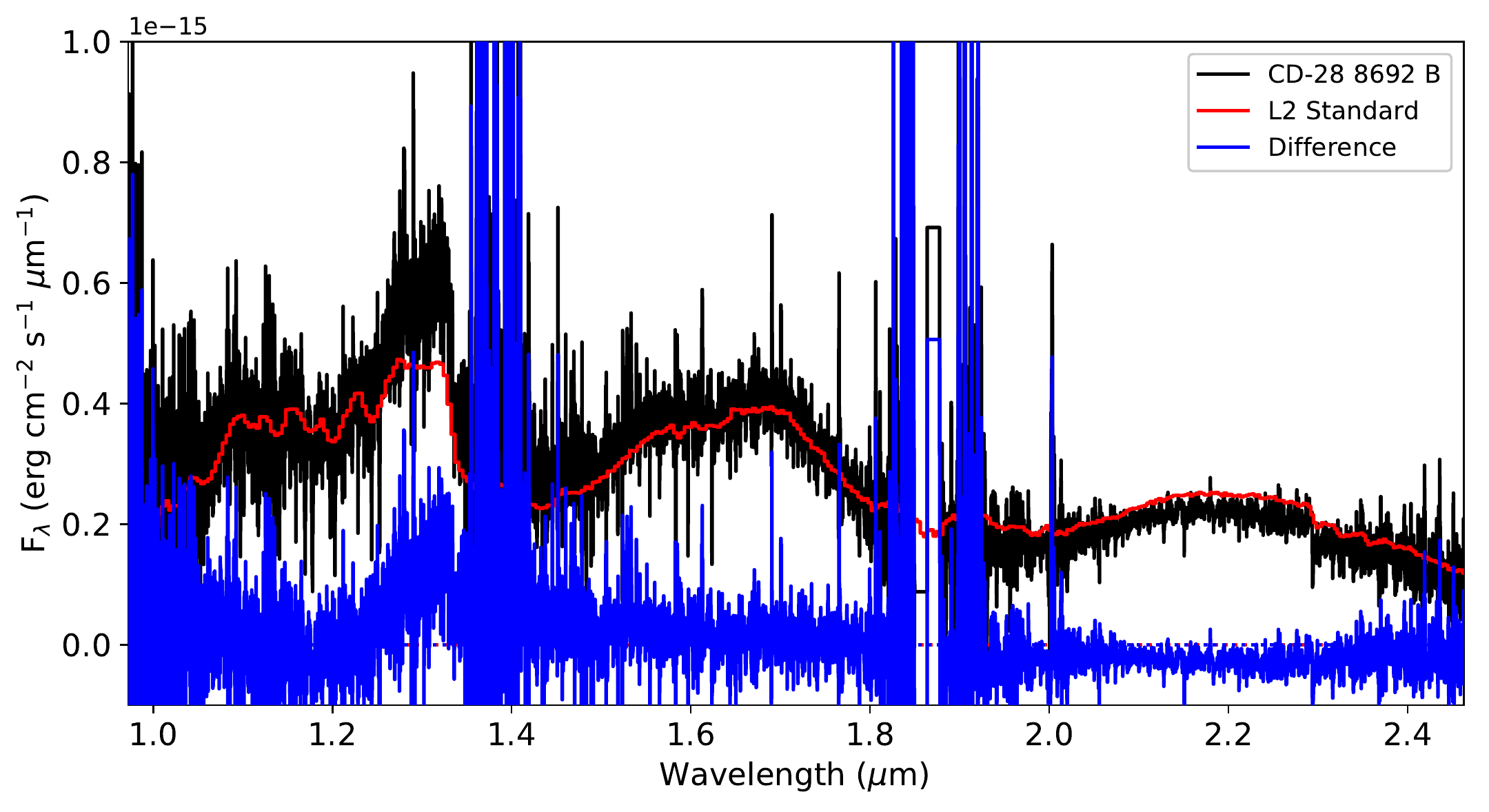}
\includegraphics[width=0.48\textwidth]{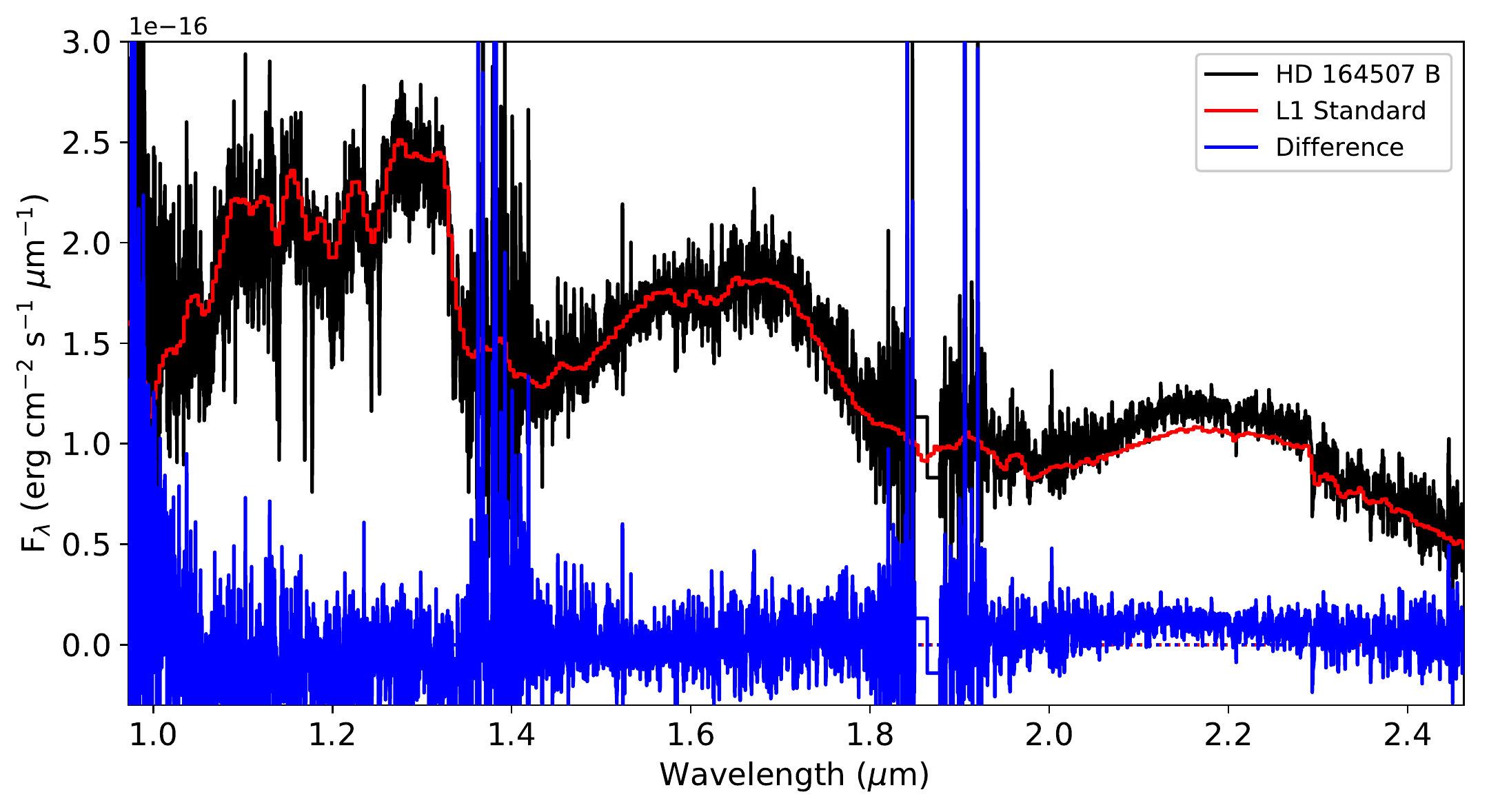}
\caption{Spectral classification of the four UCD companions to GK stars observed with TripleSpec. In each panel we show the target spectrum (black), the best-fit template from the SpeX Prism library (red) and the difference between the two (blue). Spectral typing is done with SPLAT \citep{2016AAS...22743408B}. \label{fig:spectra}}
\end{figure}

\begin{figure}
\includegraphics[width=0.48\textwidth]{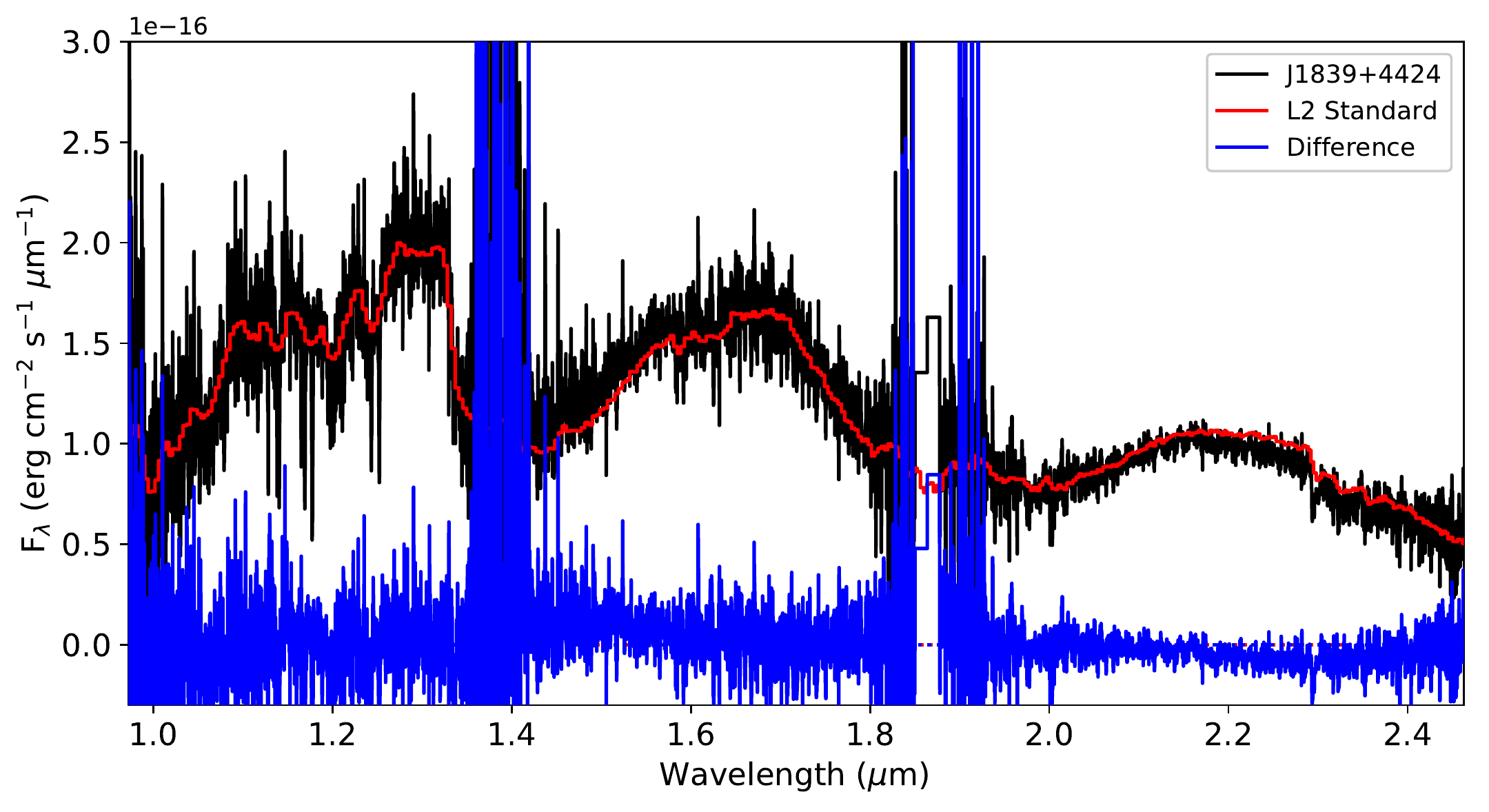}
\includegraphics[width=0.48\textwidth]{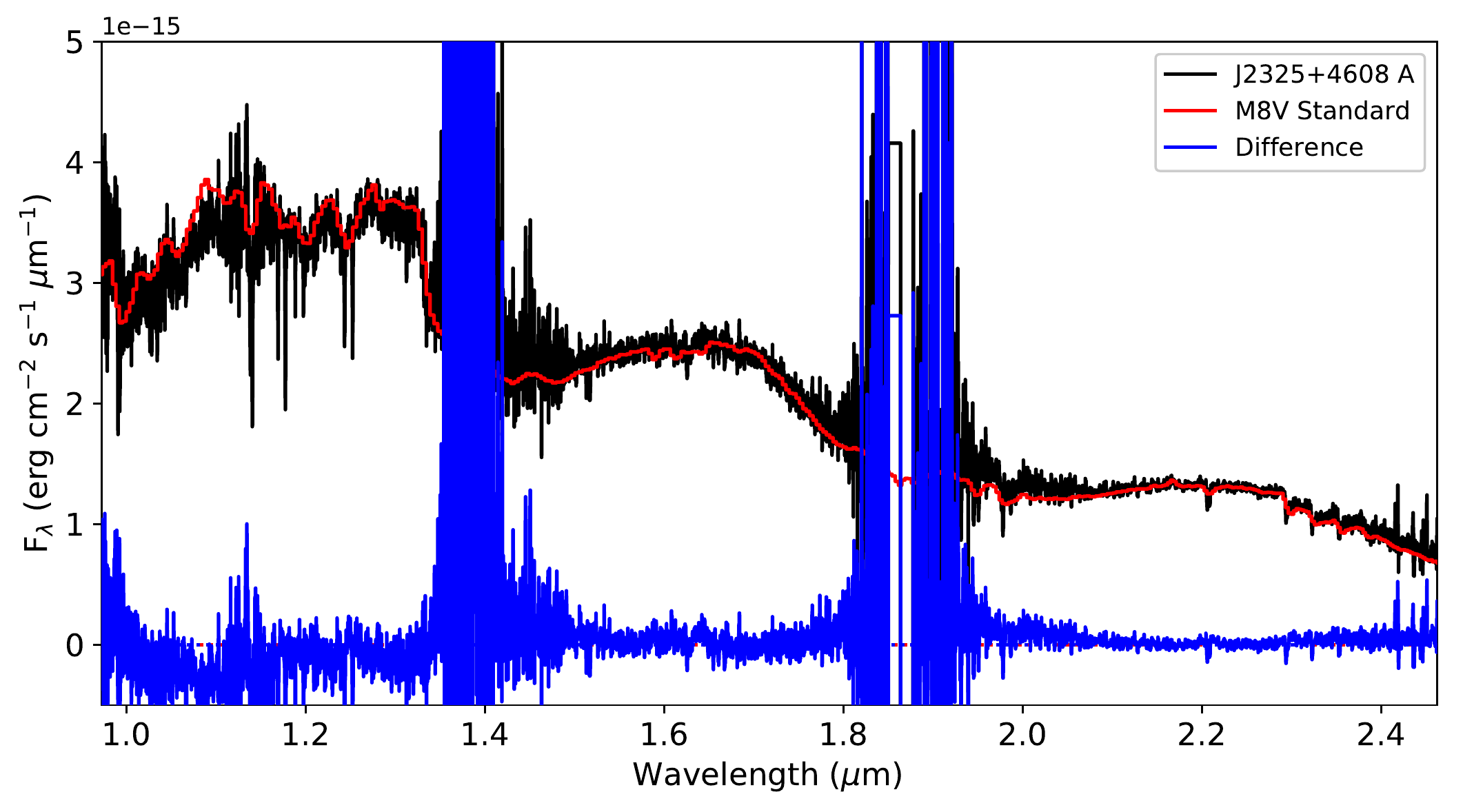}
\includegraphics[width=0.48\textwidth]{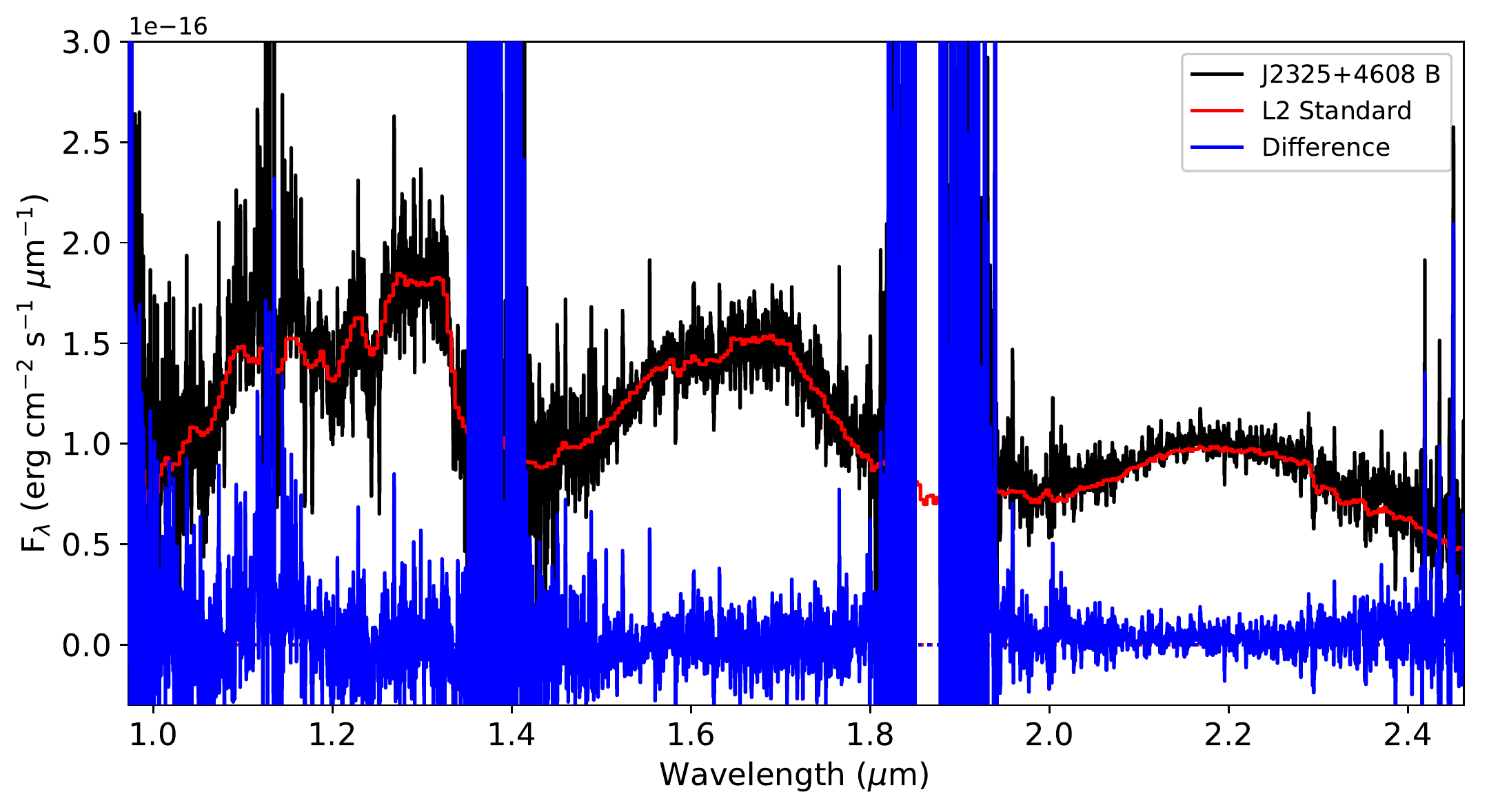}
\includegraphics[width=0.48\textwidth]{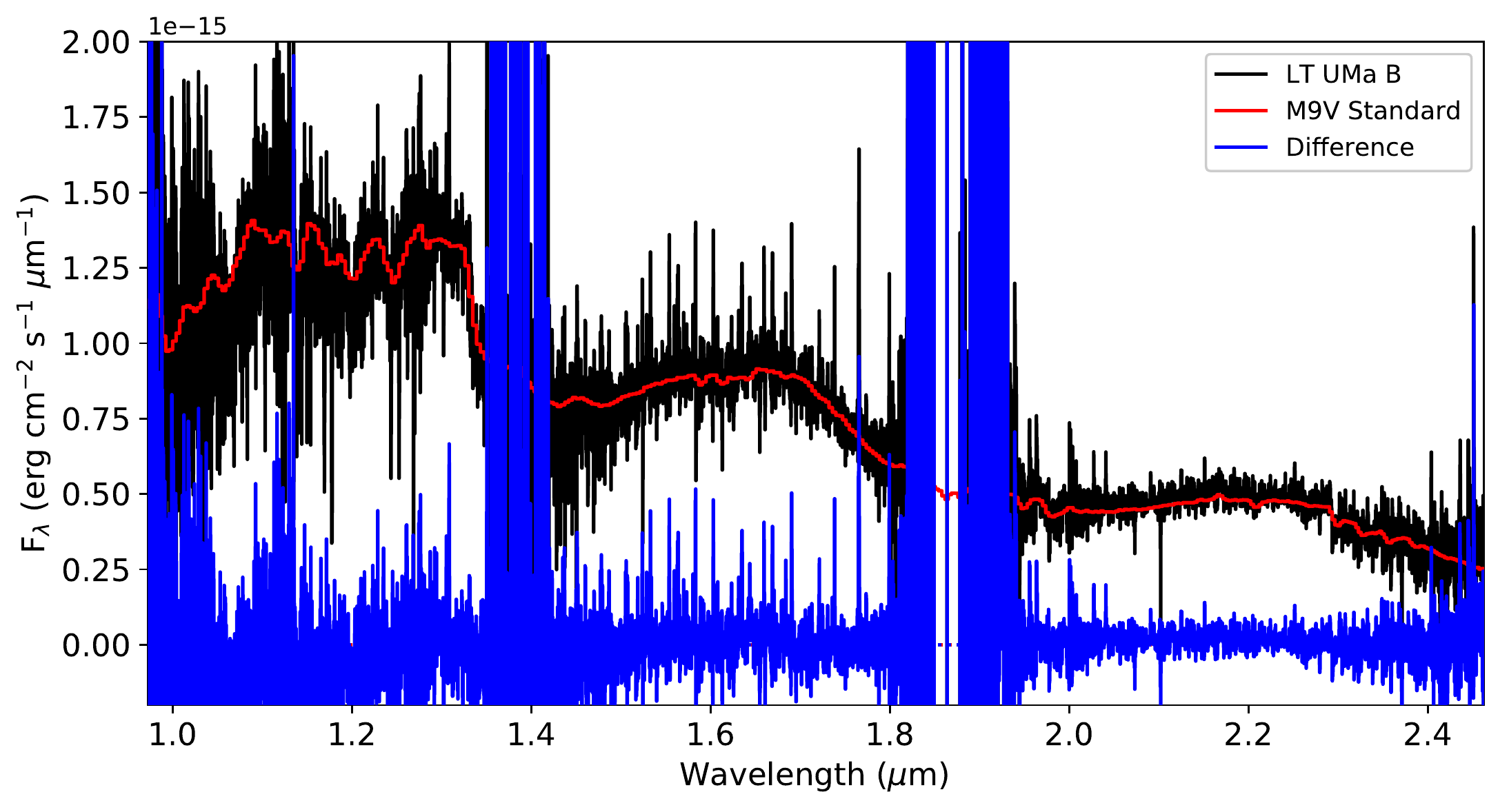}
\caption{Same as Figure~\ref{fig:spectra}, but for 2MASS~J18392740+4424510, 2MASS~J23253550+4608163, 2MASS~J23253519+4608098, and LT UMa B. \label{fig:spectra2}}
\end{figure}

The data were reduced using a modified version of the \textsc{IDL} package {\tt Spextool} \citep{2004PASP..116..362C}. The program applies basic calibration (dark subtraction and flat fielding), then pair-wise subtracts the images to remove sky background. The individual orders of the echelle spectra are traced and extracted, and wavelength calibration is achieved using the numerous OH sky lines. The individual orders are corrected for telluric absorption and flux calibrated using the observed telluric standard star, chosen 
to match the Vega spectrum used as template in {\tt Spextool}. The individual orders are then merged, using their overlap to determine flux adjustments when needed. The reduced spectra are presented in Figures~\ref{fig:spectra}, \ref{fig:spectra2}, and~\ref{fig:v478Lyr}.

We assigned a spectral type to our targets via standard template-matching using the {\tt classifyByStandard} routine in the \textsc{Python} package SPLAT\footnote{\url{http://pono.ucsd.edu/~adam/browndwarfs/splat}} \citep{2016AAS...22743408B}. The code interpolates the templates to the same wavelength grid of the observed spectra, and then minimises the $\chi^2$ of the fit, treating the scaling between the flux-calibrated target and the normalised templates as a free parameter of the fit. The {\tt classifyByStandard} routine offers the possibility to classify objects by fitting the full spectrum, as well as by fitting only the \textit{J} band, following the prescriptions of \citet{2010ApJS..190..100K}. The spectral types obtained with the two methods agree to within $\pm1$ subtype, with the exception of CD-28\,8692\,B and 2MASS~J0139+8110\,A. We discuss the discrepancies and our adopted classification in Section~\ref{sec:cd288692} and \ref{sec:ll}. We used the standard M, L, and T templates defined in \citet{2006ApJ...637.1067B} and \citet{2010ApJS..190..100K}, except in the case of V478 Lyr\,C, where standard templates gave poor fits. Further details on the spectral typing for this source are given in Section~\ref{sec:v478Lyr}. The results from template matching are presented in Figures~\ref{fig:spectra}, \ref{fig:spectra2}, and~\ref{fig:v478Lyr}, and the assigned spectral types are listed in Tables~\ref{tab:systems1} and \ref{tab:systems2}.

\section{Notes on individual systems}
\label{sec:new}

\subsection{HD 164507 AB}
\label{sec:hd164507}
The primary is a very well characterised G5\,IV star that is included in the catalogue of radial velocity standards for \G\ \citep{2013A&A...552A..64S}. Several independent estimates of the atmospheric and evolutionary parameters for this sub-giant can be found in the literature, and here we briefly summarise those based on high-resolution spectroscopy only. 

\citet{2005ApJS..159..141V} obtained $R\sim $70,000 spectroscopy for HD 164507 using the High-Resolution Echelle Spectrometer (HIRES) on the 10 m telescope at Keck Observatory \citep{1994SPIE.2198..362V}. They derived atmospheric parameters using version 2.1 of the software package Spectroscopy Made Easy \citep[SME; ][]{1996A&AS..118..595V} and the atmospheric models by \citet{1992IAUS..149..225K}. Mass and age for the star were then derived using the Y$^2$ isochrones \citep{2004ApJS..155..667D}. \citet{2007ApJS..168..297T} and \citet{2013A&A...554A..84M} derived independent age and mass using the atmospheric parameters from \citet{2005ApJS..159..141V}. \citet{2007ApJS..168..297T} employed the Yale Rotational Evolution Code (YREC) in its non-rotating mode \citep{2008Ap&SS.316...31D} to generate their set of isochrones, while \citet{2013A&A...554A..84M} used the \citet{2005ApJS..159..141V} spectroscopic $T_{\rm eff}$ and metallicity together with {\it Hipparcos} data as inputs for PARAM\footnote{\url{http://stev.oapd.inaf.it/cgi-bin/param_1.1}} \citep{2006A&A...458..609D} to derive age and mass for HD 164507. 

\citet{2015A&A...574A..50J} used high-resolution spectroscopy from SOPHIE on the 1.93 m telescope at the Observatoire de Haute-Provence \citep{2008SPIE.7014E..0JP}. The fundamental stellar parameters ($T_{\rm eff}$, $\log{g}$, [Fe/H], $\xi_{\rm t}$) were computed homogeneously using the FUNDPAR code \citep{2011RMxAA..47....3S}. The chemical abundances of 14 elements (Na, Mg, Al, Si, Ca, Sc, Ti, V, Cr, Mn, Co, Ni, Zn, and Ba) were obtained using the 2009 version of the MOOG\footnote{\url{https://www.as.utexas.edu/~chris/moog.html}} code \citep{1973ApJ...184..839S}. Rotational velocities were derived from the full width at half maximum of isolated Fe lines. Again, mass and age were derived using PARAM.

\citet{2016A&A...585A..73N} used the High Resolution Spectrograph \citep{1998SPIE.3355..387T} on the Hobby-Eberly Telescope. The $T_{\rm eff}$, $\log{g}$, $\xi_{\rm t}$, and [Fe/H] were obtained from the measured equivalent width of neutral and ionised iron absorption lines, with the TGVIT code \citep{2002PASJ...54..451T,2005PASJ...57...27T}. The stellar mass and age were determined using a Bayesian method described in \citet{2016A&A...587A.119A}, with theoretical stellar models from \citet{2012MNRAS.427..127B}. \citet{2018A&A...615A..31D} updated the age and mass derived by \citet{2016A&A...585A..73N} using the \emph{Gaia} DR2 parallax. 

\citet{2017AJ....153...21L} used spectra from The McDonald Observatory 2.1 m Telescope and Sandiford Cassegrain Echelle Spectrograph \citep{1993PASP..105..881M}. Abundances and $\xi_{\rm t}$ were calculated using measured equivalent widths and plane-parallel MARCS model atmospheres \citep{2008A&A...486..951G}, while $T_{\rm eff}$ and $\log{g}$ were computed from broad-band photometry and the photometric calibration of \citet{2010A&A...512A..54C}. Finally, \citet{2017AJ....153...21L} determined mass and age using various sets of isochrones from \citet{1994A&AS..106..275B}, \citet{2004ApJS..155..667D}, \citet{2008ApJS..178...89D}, and the 2016 version of the BaSTI isochrones \citep{2004ApJ...612..168P}.

Finally, \DR\ quotes $T_{\rm eff}=5560^{+115}_{-62}$~K \citep[see][for details on how \DR\ atmospheric parameters are derived]{2018A&A...616A...8A}, and the best-fit template used for radial velocity measurement has $T_{\rm eff}=5500$~K, log~g=3.5, and [Fe/H]=+0.2 \citep{2018A&A...616A...6S}, all in good agreement with the literature values. 

The atmospheric parameters discussed above are listed in Table~\ref{tab:hd164507}. The values derived are in general agreement with each other, and in particular point towards a slightly super solar metallicity ([Fe/H]=0.03--0.19 dex), and an age for the system in the range $3.0-5.9$~Gyr. 

More accurate age constraints on this star will be provided by {\it TESS} \citep{2015JATIS...1a4003R} via gyrochronology, making this system an exquisite benchmark for UCD models and retrieval codes testing \citep{2015ApJ...807..183L,2017MNRAS.470.1177B, 2013MNRAS.433..457B}.

The L1 companion, HD~164507~B, is an outlier in the colour-magnitude diagram of Figure~\ref{fig:hr}. With a $G-G_{RP}$ colour of $2.028\pm0.067$ mag, it is among the reddest UCDs in the \G\ sample. Objects with similar $G-G_{RP}$ colour are found in GUCDS~II to be either tight binaries or suspect tight binaries. The red $G-G_{RP}$ colour in this case would be due to the fact that $G_{RP}$ (and $G_{BP}$) magnitudes are determined by integrating the $G_{RP}$ fluxes in a $3.5\times2.1$ arcsec$^2$ window, and there is currently no treatment of multiple sources in the same window in \DR\ \citep{2018A&A...616A...4E}. Therefore, an excess in $G_{RP}$ for close binary systems is expected. However, there is no evidence for binarity of HD~164507~B. The source is not resolved by \G, and the goodness-of-fit and astrometric excess noise reported in \DR\ (2.7359 and 2.108~mas, respectively) are both consistent with the mean values for UCDs found in GUCDS~II ($5.2\pm2.6$ and $2.2\pm1.2$~mas, respectively). The primary has higher-than-solar metallicity ($0.03\,<\,{\rm [Fe/H]}\,<\, 0.19$ dex, see Section~\ref{sec:hd164507}), and higher metallicity UCDs are expected to have redder than average colours because of the enhanced dust content in their photosphere \citep[e.g.][]{2008ApJ...686..528L,2014MNRAS.439..372M}. However the near-infrared spectrum of HD~164507~B does not show obvious peculiarities (see Figure~\ref{fig:spectra}). Finally, youth is also typically associated with redder-than-usual colours \citep[see e.g.][]{2016ApJS..225...10F}, but young and suspected young objects in GUCDS~II form a relatively tight sequence with $1.6 \lesssim$ $G-G_{RP}$ $\lesssim 1.8$~mag, and the age of the system rules out youth as a cause. Optical spectroscopy for this UCD is desirable to shed light on its nature.

We derive $T_{\rm eff}$ for the companion using the \citet{2015ApJ...810..158F} spectral type to $T_{\rm eff}$ polynomial relation, and obtain $T_{\rm eff} = 2100 \pm 29$\,K. Linear interpolation of the BT-Settl isochrones\footnote{\url{https://phoenix.ens-lyon.fr/Grids/BT-Settl/CIFIST2011/ISOCHRONES/}} for solar and super-solar metallicity ([Fe/H] = +0.5 dex) in the age range 3.0--5.9\,Gyr, and for $T_{\rm eff} = 2100 \pm 29$\,K, gives a mass for the companion in the range 50--77\,$M_{\rm Jup}$, at or below the hydrogen burning limit.

To compute the bolometric luminosity ($L_{\rm bol}$) we need to determine a bolometric correction, since our TripleSpec spectrum only covers the $1.0 < \lambda < 2.4\,\mu$m range. We did this by fitting the TripleSpec spectrum with the BT-Settl atmospheric models \citep{2012RSPTA.370.2765A} with the fitting technique developed by \citet{2008ApJ...678.1372C}. The models cover the $T_{\rm eff}$ space in steps of 50\,K, the $\log{g}$ space in steps of 0.5\,dex, and the [Fe/H] space in steps of 0.5\,dex. 

We flux calibrated the target's spectrum using the measured 2MASS~$J$-band magnitude, and then allowed the scaling factor between the flux-calibrated spectrum and the models to be a parameter of the fit. The best fit scaling factor gave us a measurement of the radius ($R$) of the target via the simple geometric dilution factor $( R / d )^2$. We restricted the range of models to be considered for fitting to the $\pm\,200$\,K range around the predicted $T_{\rm eff}$ of 2100\,K and the metallicity to be within $\pm$0.5\,dex of the metallicity of the primary, for which we chose the mid point of the values quoted in the literature, i.e. 0.11\,dex.

We used the scaled best-fit atmospheric model to complete the TripleSpec spectrum at long and short wavelength ($\lambda < 1 \mu$m and $\lambda > 2.4 \mu$m). $L_{\rm bol}$ was then computed by summing the flux density over the full model+TripleSpec spectrum, and multiplying it by $4\pi\,d^2$. The uncertainty on $L_{\rm bol}$ was computed by propagating the uncertainty on the measured spectrum, as well as the uncertainty on the 2MASS magnitude used for flux calibration, and the uncertainty on the distance.

The best fit model for HD~164507~B has $T_{\rm eff} = 2300$\,K, $\log{g} = 5.0$, and [Fe/H] = +0.5\,dex. The radius corresponding to the best fit scale factor is 0.88\,$R_{\rm Jup}$, and the bolometric luminosity is 
$\log_{10}\left(L_{\rm bol}/L_\odot\right)=-3.144^{+0.039}_{-0.043}$. Approximately 17$\%$ of the bolometric luminosity reported here is outside of the TripleSpec wavelength range (1.0--2.4\,$\mu$m). This fraction decreases with spectral type, as the contribution from the optical portion of the spectral energy distribution collapses, while the longer wavelength flux does not increase significantly. The model-dependent fraction of $L_{\rm bol}$ approaches $\sim40\%$ for the late-Ms in our sample, and decreases down to $\sim8\%$ for the L2s. The best fit model for HD~164507\,B is shown in Figure~\ref{fig:model1}. The overall fit is poor: (i) the model has a triangular $H$ band, while the target has a much flatter $H$-band spectrum; (ii) the alkali lines in the $J$ band are much too shallow in the model compared to the observed ones; (iii) the $K$-band spectrum in the model is too flat, and (iv) the overall spectrum is too blue compared to our target. The best fit $T_{\rm eff}$ is 200\,K warmer than the prediction from the \citet{2015ApJ...810..158F} polynomial. 

\begin{table*}
\centering
\caption{Summary of atmospheric and evolutionary parameters for HD 164507 A. \label{tab:hd164507}}
	\begin{tabular}{l c c c c c c c c}
	\hline
    Ref.    & $T_{\rm eff}$ & $\log{g}$ & [Fe/H] & $\xi_{\rm t}$ & Instrument & $v \sin{i}$ & Age & Mass \\
	 & (K) & (cm s$^{-2}$) & (dex) & (km s$^{-1}$) & & (km s$^{-1}$) & (Gyr) & ($M_\odot$) \\
    \hline
    1 & $5650 \pm 40$ & $3.93 \pm 0.06$ & $0.19 \pm 0.03$ & \ldots & HIRES & $2.9 \pm 0.5$ & $4.2^{+1.7}_{-1.2}$ & $1.36^{0.17}_{-0.15}$ \\
    2 & \ldots & $3.83^{+0.03}_{-0.02}$ & \ldots & \ldots & \ldots & \ldots & $4.04^{+0.2}_{-0.4}$ & $1.328^{+0.048}_{-0.018}$ \\ 
    3 & \ldots & $3.78 \pm 0.02$ & \ldots & \ldots & \ldots & \ldots & $3.67 \pm 0.13$ & $1.32 \pm 0.02$ \\
    4 & $5580 \pm 20$ & $3.98 \pm 0.01$ & $0.12 \pm 0.02$ & $1.08 \pm 0.01$ & SOPHIE &	$1.02 \pm 0.23$ & $3.55 \pm 0.19$ & $1.33 \pm 0.03$ \\
    5,6 & $5534 \pm 5$ & $3.66 \pm 0.02$ & $0.13 \pm 0.01$ & $1.04 \pm 0.04$ & HRS & $2.28 \pm 0.64$ & $3.162^{+0.015}_{-0.014}$ & $1.440 \pm 0.004$ \\
    7 & $5540 \pm 60$ & $3.72$ & $0.03 \pm 0.07$ & $1.54$ & Sandiford & $5.0$ & $3.20-5.32$ & $1.17-1.37$ \\
    8,9 & $5560^{+110}_{-60}$ & 3.5 & 0.2 & \ldots & \G\ & \ldots & \ldots & \ldots \\
    \hline
	\end{tabular}
	\\ References: 1-\citet{2005ApJS..159..141V}; 2-\citet{2007ApJS..168..297T}; 3-\citet{2013A&A...554A..84M}; 4-\citet{2015A&A...574A..50J}; 5-\citet{2016A&A...585A..73N}; 6-\citet{2018A&A...615A..31D}; 7-\citet{2017AJ....153...21L}; 8-\citet{2018A&A...616A...8A}; 9-\citet{2018A&A...616A...6S}.
\end{table*}

\begin{table*}
\centering
\caption{Spectral indices and pseudo-equivalent widths for HD~164507\,B,  V478 Lyr C, CD-28 8692\,B, and LT UMa B. \label{tab:indices}}
\begin{tabular}{l c c c c c}
\hline
Index & HD~164507\,B & V478 Lyr\,C & CD-28 8692\,B & LT UMa\,B & Ref. \\
\hline
H$_2$O   & 1.20$\pm$0.01 & $1.23 \pm 0.01$ & $1.28 \pm 0.03$ & 1.16$\pm$0.02 & 1\\
H$_2$OD  & 0.915$\pm$0.008 & $0.958 \pm 0.006$ & $0.88 \pm 0.02$ & 1.07$\pm$0.02 & 2 \\
H$_2$O-1 & 0.626$\pm$0.004 & $0.648 \pm 0.006$ & $0.63 \pm 0.01$ & 0.71$\pm$0.01 & 3 \\
H$_2$O-2 & 0.850$\pm$0.009 & $0.841 \pm 0.006$ & $0.84 \pm 0.02$ & 0.94$\pm$0.02 & 3 \\
FeH$_z$  & 1.12$\pm$0.07 & $1.34 \pm 0.07$ & $1.2 \pm 0.2$ & 1.2$\pm$0.1 & 4 \\
FeH$_J$  & 1.16$\pm$0.02 & $1.22 \pm 0.02$ & $1.23 \pm 0.07$ & 1.22$\pm$0.04 & 4 \\
VO$_z$   & 1.18$\pm$0.01 & $1.13 \pm 0.02$ & $1.17 \pm 0.03$ & 1.10$\pm$0.02 & 4 \\
KI$_J$   & 1.158$\pm$0.006 & $1.157 \pm 0.008$ & $1.19 \pm 0.02$ & 1.12$\pm$0.01 & 4 \\
$H$-cont & 0.931$\pm$0.005 & $0.910 \pm 0.005$ & $0.92 \pm 0.02$ & 0.90$\pm$0.01 & 4 \\
H$_2$O-$J$ & 0.882$\pm$0.005 & $0.999 \pm 0.006$ & $0.66 \pm 0.01$ & 0.93$\pm$0.02 & 5 \\
H$_2$O-$H$ & 0.851$\pm$0.004 & $0.812 \pm 0.004$ & $0.79 \pm 0.01$ & 0.88$\pm$0.01 & 5 \\
H$_2$O-$K$ & 1.008$\pm$0.006 & $1.036 \pm 0.005$ & $0.98 \pm 0.01$ & 1.14$\pm$0.01 & 5 \\
CH$_4$-$J$ & 0.853$\pm$0.004 & $0.875 \pm 0.005$ & $0.99 \pm 0.01$ & 0.91$\pm$0.01 & 5 \\
CH$_4$-$H$ & 1.048$\pm$0.004 & $1.142 \pm 0.004$ & $1.05 \pm 0.01$ & 1.020$\pm$0.009 & 5 \\
CH$_4$-$K$ & 1.018$\pm$0.005 & $1.036 \pm 0.003$ & $1.042 \pm 0.009$ & 1.046$\pm$0.009 & 5 \\
$K/J$      & 0.456$\pm$0.002 & $0.492 \pm 0.002$ & $0.380 \pm 0.004$ & 0.356$\pm$0.004 & 5 \\
$H$-dip    & 0.484$\pm$0.002 & $0.502 \pm 0.002$ & $0.486 \pm 0.006$ & 0.487$\pm$0.004 & 6 \\
\hline
Line & HD~164507\,B & V478 Lyr\,C & CD-28 8692\,B & LT UMa\,B & Ref. \\
 & (\AA) & (\AA) & (\AA) & (\AA) & \\
\hline
\ion{Na}{i} $1.138\,\mu$m & 10.3$\pm$0.3 & $9.6 \pm 0.3$ & $8.6 \pm 0.7$  & 7.7$\pm$0.6 & 4 \\
\ion{K}{i} $1.169\,\mu$m & 6.0$\pm$0.3 & $5.5 \pm 0.3$ & $8.0 \pm 0.7$  & 2.8$\pm$0.5 & 4 \\
\ion{K}{i} $1.177\,\mu$m & 6.9$\pm$0.2 & $8.2 \pm 0.3$ & $11.0 \pm 0.6$ & 3.9$\pm$0.4 & 4 \\
\ion{K}{i} $1.244\,\mu$m & 5.3$\pm$0.2 & $5.0 \pm 0.3$ & $6.9 \pm 0.7$  & 5.4$\pm$0.5 & 4 \\ 
\ion{K}{i} $1.253\,\mu$m & 5.0$\pm$0.2 & $4.9 \pm 0.2$ & $6.7 \pm 0.6$  & 4.4$\pm$0.5 & 4 \\
\hline
\end{tabular} \\
References: 1-\citet{2007ApJ...657..511A}; 2-\citet{2003ApJ...596..561M}; 3-\citet{2004ApJ...610.1045S}; 4-\citet{2013ApJ...772...79A}; 5-\citet{2006ApJ...637.1067B} 6-\citet{2010ApJ...710.1142B}.
\end{table*}

\subsection{V478 Lyr ABC}
\label{sec:v478Lyr}
The primary is a chromospherically active G8\,V single-lined spectroscopic binary with a period of about 2.13\,d \citep{1988AJ.....95..215F}. This star was found to have strong ultraviolet emission features and a filled-in H$\alpha$ absorption line that is variable in strength. Therefore, \citet{1988AJ.....95..215F} classified it as an early-type BY Draconis system. The secondary had its mass estimated to be about 0.3 $M_\odot$ and to be probably an M2--M3 dwarf. The inclination of the system was measured to be $67 \pm 12\deg$. The lithium abundance of the G8 dwarf, estimated from the equivalent width of the \ion{Li}{i} 6707.8 \AA\ line (47 m\AA), led \citet{1988AJ.....95..215F} to propose an age for the system that is somewhat less than that of the Hyades cluster \citep[680 Myr;][]{2018ApJ...863...67G}.

Using the BANYAN $\Sigma$ on-line tool, the \DR\ astrometry, and the mean radial velocity from \citet{2004A&A...418..989N}, we find a probability of 0\% for the object to be a member of any of the young moving groups considered in BANYAN $\Sigma$ (including the Hyades).

Nevertheless, the UCD companion, dubbed V478 Lyr\,C, shows a somewhat triangular $H$-band spectrum, a feature previously associated with youth \citep{2001MNRAS.326..695L,2013ApJ...772...79A}. Gravity-sensitive spectral indices and pseudo-equivalent width defined in \citet{2013ApJ...772...79A} however lead to a L1 field surface gravity (FLD-G) classification for the companion. Intermediate surface gravity (INT-G) and very-low surface gravity (VL-G) objects in the \citet{2013ApJ...772...79A} sample have typical age $<$200 Myr and, according to a more recent study conducted by \citet{2017ApJ...838...73M}, the reliability of the gravity classification drops significantly for objects with age $>$100 Myr. On the other hand, the L1 companion to the young A3V star $\beta$ Circini has a flat $H$-band spectrum \citep[and no low gravity features, see][]{2015MNRAS.454.4476S}. The age of the $\beta$ Circini system has been estimated to be in the 370--500 Myr range. We would therefore expect the V478 Lyr system to be somewhat younger than the $\beta$ Circini system, but likely older than $\sim$100 Myr.

\begin{figure}
\includegraphics[width=0.48\textwidth]{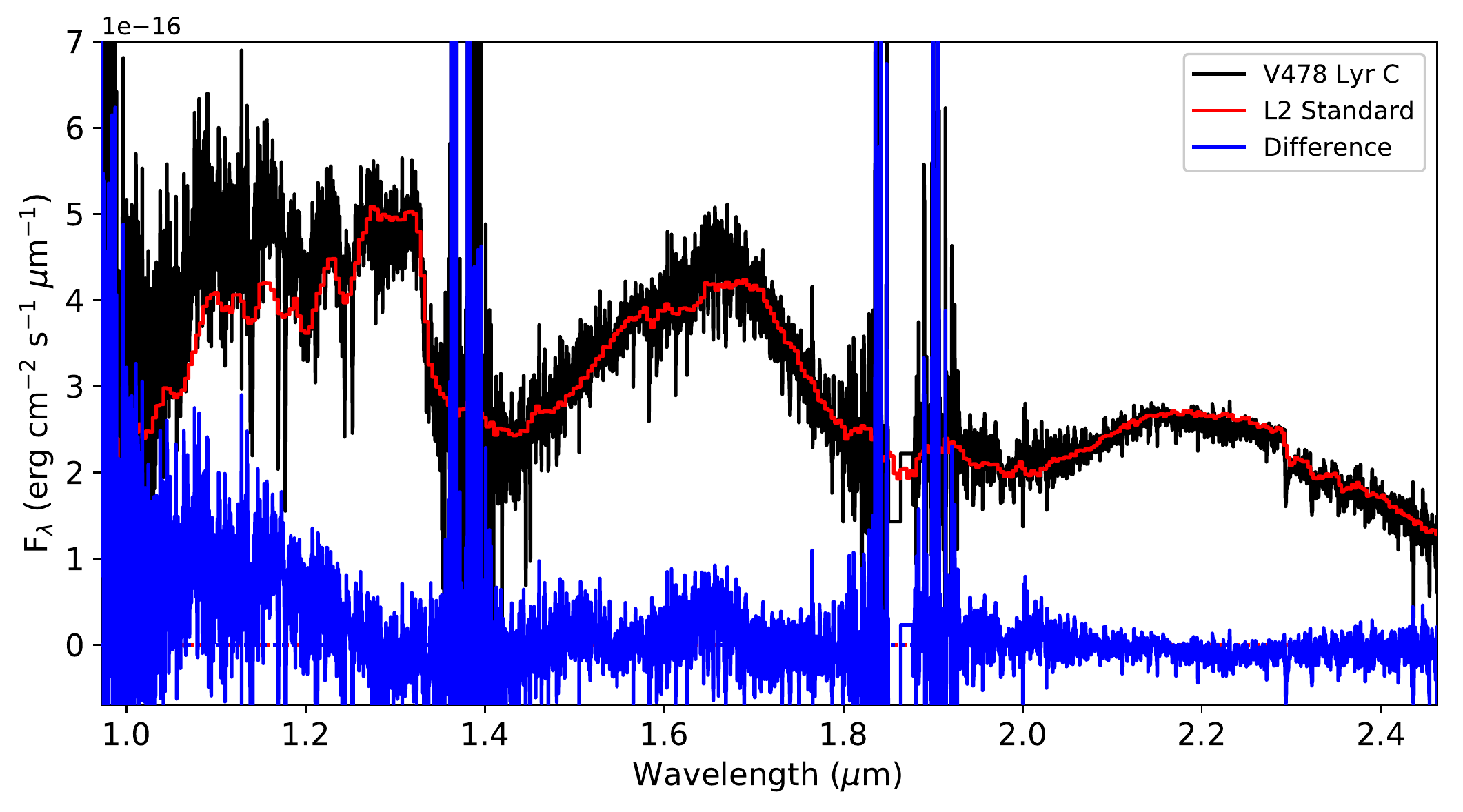}
\includegraphics[width=0.48\textwidth]{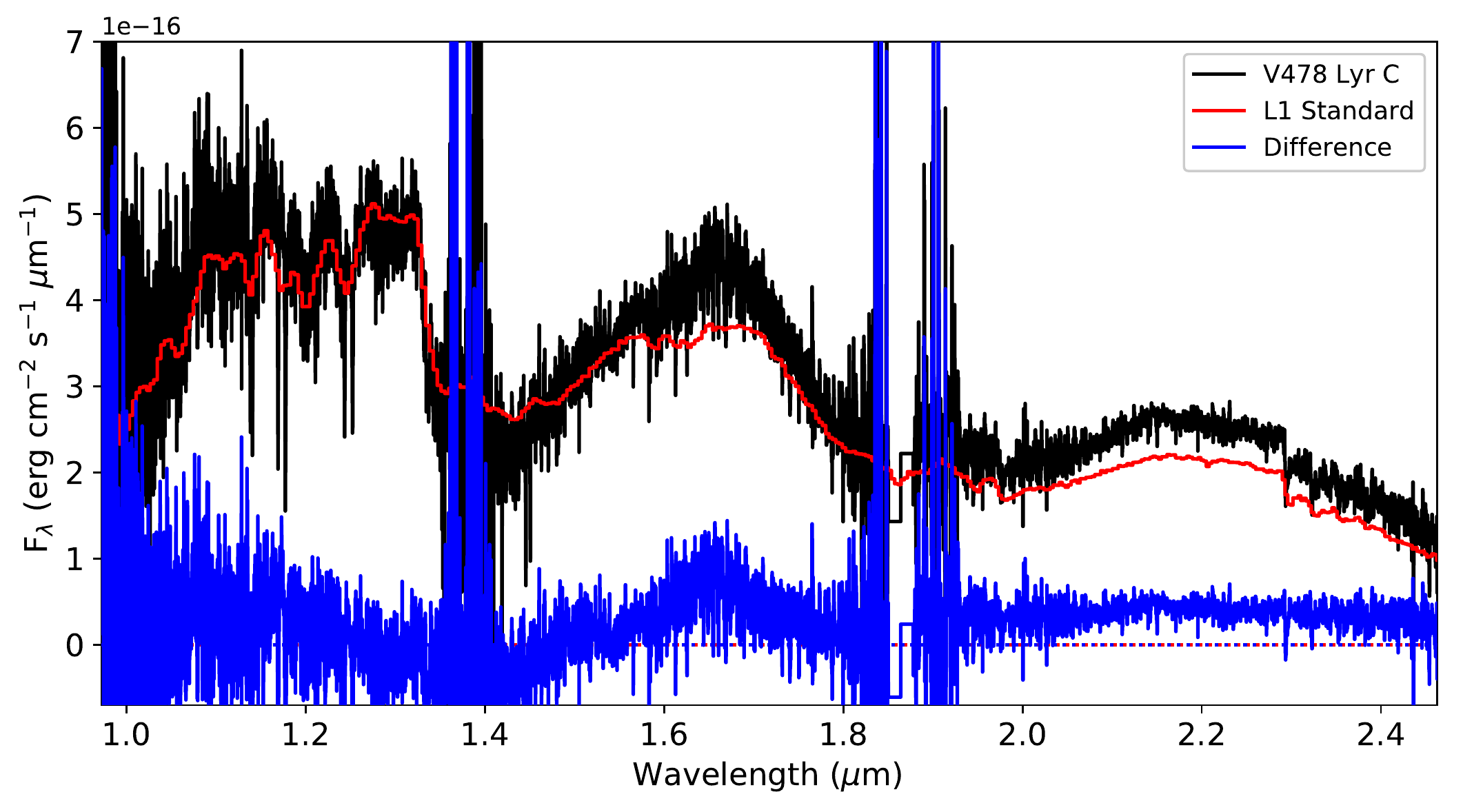}
\includegraphics[width=0.48\textwidth]{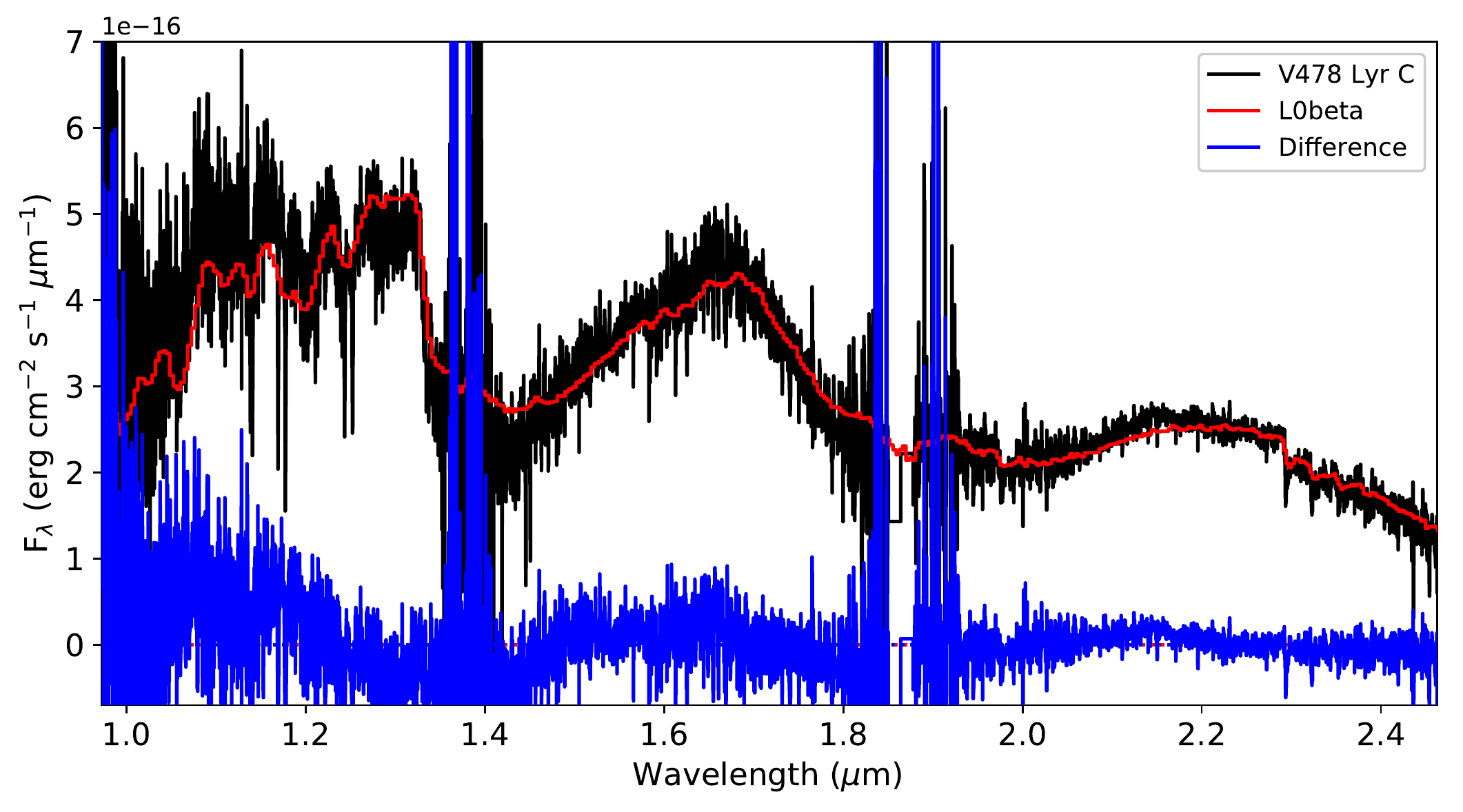}
\caption{Spectral classification for V478 Lyr C. The top panel shows the best fit standard template fitting the whole spectrum, the middle panel shows the best fit standard template using the \citet{2010ApJS..190..100K} method, while the bottom panel shows the best fit INT-G template \citep[defined in][]{2013ApJ...772...79A}. The colour-coding of spectra is the same as Figure~\ref{fig:spectra}. \label{fig:v478Lyr}}
\end{figure}

Spectral typing via standard template matching leads to an L2 type. However the fit in Figure~\ref{fig:v478Lyr} is poor, with the standard not only failing to match the $H$ band shape, but also underestimating the flux at the blue end of the spectrum (up to $\sim1.2 \mu$m). Using the \citet{2010ApJS..190..100K} method, i.e. fitting only the 0.9--1.4\,$\mu$m range, the best fit template is the L1 standard, but the target shows flux excess at the longer wavelength, as expected for a low surface gravity object. 

If we fit V478 Lyr C with the low gravity templates defined in \citet{2013ApJ...772...79A}, the best fit is the L0$\beta$ standard. The fit to the $H$ band is much more accurate, and the flux in the $J$ band is less underestimated, but at the same time the fit to the H$_2$O band at $\sim1.4\,\mu$m is poorer. Given all of the above, we assign V478 Lyr C a spectral type of L1:.

\citet{2015ApJ...810..158F} derived a $M_H$ to $T_{\rm eff}$ polynomial relation for young objects, but the available near-infrared photometry for V478\,Lyr\,C is heavily contaminated by the parent star (at $\rho\sim17$\,arcsec). We computed a synthetic $H$-band magnitude using our flux-calibrated TripleSpec spectrum and the 2MASS $H$-band response curve \citep{2003AJ....126.1090C}. We estimated the accuracy of our synthetic $H$ magnitude by comparing the synthetic magnitudes obtained for the other objects observed as part of our TripleSpec run, against their measured 2MASS $H$ (for all except HD\,164507\,B, since its photometry is also contaminated). The mean offset between our synthetic magnitudes and the measured ones is --0.007\,mag and the $1\sigma$ dispersion around the mean is 0.44\,mag. We therefore adopted 13.74$\pm$0.44\,mag as our synthetic measurement, and obtain $T_{\rm eff} = 1740 \pm 130$\,K for V478\,Lyr\,C as a result. Linear interpolation of the BT-Settl isochrones for solar metallicity in the age range $0.10-0.37$\,Gyr gives a mass for this object in the range 10--28\,$M_{\rm Jup}$, straddling the deuterium fusion mass limit.

We determined $L_{\rm bol}$ for V478\,Lyr\,C following the same procedure described in Section~\ref{sec:hd164507}. The best fit model has $T_{\rm eff} = 1800$\,K, $\log{g} = 5.0$, and solar metallicity. The $\log{g} = 5.0$ is somewhat higher than one might expect, given the age of the system, and the fact that this object shows signs of youth. The radius resulting from the best-fit scaling factor is 1.31\,$R_{\rm Jup}$ and the bolometric luminosity is 
$\log_{10}\left(L_{\rm bol}/L_\odot\right)=-3.33^{+0.26}_{-0.78}$. The best fit model can be seen in Figure~\ref{fig:model1}. The overall fit is good, with the model only slightly underpredicting the flux at the shortest wavelength ($\lambda < 1.2\,\mu$m) but that is the region of lowest signal-to-noise-ratio.

\citet{2017AJ....153..257O} found the SB1 primary to form a very wide co-moving pair with the G6V HD~171067, with a projected separation of $\sim$8 pc. The \citet{2017AJ....153..257O} analysis however did not take into account radial velocity (RV). The measured system RV for V478 Lyr AB is $-25.2 \pm 4.8$ km s$^{-1}$ \citep{2004A&A...418..989N}, and is discrepant from the RV of HD~171067 ($-46.197 \pm 0.002$ km s$^{-1}$; \citealt{2013A&A...552A..64S}). As a result, the G6V is unlikely to be associated with the V478 Lyr triple system.

V478 Lyr ABC joins the rank of triple systems consisting of a spectroscopic binary with a wide, low-mass tertiary component \citep[see][and references therein]{2012AJ....144...62A}. These systems are precious for testing formation simulations of very close separation binaries, which require a mechanism to draw angular momentum away from an already close pair of objects. One proposed mechanism is through three-body interactions with cool dwarfs \citep[see e.g.][]{2003A&A...400.1031S,2004MNRAS.351..617D,2005ApJ...623..940U}, and a key observable to test such scenario is the fraction of tight spectroscopic binaries that have a wide additional companion. Towards this goal, V478 Lyr AB was among the stars targeted by \citet{2012AJ....144...62A}, who conducted a deep near-infrared survey looking for low-mass tertiary components around 118 known spectroscopic binaries within 30~pc of the Sun. However, V478 Lyr C was missed probably because of the combination of its tight angular separation from the binary (17.05\,arcsec, close to the \citealt{2012AJ....144...62A} survey limit of 10--15\,arcsec), the large magnitude difference between SB1 primary and L dwarf companion, and the large contamination by reddened background sources resulting from its proximity to the Galactic plane ($b$ =  
10.1\,deg).

Finally, the estimated orbital period for this system is $\gtrsim$8,000 yr, despite this being the most favorable configuration among the seven systems presented here -- i.e. a relatively massive primary, with a relatively tight separation, and assuming a face-on circular orbit. If instead we assume the wide L1: companion is coplanar with the SB1, i.e. that the inclination angle is $67\pm12$\,deg, then the orbital period would be $\sim$9,700\,yr. In either case, no dynamical mass measurement is possible for the UCD. The other systems presented here have even longer estimated orbital periods. 

\subsection{CD$-$28~8692 AB}
\label{sec:cd288692}
The primary is a slightly metal poor K5\,V star. It has been monitored with HARPS for planets by \citet{2011A&A...526A..99S}, who found no evidence for RV variations. \citet{2011A&A...526A..99S} also used the HARPS spectra to determine atmospheric parameters, and obtained $T_{\rm eff} = 4799 \pm 90$~K, log~g = $4.43 \pm 0.18$, and [Fe/H] = $-0.22\pm0.06$~dex. They then estimated a mass of $0.715 \pm 0.014 M_\odot$ for the star using the measured atmospheric parameters and the Padova isochrones. \citet{2012A&A...545A..32A} used the atmospheric parameters estimated by \citet{2011A&A...526A..99S} and the HARPS spectra to measure detailed abundances of 12 chemical species, with typical precision in the $0.035-0.260$\,dex range. 

Later, \citet{2015A&A...576A..69D} used the HARPS data to estimate atmospheric parameters and combined them with the \ion{Li}{i} abundance to infer an age of 4.48~Gyr for this star. 

The \DR\ effective temperature for this star is $4742^{+138}_{-116}$~K \citep{2018A&A...616A...8A}, while the best-fit template used for radial velocity measurement has $T_{\rm eff}=4750$~K, log~g=4.5, [Fe/H]=--0.2~dex \citep{2018A&A...616A...6S}. All \DR\ values are in good agreement with the literature measurements.

The companion presented here is classified as L2, with a projected separation of 2026 au (50.91\,arcsec). The L2 template is a good fit to the spectrum of the target, with the exception of a slightly suppressed $K$ band \citep[typical of metal-poor and high surface gravity dwarfs;][]{2008ApJ...674..451B,2017MNRAS.468..261Z}, and a flux excess at $\sim1.3\,\mu$m. Scatter in the strength of the $\sim1.3\,\mu$m peak among objects of a given spectral type has been observed before \citep{2018AJ....155...34C}. The \citet{2010ApJS..190..100K} method yielded a very different classification of L6. While the L6 template does indeed provide a slightly better fit to the \textit{J} band reducing the overluminosity at $\sim1.3\mu$m, the target is much bluer than the L6 standard at longer wavelength. Low metallicity L dwarfs are indeed slightly bluer compared to their solar metallicity counterparts, but this system is only slightly metal poor, and therefore a large suppression of the \textit{H}- and \textit{K}-band flux is unlikely. Moreover, the absolute $G$ magnitude for CD-28\,8692\,B is 17.406$\pm$0.004\,mag, which is consistent with the median value for L2s \citep[17.24$\pm$0.41\,mag;][]{2019MNRAS.485.4423S}, but nearly two magnitudes overluminous compared to typical L6s \citep[19.25$\pm$0.60\,mag;][]{2019MNRAS.485.4423S}. Therefore, we retain a classification of L2 for this object. 

Somewhat counterintuitively, the spectral indices for CD-28\,8692\,B are consistent with an INT-G classification. This is unexpected, since a relatively old, metal-poor object should exhibit surface gravity typical of standard field L dwarfs, or at most slightly higher. The transition between INT-G and FLD-G however is not very sharp, and scatter around the dividing line has been previously noted \citep{2017ApJ...838...73M}. The unusual metallicity of the CD-28 8692 AB system further affects the reliability of the gravity classification, as first noticed by \citet{2016AJ....151...46A} for the M9.5 companion to the metal poor M1\,V GJ~660.1A ([Fe/H]\,=\,--0.63$\pm$0.06\,dex). We therefore conclude that our INT-G classification for CD-28~8692\,B is incorrect.

The solar metallicity BT-Settl isochrones at $T_{\rm eff} = 1960\pm29$\,K (as given by the \citealt{2015ApJ...810..158F} polynomial relations) and age = 4.48\,Gyr gives a mass of $\sim$70\,$M_{\rm Jup}$. Although the system is slightly metal poor, we cannot use the publicly available BT-Settl isochrones for low metallicity ([Fe/H] = --0.5\,dex), since they do not extend below 75\,$M_{\rm Jup}$ and $T_{\rm eff} \sim 3000$\,K. 

We determined $L_{\rm bol}$ for CD-28~8692\,B following the same procedure described in Section~\ref{sec:hd164507}. The best fit model has $T_{\rm eff} = 1800$\,K, $\log{g} = 5.0$, and solar metallicity. We determine a radius of 0.87\,$R_{\rm Jup}$, and 
$\log_{10}\left(L_{\rm bol}/L_\odot\right)=-3.688^{+0.047}_{-0.053}$. The best fit model is shown in Figure~\ref{fig:model1}. The model fit is of good quality, the main discrepancies being in the blue wing of the $H$ band (the model underpredicting the observed flux) and at $\sim$1.3\,$\mu$m, where the model does not correctly reproduce the sharp observed peak (see above).

\subsection{2MASS~J23253550+4608163 + 2MASS~J23253519+4608098}
\label{sec:2m2325}

2MASS~J23253550+4608163 is overluminous compared to objects of similar $G-G_{RP}$ colour and spectral type. Typical M8 dwarfs have $M_G = 15.24\pm0.63$ mag (see GUCDS~II), while our target has $M_G = 12.850\pm0.004$ mag\footnote{Absolute magnitudes throughout this paper are computed using $1/\varpi$ as the distance, since for all targets $\varpi/\sigma_\varpi > 10$.}. The overluminosity cannot be explained by unresolved binarity alone, since an equal-mass binary would at most be 0.75\,mag overluminous, while the target is almost 2.4\,mag overluminous. Young objects can also be redder and overluminous compared to field-age objects. However, 2MASS~J23253550+4608163 does not show any indication of youth in its near-infrared spectrum (see Figure~\ref{fig:spectra2}, middle panel) and its kinematics are inconsistent with membership to any of the young moving groups using the BANYAN~$\Sigma$ on-line tool\footnote{\url{http://www.exoplanetes.umontreal.ca/banyan/banyansigma.php}} \citep{2018ApJ...856...23G}. Contamination by a background object could be another possibility, and this source is indeed flagged as duplicate ({\tt duplicated$\_$source} = 1), however the background object would need to have the $G-G_{RP}$ colour of a late-M dwarf, since the $G-G_{RP}$ of 2MASS~J23253550+4608163 is in line with the median colour of M8 dwarfs ($1.592\pm0.005$ mag vs. $1.61\pm0.95$ mag, see GUCDS~II). External photometry from 2MASS, PanSTARRS-1 and AllWISE does not show evidence of contamination nor peculiar colours, but all absolute magnitudes are similarly overluminous when compared with M8 dwarfs.  

An indication of possible problems is the relatively large goodness-of-fit ({\tt astrometric$\_$gof$\_$al}) of 132 (cf. the mean value of $5.2\pm2.6$ for objects in GUCDS~II), which may indicate that the parallax for this source is spuriously large. 
The companion, 2MASS~J23253519+4608098, does not show any sign of peculiarity, neither photometric nor spectroscopic. This could therefore be an unfortunate case of chance alignment, with 2MASS~J23253550+4608163 being a background M dwarf whose spurious astrometry is consistent, by chance, with being a companion to 2MASS~J23253519+460898. The astrometry for 2MASS~J23253519+46089 would instead be correct. The chance of such an unfortunate alignment is however extremely low, given the tight separation of the pair on the sky (7.24''). We therefore have no conclusive explanation for the overluminosity of this object.

We determined $L_{\rm bol}$ for both components of this system following the same procedure described in Section~\ref{sec:hd164507}. The best fit model for the A component has $T_{\rm eff} = 2400$\,K, $\log{g} = 5.0$, and [Fe/H] = +0.5\,dex. The radius is 3.14\,$R_{\rm Jup}$ (c.f. model-predicted value of 2.33\,$R_{\rm Jup}$), which is unusually large for an ultra-cool dwarf, but probably a consequence of the overluminosity discussed above. The result is $\log_{10}\left(L_{\rm bol}/L_\odot\right)=-1.928^{+0.026}_{-0.027}$.  

The best fit model for the B component has $T_{\rm eff} = 1800$\,K, $\log{g} = 5.5$, and solar metallicity. The radius corresponding to the best-fit scaling factor is $R = 1.39\,R_{\rm Jup}$ which is somewhat large for an object with this temperature and surface gravity (the BT-Settl models predict $R \sim 0.9\,R_{\rm Jup}$). The bolometric luminosity is 
$\log_{10}\left(L_{\rm bol}/L_\odot\right)=-3.265^{+0.053}_{-0.060}$. The best fit models for both components are shown in Figure~\ref{fig:model2}. The fit to the spectrum of 2MASS~J23253550+4608163 is overall poor. The model appears too blue compared to the observed spectrum with the flux at $\lambda < 1.3 \mu$m being overestimated and the flux in the $K$ band being underestimated. The shape of the $H$ band is also poorly reproduced, with the model having a more pronounced peak, while the observed spectrum appears flatter. The fit to the L dwarf component, 2MASS~J23253519+4608098, is good, with the model only slightly underpredicting the flux at $\lambda < 1.25 \mu$m.

\subsection{2MASS~J01390902+8110003 + 2MASS~J01385969+8110084}
\label{sec:ll}
With a projected separation of 959\,au, this system is to our knowledge the widest L+L dwarf binary known to date. 

The primary is an L1 based on the template fitting to the whole spectrum, while a fit to the \textit{J} band alone results in a significantly earlier spectral type, M8. The discrepancy is mostly driven by the slightly overluminous blue end of the TripleSpec spectrum ($\lambda < 1.1 \mu$m, see Figure~\ref{fig:spectra}). The L1 standard gives a good fit to the overall spectrum except for this wavelength range, which is however also the lowest signal-to-noise-ratio portion of the spectrum. On the other hand, the M8 template reproduces better this part of the spectrum, but starts to diverge from the observations at wavelength longer than $\sim1.3 \mu$m, with the target being overall redder than the template. While in principle this could be evidence of youth, the morphology of the \textit{H} band, and the depth of the \ion{Na}{i} and \ion{K}{i} absorption lines suggest that the object is not particularly young. We assume a spectral type of L1 for this object in the rest of the analysis. The companion is classified as L2 by both methods. 

Various authors have focused on the identification of wide low-mass binaries. Recent examples include SLoWPoKES \citep[Sloan Low-mass Wide Pairs Of Kinematically Equivalent Stars;][]{2010AJ....139.2566D}, \citet{2015ApJ...802...37B}, and \citet{2017MNRAS.466.2983G}. Extremely wide low mass binaries do exist, with separations out to tens of thousands of au, and are found in young clusters and moving groups (see e.g. GUCDS~II, \citealt{2015A&A...583A..85A}) as well as in the field (\citealt{2012Obs...132....1C,2012Obs...132..176C,2012Obs...132..252C}, and \citealt{2010AJ....139.2566D}). These systems are rare, with an estimated fraction of wide low-mass binaries in the field of 1--2\% \citep{2009AJ....138.1563B}. Their paucity may be explained via Galactic dynamical evolution, with subsequent stellar encounters in the Galactic disk progressively increasing the separation between the low-mass binary components, eventually leading to its dissolution \citep{1987ApJ...312..367W}. This sets a hard lower limit on the binding energy \citep[see e.g.][]{2003ApJ...586..512B,2009A&A...507..251C}.

However rare, these systems pose a challenge to the formation models of low-mass stars and brown dwarfs. In particular, \citet{2010MNRAS.404.1835K,2011ASPC..451....9K} argued that systems with separation $>1000$\,au are unlikely to have been formed as primordial binaries (since their orbital separation would be comparable to the size of an embedded cluster), but instead originated during the cluster dissolution process. \citet{2010AJ....139.2566D} observed a bimodal binary separation \citep[also observed by][]{2010MNRAS.404.1835K}, suggesting the presence of two populations, one old and tightly bound, and another young and weakly bound, recently formed and unlikely to survive more than a few Gyr.

For us to determine how strongly bound this system is, we need to constrain the mass of the components. The spectra, presented in Figure~\ref{fig:spectra}, do not present any obvious peculiarity, and both give a good fit to the standard templates. We can therefore reasonably assume that these two L dwarfs are of solar metallicity, and with age $>$\,0.37\,Gyr (following the same reasoning used in Section~\ref{sec:v478Lyr}). We estimate the effective temperature for the two components using the \citet{2015ApJ...810..158F} polynomial relation, and obtain 2100$\pm$29\,K and 1960$\pm$29\,K for the L1 and L2, respectively. Given these temperatures, interpolation of the BT-Settl isochrones in the $0.37-13$\,Gyr range gives a mass of $44-82\,M_{\rm Jup}$ and $42-80\,M_{\rm Jup}$, respectively, corresponding to a total system mass in the $0.08-0.15\,M_\odot$ range. The corresponding binding energy for the pair is $3\times10^{33}<|U_g^*|<1\times10^{34}\,J$, just above the $|U_g^*|>10^{33}\,J$ limit proposed by \cite{2009A&A...507..251C}.

We can finally estimate how long the 2MASS~J01390902+8110003 + 2MASS~J01385969+8110084 system is likely to survive stellar encounters in the Galactic disk, using the method described in \citet{2010AJ....139.2566D}. Rearranging their equation 18, and assuming a lower limit on the total mass for this system of 0.08\,$M_\odot$, we find that the expected lifetime would be $>$22\,Gyr. Alternatively, we can compute the maximum separation for a binary of given total mass to remain bound for at least 10~Gyr, rearranging equation 28 from \citet{1987ApJ...312..367W} and following their assumption of an average Galactic stellar density of 0.16\,pc$^{-3}$, an average stellar mass of 0.7\,$M_\odot$, and a relative velocity for the stellar encounters of $\sim$20\,km\,s$^{-1}$. We find the maximum separation for a system of total mass $>\,0.08\,M_\odot$ to be $>\,1.5\,\times\,10^3$\,au. The system is therefore bound. 

We determined $L_{\rm bol}$ for both components of the system following the same procedure described in Section~\ref{sec:hd164507}. The best fit model for component A has $T_{\rm eff} = 1950$\,K, $\log{g} = 5.5$, and solar metallicity. We determine a radius of 2.16\,$R_{\rm Jup}$ and 
$\log_{10}\left(L_{\rm bol}/L_\odot\right)=-2.753^{+0.029}_{-0.031}$. The radius is unusually large, and inconsistent with the model-predicted radius for an object of such atmospheric properties (1.14\,$R_{\rm Jup}$).

The best fit model for the B component has $T_{\rm eff} = 1800$\,K, $\log{g} = 5.0$, and solar metallicity. The resulting radius is 1.58\,$R_{\rm Jup}$ and 
$\log_{10}\left(L_{\rm bol}/L_\odot\right)=-3.158^{+0.034}_{-0.037}$. The radius is once again unusually large, and even more inconsistent with the model-predicted radius for an object of such atmospheric properties (0.91\,$R_{\rm Jup}$).The best fit models for both components can be seen in Figure~\ref{fig:model2}. The fit to the L1 (2MASS~J01390902+8110003) is overall good, while the fit to the L2 (2MASS~J01385969+8110084) is of slightly lower quality. Main discrepacies are an overall underestimated flux in the blue wing of the H band, as well as in the K band and at $\sim 1.3 \mu$m.

\subsection{2MASS~J18392917+4424386 + 2MASS~J18392740+4424510}
This is a very wide (811\,au) M+L binary, akin to the 2MASS~J01390902+8110003 + 2MASS~J01385969+8110084 system. 

The primary is the only previously known UCD discussed in this paper, and was classified M9\,V using IRTF/SpeX spectroscopy in \citet{2014ApJ...794..143B}. The TripleSpec spectrum for the companion is presented in Figure~\ref{fig:spectra}, and does not present any obvious peculiarity. We classify it as L2 via template matching.

Following the same method described above, we estimate the $T_{\rm eff}$ for the two components to be $2400\pm29$\,K and $1960\pm29$\,K respectively, leading to masses of $49-88\,M_{\rm Jup}$ and $42-80\,M_{\rm Jup}$. The binding energy of the system is therefore $4\times10^{33}<|U_g^*|<1\times10^{34}\,J$. The expected lifetime (computed using the same procedure described in Section~\ref{sec:ll}) is $>28$\,Gyr and the separation limit $>1.6\,\times\,10^3$\,au. The system is therefore bound. 

We determined $L_{\rm bol}$ for the L dwarf following the same procedure described in Section~\ref{sec:hd164507}. The best fit model has $T_{\rm eff} = 1800$\,K, $\log{g} = 5.5$, and solar metallicity. We determine a radius of 1.19\,$R_{\rm Jup}$ and 
$\log_{10}\left(L_{\rm bol}/L_\odot\right)=-3.440^{+0.042}_{-0.046}$. The best fit model is presented in Figure~\ref{fig:model1}.

\subsection{LT UMa AB}
\label{sec:ltuma}
LT UMa is a variable star of BY Dra type, with an amplitude of 0.03\,mag (no period listed) in The International Variable Star Index\footnote{\url{https://www.aavso.org/vsx/index.php}}, based on 11 observations by \citet{2000A&AS..142..275S}.

The companion was first identified by \citet{2006MNRAS.368.1281P} based on motion and colour, but no spectroscopy was presented there. The Washington Double Star Catalog lists the pair as WDS~J08448+5532.
The spectral types are reported as ``K0\,III\,+\,L?'', following the primary classification presented in \citet{1961ApJ...134..809Y} and the companion estimated spectral type derived in \citet{2006MNRAS.368.1281P}. The primary was however reclassified as K0\,V in \citet{2000A&AS..142..275S} and \citet{2008AstL...34...17T}. 

\citet{2000A&AS..142..275S} determined the effective temperature for LT UMa using the $B$ and $V$ magnitudes taken from the Tycho catalogue \citep{1997A&A...323L..57H}, and the $B-V$ calibration from \citet{1996ApJ...469..355F} to obtain $T_{\rm eff}\,=\,5290$\,K. More recently \citet{2017AJ....154..259S} combined optical and near-infrared photometry, and derived $T_{\rm eff}\,=\,5324\pm26$\,K. They combined this photometric temperature with the parallax from \G\ DR1 \citep{2016A&A...595A...2G,2016A&A...595A...4L} and estimated the angular diameter, finding $\theta\,=\,174.5\,\pm\,1.9\,\mu$as. \citet{2018AJ....155...22S} combined literature photometry, \DR\ astrometry, and various colour-$T_{\rm eff}$, $T_{\rm eff}$-radius, and $T_{\rm eff}$-mass empirical relations to determine the basic properties of LT UMa. They found $T_{\rm eff}\,=\,5351$\,K, $\log{g}\,=\,4.51\,\pm\,0.28$, $R_*\,=\,0.88\,\pm\,0.11\,R_\odot$, and $M_*\,=\,0.92\,\pm\,0.12\,M_\odot$. \DR\ quotes $T_{\rm eff}=5342^{+92}_{-58}$~K, and the best-fit template used for radial velocity measurement has $T_{\rm eff}=5250$~K, $\log{g}=4.5$, and [Fe/H]=0.0, all in good agreement with the literature values. Finally, we determined $T_{\rm eff}$ through SED fitting, using the Virtual Observatory SED Analyzer\footnote{\url{http://svo2.cab.inta-csic.es/theory/vosa/index.php}} \citep[VOSA;][]{2008A&A...492..277B}. Given its brightness and relative proximity, LT UMa has photometric data covering the full range from far-UV to mid-IR. We fit this SED with the BT-Settl models \citep{2012RSPTA.370.2765A}, available through VOSA, and found $T_{\rm eff}\,=\,5300$\,K. Combining our VOSA-based estimate with all the values found in the literature, we adopted $T_{\rm eff}\,=\,5300\,\pm\,50$\,K. VOSA measures 
$\log_{10}\left(L_{\rm bol}/L_\odot\right)=-0.3091^{+0.0049}_{-0.0050}$ implying a radius $R_*\,=\,0.837\,\pm\,0.016\,R_\odot$.

The primary was found to be active by \citet{2000A&AS..142..275S}, who measured the strength of the \ion{Ca}{ii} H and K lines. \citet{2013A&A...551L...8P} used the \citet{2000A&AS..142..275S} measurements and derived an equivalent of the S-index in the Mount Wilson scale, and then used the procedure of \citet{1984ApJ...279..763N} to convert the S-index into $R'_{\rm HK}$, and measured $\log{R'_{\rm HK}}\,=\, -4.443$. We used this value together with the calibrations of \citet{2008ApJ...687.1264M} to estimate the age of this system. Equation 3 from \citet{2008ApJ...687.1264M}, based on chromospheric activity, leads to and age of 0.41\,Gyr. We also used the activity to Rossby number correlation from \citet[their equation 7]{2008ApJ...687.1264M} and their recalibrated colour-dependent version of the Skumanich law \citep{1972ApJ...171..565S}, to derive a gyrochronology age of 0.70\,Gyr.

\begin{figure}
    \centering
    \includegraphics[width=0.48\textwidth, trim={0 3cm 3cm 4cm}, clip]{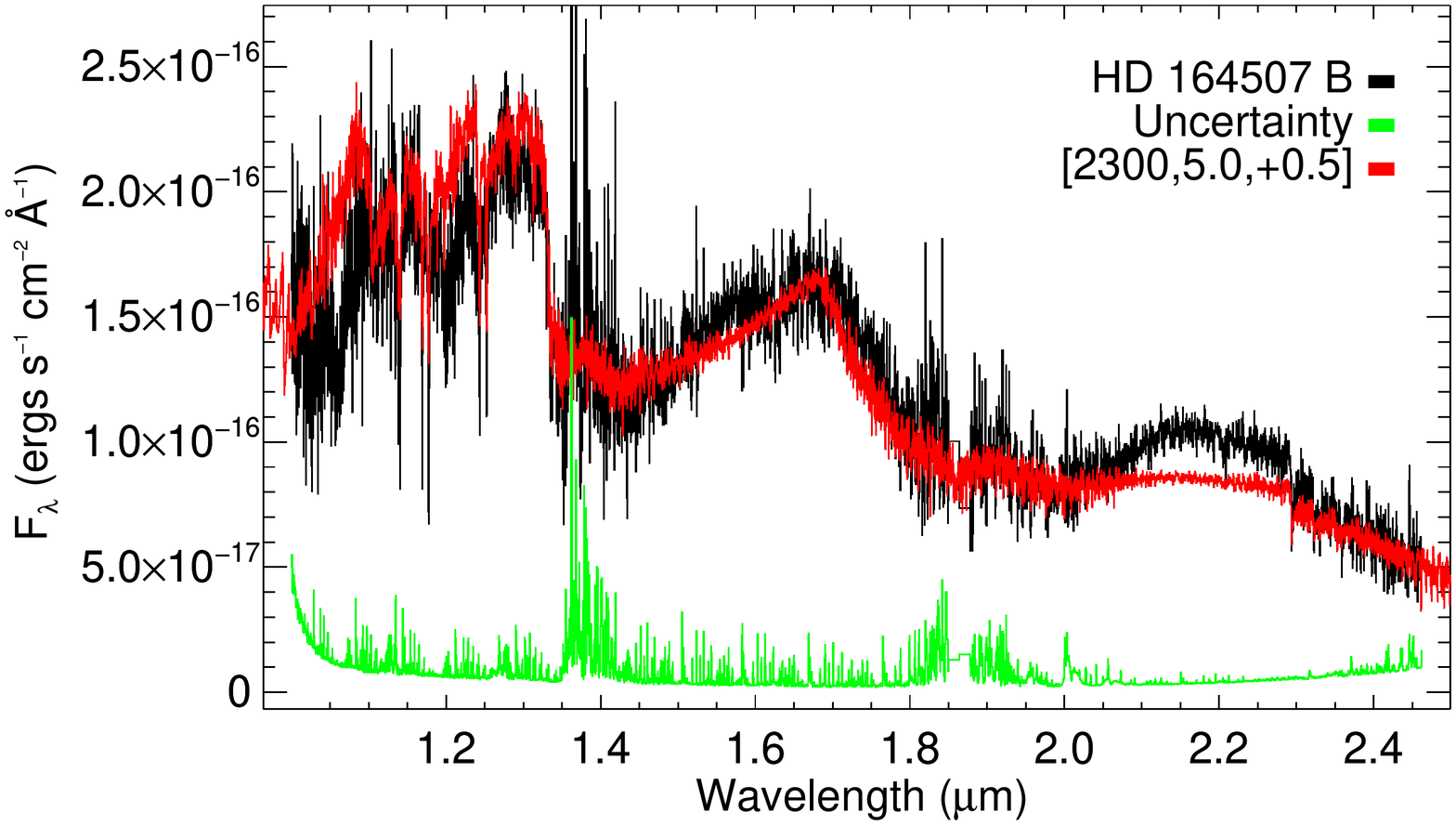}
    \includegraphics[width=0.48\textwidth, trim={0 3cm 3cm 4cm}, clip]{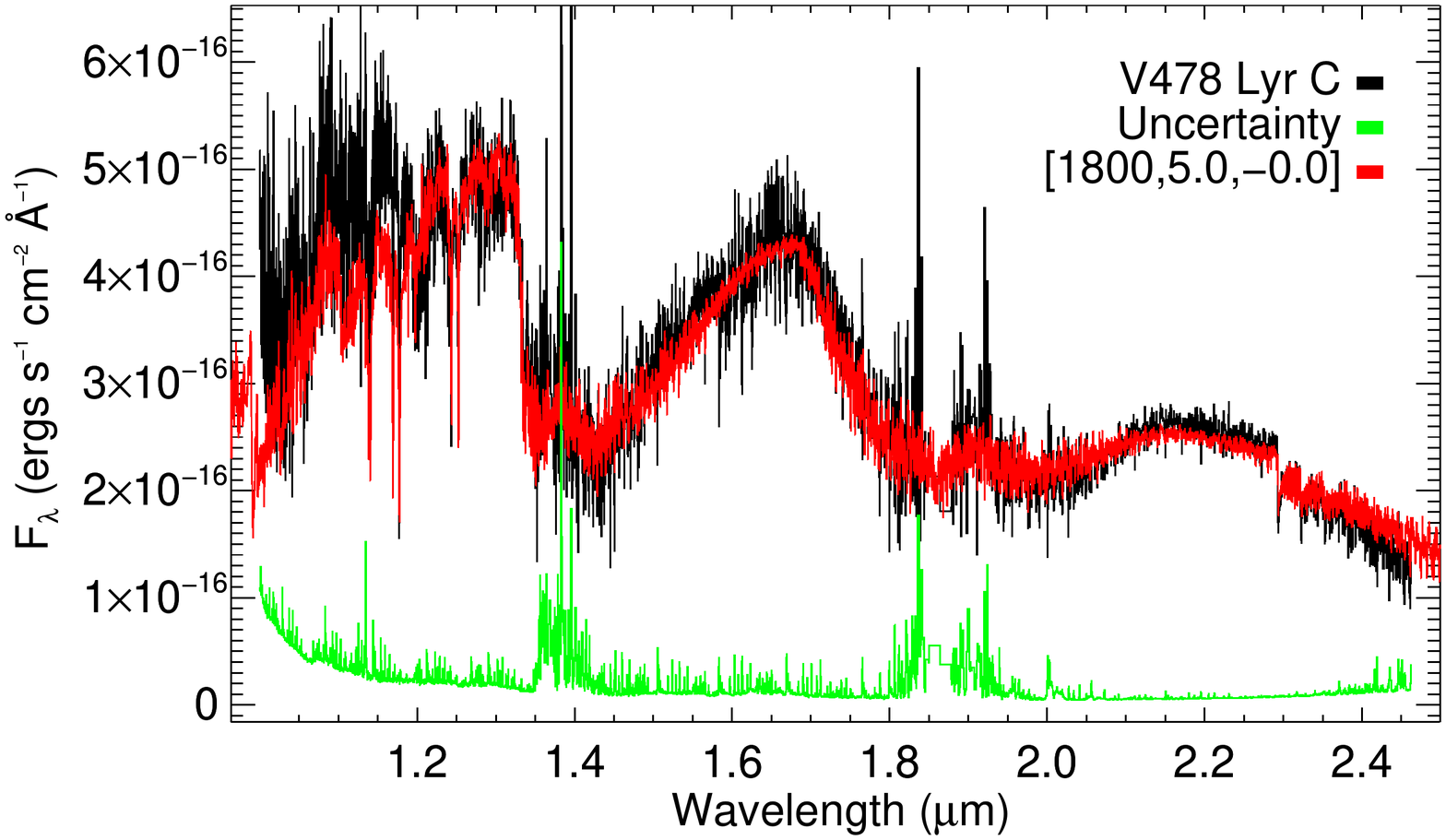}
    \includegraphics[width=0.48\textwidth, trim={0 3cm 3cm 4cm}, clip]{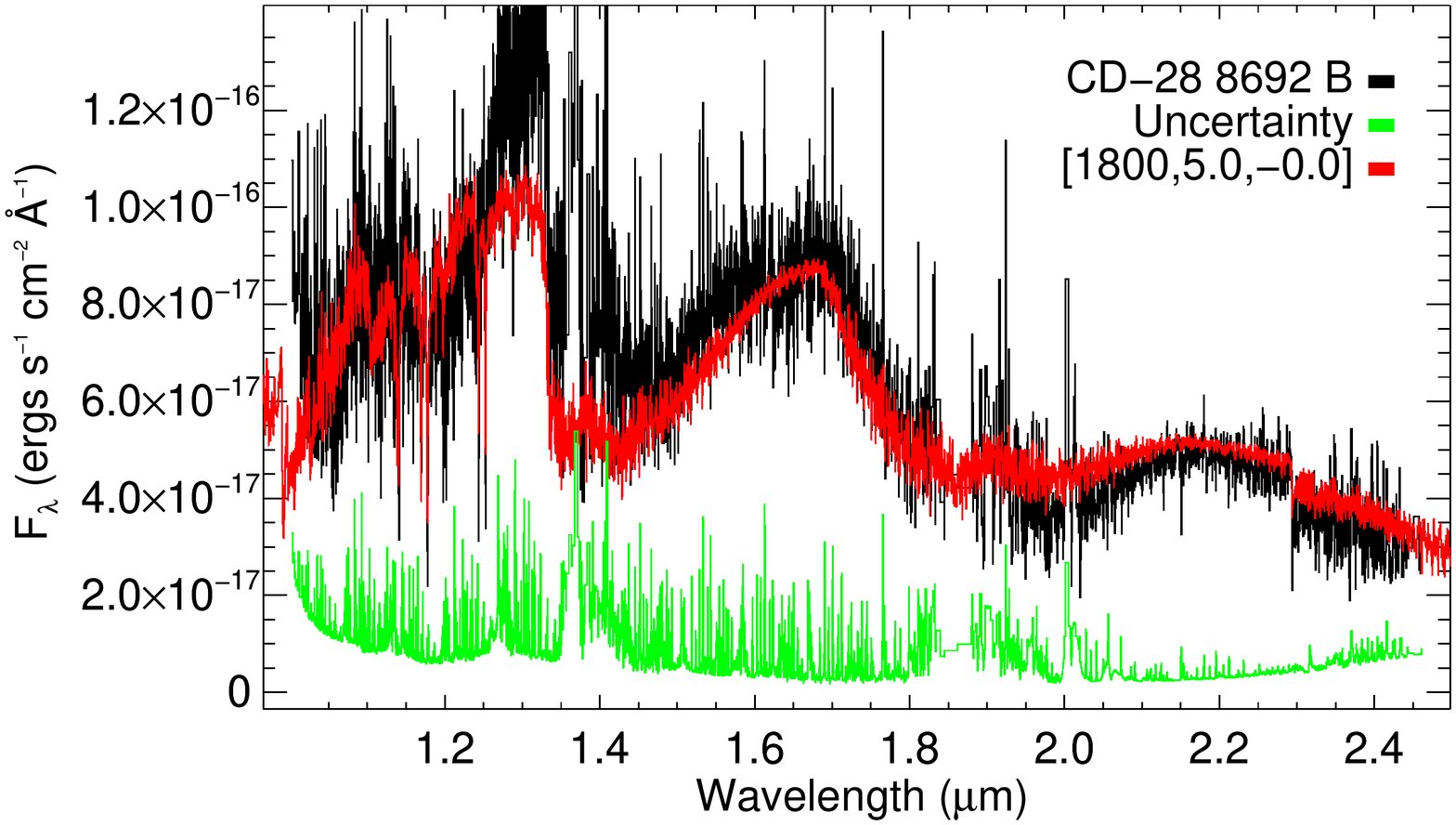}
    \includegraphics[width=0.48\textwidth, trim={0 3cm 3cm 4cm}, clip]{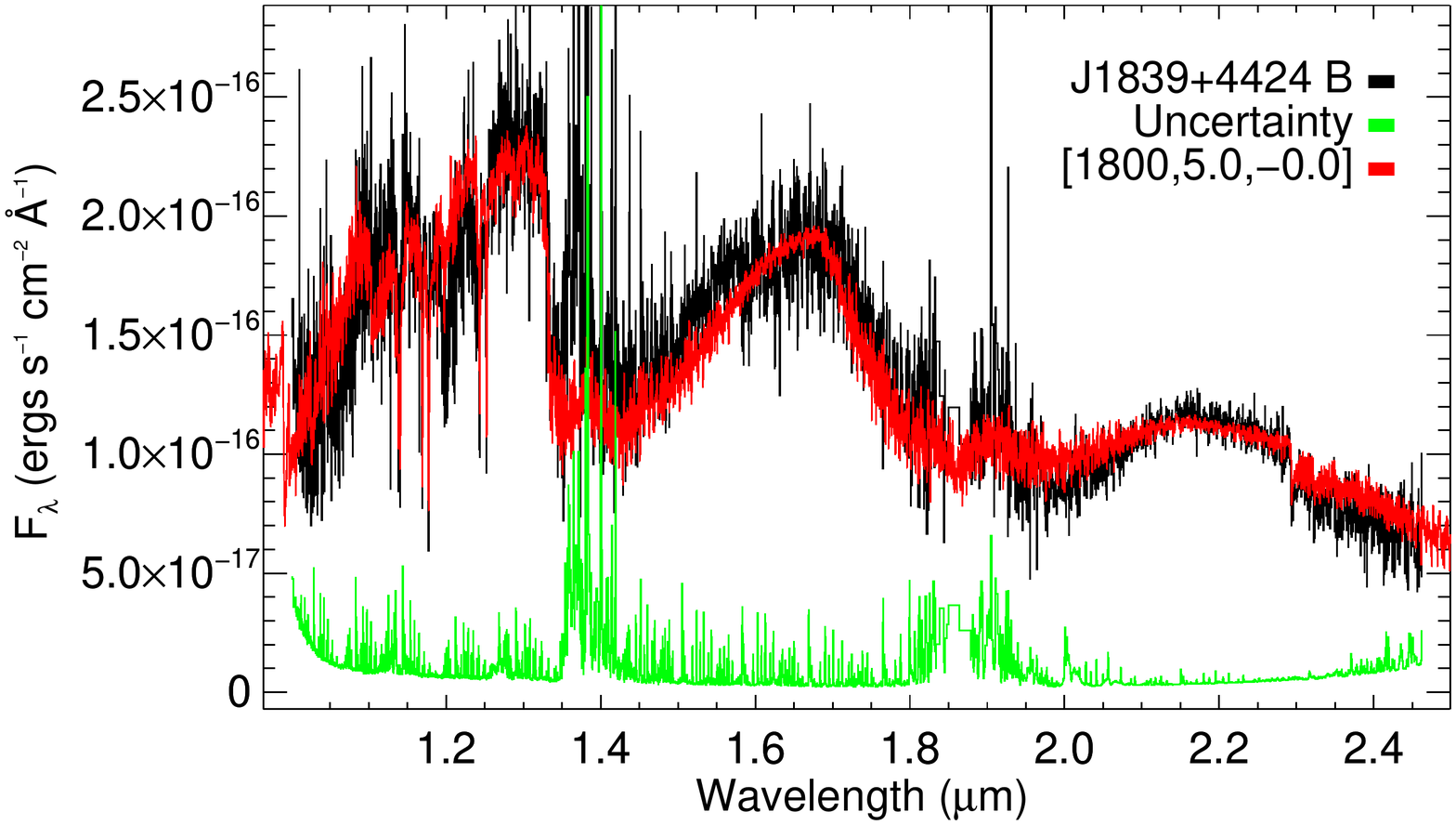}
    \caption{The spectra of HD~164507~B, V478~Lyr~C, CD-28~8692~B, and 2MASS J18392740+4424510 (black) with the measured flux uncertainty (green) and the best-fit BT-Settl atmospheric model (red). The best-fit $T_{\rm eff}$, $\log{g}$, and [Fe/H] are indicated on the plot. For details on the fitting procedure, see Sections~\ref{sec:hd164507}--\ref{sec:ltuma}.}
    \label{fig:model1}
\end{figure}

\begin{figure}
    \centering
    \includegraphics[width=0.48\textwidth, trim={0 3cm 3cm 4cm}, clip]{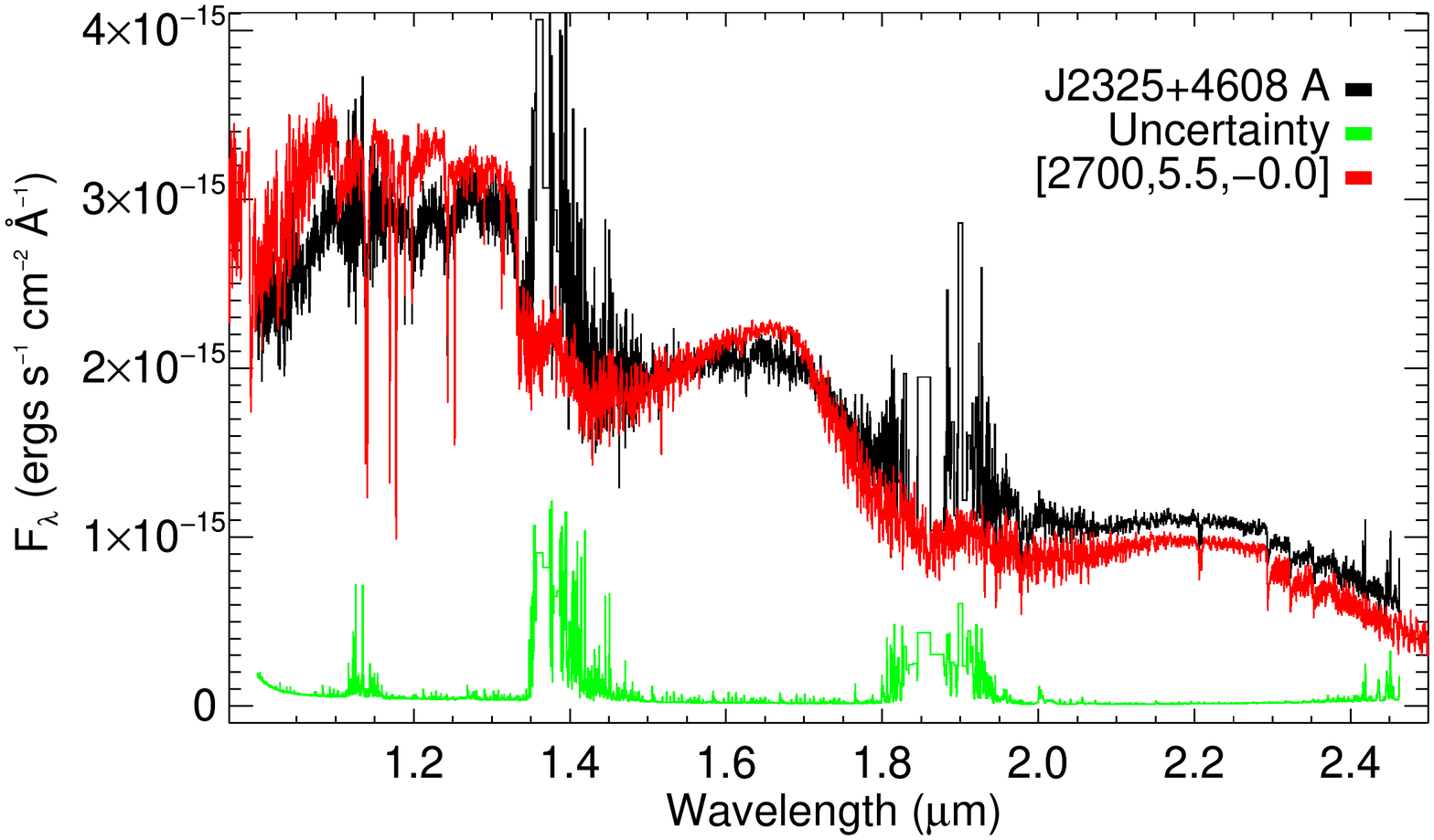}
    \includegraphics[width=0.48\textwidth, trim={0 3cm 3cm 4cm}, clip]{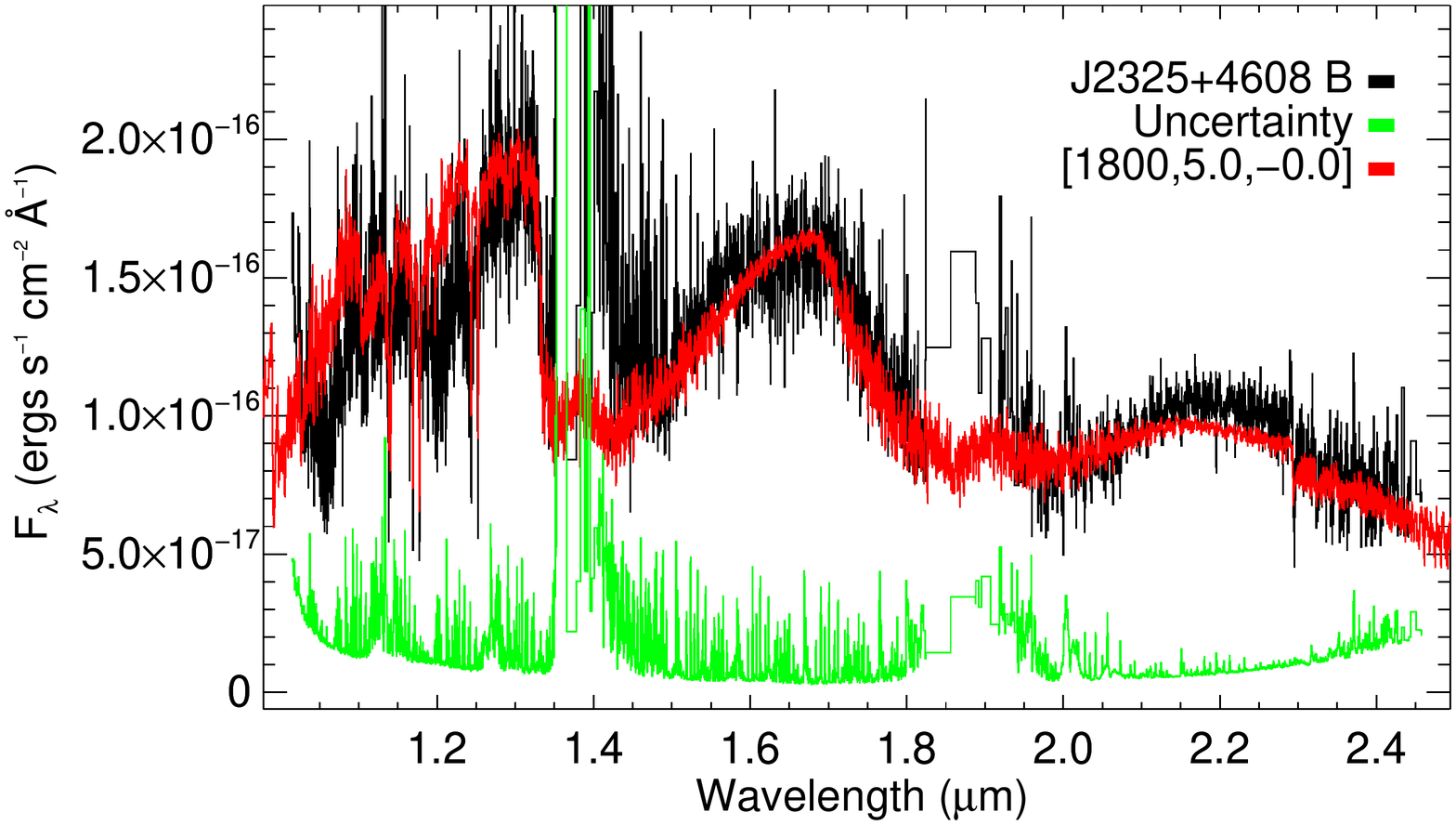}
    \includegraphics[width=0.48\textwidth, trim={0 3cm 3cm 4cm}, clip]{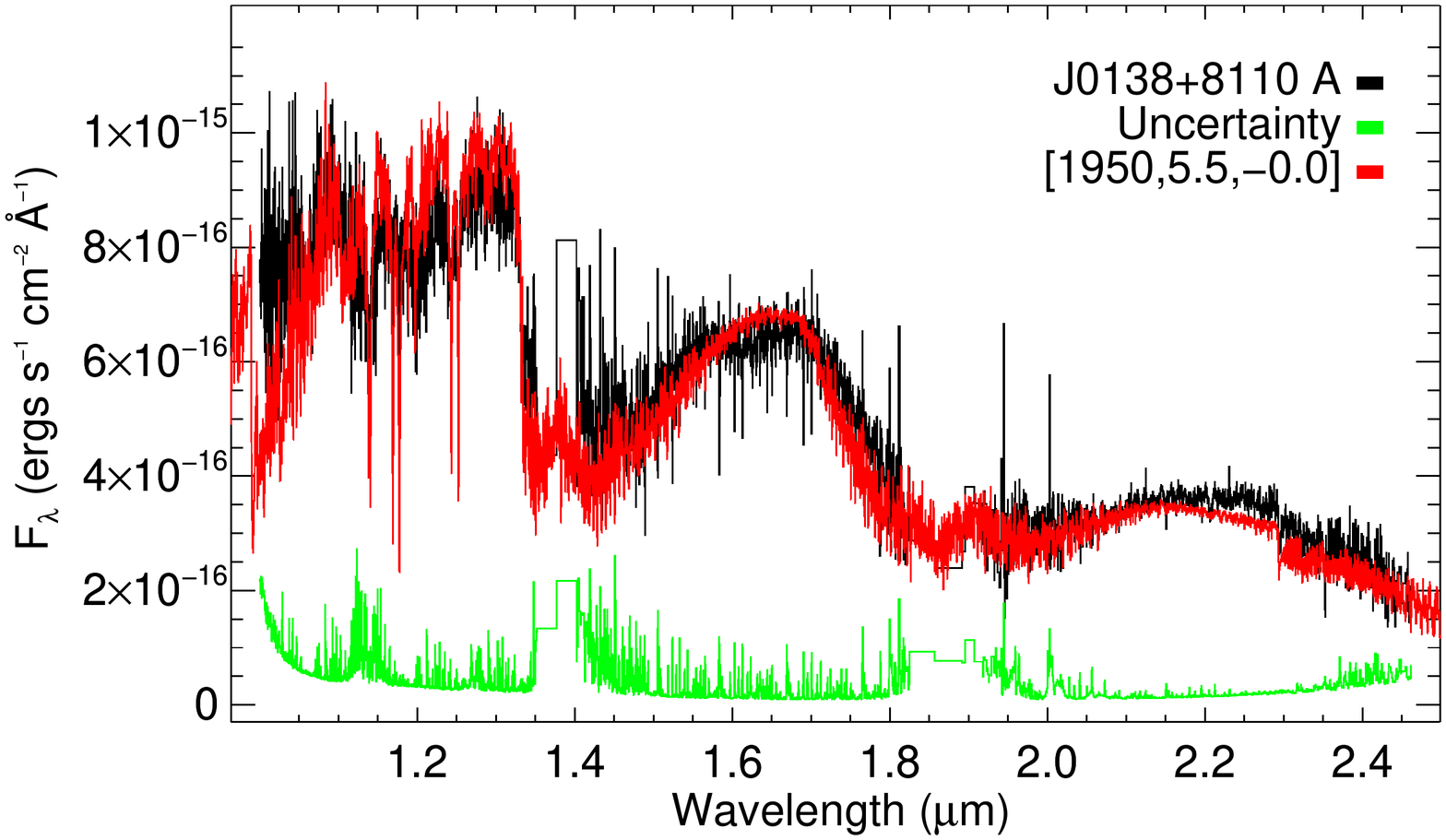}
    \includegraphics[width=0.48\textwidth, trim={0 3cm 3cm 4cm}, clip]{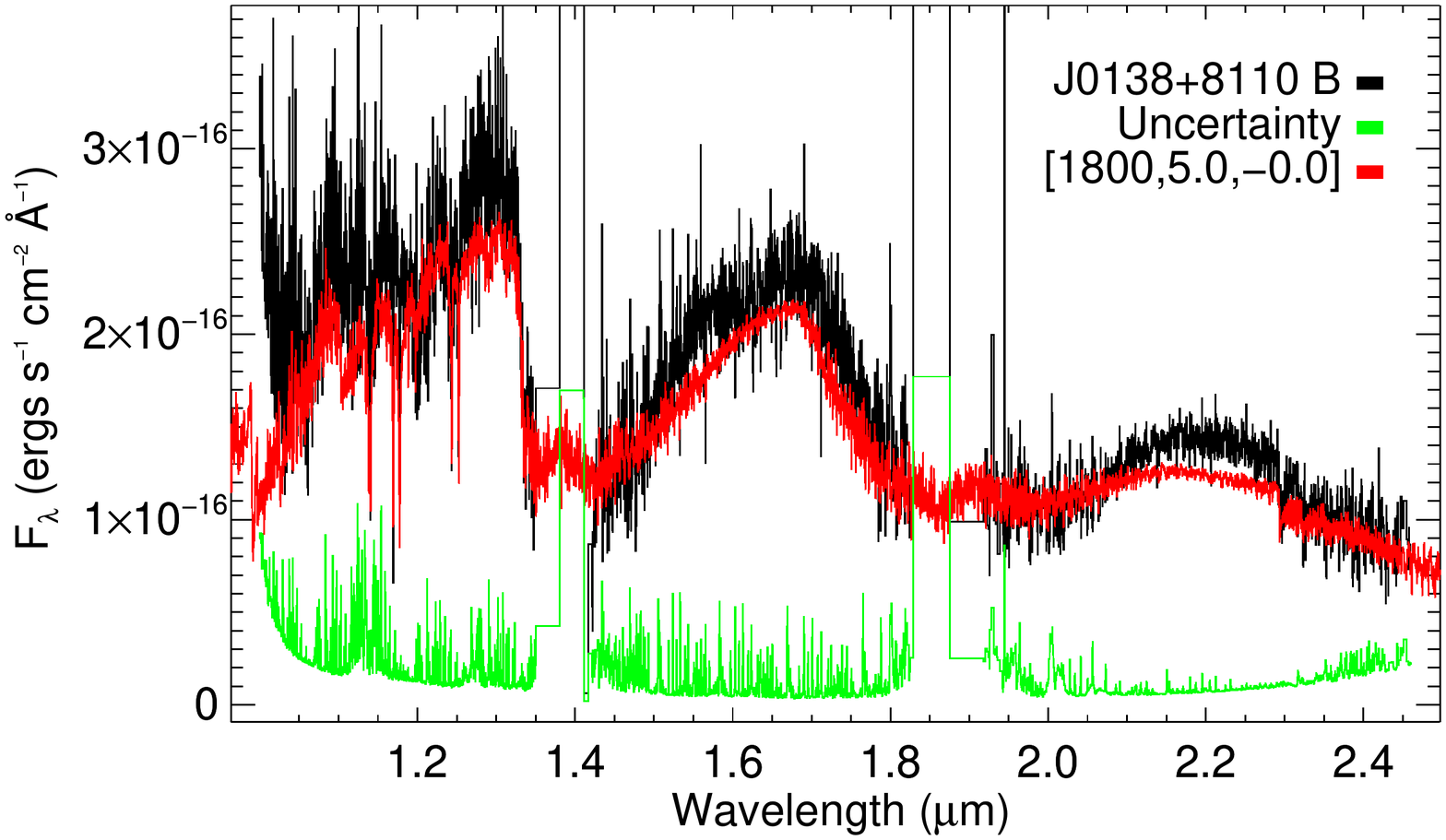}
    \caption{Same as Figure~\ref{fig:model1}, but for 2MASS  J23253550+4608163,  2MASS  J23253519+4608098, 2MASS J01390902+8110003, and 2MASS J01385969+8110084.}
    \label{fig:model2}
\end{figure}

The TripleSpec spectrum of the companion is presented in Figure~\ref{fig:spectra2}, and we classify it as M9\,V via template fitting. The spectrum does not show signs of youth (i.e. low surface gravity), and the gravity-sensitive spectral indices give a classification of FLD-G. As discussed in Section~\ref{sec:v478Lyr}, low-gravity features tend to disappear by the time the object reaches $\sim$400\,Myr. The absence of low-gravity features from the spectrum of LT UMa B is therefore consistent with the age of the system (0.41--0.70\,Gyr) and its solar metallicity. Using the \citet{2015ApJ...810..158F} relation we obtain $T_{\rm eff}\,=\,2395\pm29$\,K, which implies a mass in the $48-77\,M_{\rm Jup}$ range. 
We determined $L_{\rm bol}$ for the M dwarf following the same procedure described in Section~\ref{sec:hd164507}. The best fit model has $T_{\rm eff} = 2300$\,K, $\log{g} = 5.0$, and [Fe/H] = +0.5\,dex. The radius is 1.13\,$R_{\rm Jup}$, in good agreement with the model-predicted radius (1.18\,$R_{\rm Jup}$). The bolometric luminosity is 
$\log_{10}\left(L_{\rm bol}/L_\odot\right)=-2.968^{+0.028}
_{-0.029}$. The best fit model can be seen in Figure~\ref{fig:model1}. The quality of the fit is poor. The model has a triangular-shaped $H$ band that is not present in the target, which instead displays a flat $H$-band spectrum. The alkali lines in the $J$ band are also weaker in the model compared to the observed ones. 

\begin{figure}
    \centering
    \includegraphics[width=0.48\textwidth, trim={0 3cm 3cm 4cm}, clip]{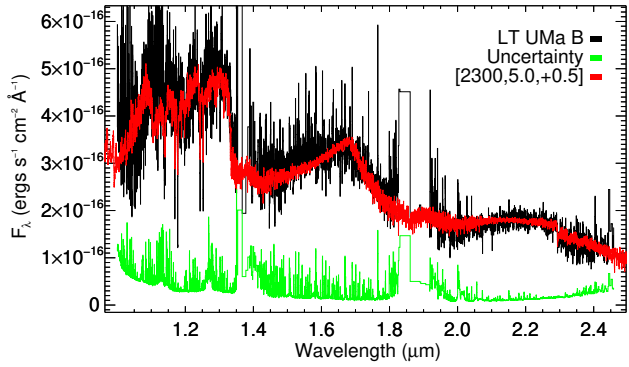}
    \caption{Same as Figure~\ref{fig:model1}, but for LT UMa B.}
    \label{fig:model3}
\end{figure}

\section{Comparison of L dwarf spectral features}
\label{sec:features}

\begin{table}
    \centering
    \caption{Summary of the inferred properties for HD~164507\,B, V478~Lyr\,C, and CD-28\,8692\,B. \label{tab:summary3}}
    \begin{tabular}{l c c c c}
    \hline
        Name & Sp. Type & $T_{\rm eff}$ & Age & [Fe/H] \\
         & & (K) & (Gyr) & (dex) \\
    \hline
        HD~164507\,B & L1 & 2100$\pm$30 & 3.0--5.9 & 0.03--0.19 \\
        V478~Lyr\,C & L1: & 1740$\pm$60 & 0.1--0.37 & \ldots \\
        CD-28\,8692\,B & L2 & 1960$\pm$30 & 4.5 & --0.22 \\
    \hline
    \end{tabular}
    \\ $T_{\rm eff}$ for HD~164507\,B and CD-28\,8692\,B are computed using the spectral type to $T_{\rm eff}$ polynomial relations for field-age objects derived in \citet{2015ApJ...810..158F}, while for V478~Lyr\,C we used the $M_H$ to $T_{\rm eff}$ polynomial relation for young objects presented in the same paper.
\end{table}

\begin{figure*}
	\includegraphics[width=\textwidth]{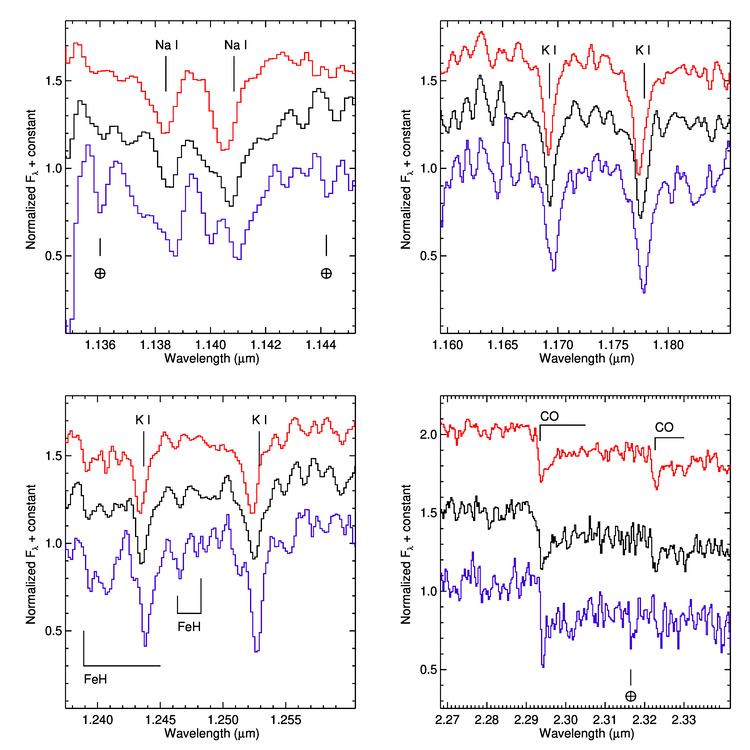}
    \caption{A direct comparison of the main absorption features in the spectra of V478 Lyr C (red), HD~164507\,B (black), and CD-28~8692\,B (blue). Features likely due to telluric absorption are labelled with the symbol $\oplus$. All spectra are smoothed down to a resolution of 3 \AA\ pix$^{-1}$ to reduce the noise. The alkali lines in V478 Lyr\,C and HD~164507\,B show remarkable similarity, while those in CD-28~8692\,B are deeper and broader, confirming the known trend with metallicity \citep[see e.g.][]{2010ApJS..190..100K}. The CO band head at 2.293~$\mu$m appears deeper in CD-28~8692\,B as well, while the CO band head at 2.322~$\mu$m is in a region of too low signal-to-noise-ratio. \label{fig:features}}
\end{figure*}

Despite the relatively small sample size, it is nonetheless interesting to compare the spectroscopic features in our newly discovered L companions. In particular, V478 Lyr C, HD~164507~B, and CD-28~8692~B offer an interesting comparison set. With very similar spectral type (L1:, L1, and L2, respectively), but different ages and metallicity, these three objects can be used to qualitatively determine the dependence of spectral features on these parameters. Properties for these three UCDs relevant to this analysis are summarised in Table~\ref{tab:summary3}. Figure~\ref{fig:features} shows the normalised IR spectra, centred around four of the main absorption features in the spectra of early L dwarfs: the \ion{Na}{i} doublet at $\sim 1.139\,\mu$m, the \ion{K}{i} doublets at $\sim 1.173\,\mu$m and $\sim 1.248\,\mu$m, and the CO band head at 2.30\,$\mu$m. 

The alkali lines in V478 Lyr C and HD~164507~B show remarkable similarity, while those in CD-28\,8692\,B are deeper and broader. FeH absorption in the 1.24--1.25$\mu$m range is also stronger in CD-28\,8692\,B, as expected from its age, which confirms the known trend of alkali lines and hydride bands with metallicity \citep[see e.g.][]{2010ApJS..190..100K}. Surprisingly, the CO band at 2.293~$\mu$m appears deeper in CD-28\,8692\,B as well, while the band at 2.322~$\mu$m is in a region of too low signal-to-noise ratio. While the strength of this CO band is relatively insensitive to changes in effective temperature in the L0--4 range \citep{2018AJ....155...34C}, blue L dwarfs and L subdwarfs have weaker CO bands than their solar-metallicity counterparts \citep[see e.g.][]{2017MNRAS.468..261Z}. A strong CO band has been previously observed in the blue L1 dwarf 2MASS~J17561080+2815238 \citep{2010ApJS..190..100K}. 

Comparison of the spectral indices and equivalent widths presented in Table~\ref{tab:indices} as a function of the age and metallicity for these three systems leads to some preliminary considerations:

\begin{itemize}
    \item the ``water-based'' indices H$_2$O, H$_2$O-$J$, H$_2$O-$H$ and, to a lesser extent, H$_2$OD and H$_2$O-$K$ appear sensitive to metallicity -- e.g. H$_2$O-$J = 0.659\pm0.013$ at Fe/H = --0.22 dex vs. H$_2$O-$J = 0.8819\pm0.0054$ at Fe/H = 0.03--0.19 dex;
    \item the H$_2$OD and H$_2$O-$J$ indices seem sensitive to age (i.e. surface gravity) too; 
    \item the \ion{K}{i} lines are sensitive to age (i.e. surface gravity) but also metallicity, becoming stronger (i.e. having larger equivalent width) as age increases, but weaker at higher metallicity. As a result, the young ($\approx$100--370 Myr) L1: V478~Lyr~C has \ion{K}{i} lines of roughly equal strength as the older (3.0--5.9 Gyr) but metal rich L1 HD~164507~B (5.52, 8.20, 5.05, and 4.89 \AA\ for V478\,Lyr\,C vs. 5.97, 6.87, 5.28, and 5.03 \AA\ for HD\,164507\,B).
\end{itemize}

Followup of a larger sample of benchmark L dwarfs is fundamental to better identify/quantify possible dependencies of the above spectral features on age and metallicity.

\section{Conclusions}
\label{sec:conclusions}
We have presented seven multiple systems discovered in \DR\ data, identified as part of our GUCDS project. The systems presented here include an L1 companion to the G5\,IV star HD~164507, an L1: companion to the RS CVn star V478 Lyr, three low-mass binaries consisting of late Ms and early Ls, an L2 companion to the metal-poor K5\,V star CD-28~8692, and an M9\,V companion to the young variable K0\,V star LT UMa. The HD~164507 and CD-28~8692 systems are particularly important benchmarks, because the primaries are very well characterised and offer excellent constraints on the atmospheric parameters of the companion. While the HD~164507 AB system is slightly metal rich, the CD-28~8692 AB system is slightly metal poor, and therefore cover an exotic region of the parameter space, where observational constraints on theoretical models is currently scarce. The V478 Lyr ABC system is a nice addition to the sample of wide low-mass tertiary components to tight binaries, a population of crucial importance to validate formation theories for tight binaries.

We have also reported the discovery of the currently widest L+L binary known -- the 2MASS~J01390902+8110003 + 2MASS~J01385969+8110084 system, with a projected separation of about 960\,au. This system, together with the other two wide low-mass wide binaries presented here, pose an increasing challenge to models of formation and evolution of wide low-mass binaries.

A first, qualitative analysis of the sample reveals tentative correlations between spectral indices, equivalent widths, and age and metallicity for the ultra-cool dwarfs presented here. Analysis of a larger sample of benchmarks will provide stronger constraints on such correlations, and \DR\ will play a cornerstone role in shaping our understanding of ultra-cool atmospheres.

\section*{Acknowledgements}
We thank the anonymous referee for helpful suggestions that improved the quality of this paper. 

We thank the staff of Palomar Observatory for successful nights of observing with TripleSpec, in particular Carolyn Heffner, Paul Nied, Joel Pearman, Kajsa Peffer, and Kevin Rykoski.

FM acknowledges support by the NASA Postdoctoral Program at the Jet Propulsion Laboratory, administered by Universities Space Research Association under a contract with NASA.

Part of this research was carried out at the Jet Propulsion Laboratory, California Institute of Technology, under a contract with NASA. 

This work has made use of data from the European Space Agency (ESA) mission \G\ (\url{https://www.cosmos.esa.int/gaia}), processed by the \G\ Data Processing and Analysis Consortium (DPAC, \url{https://www.cosmos.esa.int/web/gaia/dpac/consortium}). Funding for the DPAC has been provided by national institutions, in particular the institutions participating in the \G\ Multilateral Agreement.

SPLAT is a collaborative project of research students in UCSD Cool Star Lab. Contributors to SPLAT have included Christian Aganze, Jessica Birky, Daniella Bardalez Gagliuffi, Adam Burgasser (PI), Caleb Choban, Andrew Davis, Ivanna Escala, Aishwarya Iyer, Yuhui Jin, Mike Lopez, Alex Mendez, Gretel Mercado, Elizabeth Moreno Hilario, Johnny Parra, Maitrayee Sahi, Adrian Suarez, Melisa Tallis, Tomoki Tamiya, Chris Theissen and Russell van Linge. This project is supported by the National Aeronautics and Space Administration under Grant No. NNX15AI75G.

This publication makes use of VOSA, developed under the Spanish Virtual Observatory project supported by the Spanish MINECO through grant AyA2017-84089.



\bibliographystyle{mnras}
\bibliography{refs} 

\begin{thebibliography}{}
\makeatletter
\relax
\def\mn@urlcharsother{\let\do\@makeother \do\$\do\&\do\#\do\^\do\_\do\%\do\~}
\def\mn@doi{\begingroup\mn@urlcharsother \@ifnextchar [ {\mn@doi@}
  {\mn@doi@[]}}
\def\mn@doi@[#1]#2{\def\@tempa{#1}\ifx\@tempa\@empty \href
  {http://dx.doi.org/#2} {doi:#2}\else \href {http://dx.doi.org/#2} {#1}\fi
  \endgroup}
\def\mn@eprint#1#2{\mn@eprint@#1:#2::\@nil}
\def\mn@eprint@arXiv#1{\href {http://arxiv.org/abs/#1} {{\tt arXiv:#1}}}
\def\mn@eprint@dblp#1{\href {http://dblp.uni-trier.de/rec/bibtex/#1.xml}
  {dblp:#1}}
\def\mn@eprint@#1:#2:#3:#4\@nil{\def\@tempa {#1}\def\@tempb {#2}\def\@tempc
  {#3}\ifx \@tempc \@empty \let \@tempc \@tempb \let \@tempb \@tempa \fi \ifx
  \@tempb \@empty \def\@tempb {arXiv}\fi \@ifundefined
  {mn@eprint@\@tempb}{\@tempb:\@tempc}{\expandafter \expandafter \csname
  mn@eprint@\@tempb\endcsname \expandafter{\@tempc}}}

\bibitem[\protect\citeauthoryear{{Adamczyk}, {Deka-Szymankiewicz}  \&
  {Niedzielski}}{{Adamczyk} et~al.}{2016}]{2016A&A...587A.119A}
{Adamczyk} M.,  {Deka-Szymankiewicz} B.,   {Niedzielski} A.,  2016, \mn@doi
  [\aap] {10.1051/0004-6361/201526628}, \href
  {http://adsabs.harvard.edu/abs/2016A%26A...587A.119A} {587, A119}

\bibitem[\protect\citeauthoryear{{Adibekyan}, {Sousa}, {Santos}, {Delgado
  Mena}, {Gonz{\'a}lez Hern{\'a}ndez}, {Israelian}, {Mayor}  \&
  {Khachatryan}}{{Adibekyan} et~al.}{2012}]{2012A&A...545A..32A}
{Adibekyan} V.~Z.,  {Sousa} S.~G.,  {Santos} N.~C.,  {Delgado Mena} E.,
  {Gonz{\'a}lez Hern{\'a}ndez} J.~I.,  {Israelian} G.,  {Mayor} M.,
  {Khachatryan} G.,  2012, \mn@doi [\aap] {10.1051/0004-6361/201219401}, \href
  {https://ui.adsabs.harvard.edu/#abs/2012A&A...545A..32A} {545, A32}

\bibitem[\protect\citeauthoryear{{Aganze} et~al.,}{{Aganze}
  et~al.}{2016}]{2016AJ....151...46A}
{Aganze} C.,  et~al., 2016, \mn@doi [\aj] {10.3847/0004-6256/151/2/46}, \href
  {http://adsabs.harvard.edu/abs/2016AJ....151...46A} {151, 46}

\bibitem[\protect\citeauthoryear{{Allard}, {Homeier}, {Freytag}  \&
  {Sharp}}{{Allard} et~al.}{2012a}]{2012EAS....57....3A}
{Allard} F.,  {Homeier} D.,  {Freytag} B.,   {Sharp} C.~M.,  2012a, in
  {Reyl{\'e}} C.,  {Charbonnel} C.,   {Schultheis} M.,  eds,  EAS Publications
  Series Vol. 57, EAS Publications Series. pp 3--43 (\mn@eprint {arXiv}
  {1206.1021}), \mn@doi{10.1051/eas/1257001}

\bibitem[\protect\citeauthoryear{{Allard}, {Homeier}  \& {Freytag}}{{Allard}
  et~al.}{2012b}]{2012RSPTA.370.2765A}
{Allard} F.,  {Homeier} D.,   {Freytag} B.,  2012b, \mn@doi [Philosophical
  Transactions of the Royal Society of London Series A]
  {10.1098/rsta.2011.0269}, \href
  {http://cdsads.u-strasbg.fr/abs/2012RSPTA.370.2765A} {370, 2765}

\bibitem[\protect\citeauthoryear{{Allard}, {Homeier}  \& {Freytag}}{{Allard}
  et~al.}{2013}]{2013MmSAI..84.1053A}
{Allard} F.,  {Homeier} D.,   {Freytag} B.,  2013, \memsai, \href
  {http://cdsads.u-strasbg.fr/abs/2013MmSAI..84.1053A} {84, 1053}

\bibitem[\protect\citeauthoryear{{Allen}, {Burgasser}, {Faherty}  \&
  {Kirkpatrick}}{{Allen} et~al.}{2012}]{2012AJ....144...62A}
{Allen} P.~R.,  {Burgasser} A.~J.,  {Faherty} J.~K.,   {Kirkpatrick} J.~D.,
  2012, \mn@doi [\aj] {10.1088/0004-6256/144/2/62}, \href
  {https://ui.adsabs.harvard.edu/#abs/2012AJ....144...62A} {144}

\bibitem[\protect\citeauthoryear{{Allers} \& {Liu}}{{Allers} \&
  {Liu}}{2013}]{2013ApJ...772...79A}
{Allers} K.~N.,  {Liu} M.~C.,  2013, \mn@doi [\apj]
  {10.1088/0004-637X/772/2/79}, \href
  {http://adsabs.harvard.edu/abs/2013ApJ...772...79A} {772, 79}

\bibitem[\protect\citeauthoryear{{Allers} et~al.,}{{Allers}
  et~al.}{2007}]{2007ApJ...657..511A}
{Allers} K.~N.,  et~al., 2007, \mn@doi [\apj] {10.1086/510845}, \href
  {http://adsabs.harvard.edu/abs/2007ApJ...657..511A} {657, 511}

\bibitem[\protect\citeauthoryear{{Alonso-Floriano}, {Caballero},
  {Cort{\'e}s-Contreras}, {Solano}  \& {Montes}}{{Alonso-Floriano}
  et~al.}{2015}]{2015A&A...583A..85A}
{Alonso-Floriano} F.~J.,  {Caballero} J.~A.,  {Cort{\'e}s-Contreras} M.,
  {Solano} E.,   {Montes} D.,  2015, \mn@doi [\aap]
  {10.1051/0004-6361/201526795}, \href
  {http://adsabs.harvard.edu/abs/2015A%26A...583A..85A} {583, A85}

\bibitem[\protect\citeauthoryear{{Andrae} et~al.,}{{Andrae}
  et~al.}{2018}]{2018A&A...616A...8A}
{Andrae} R.,  et~al., 2018, \mn@doi [\aap] {10.1051/0004-6361/201732516}, \href
  {https://ui.adsabs.harvard.edu/\#abs/2018A&A...616A...8A} {616, A8}

\bibitem[\protect\citeauthoryear{{Bardalez Gagliuffi} et~al.,}{{Bardalez
  Gagliuffi} et~al.}{2014}]{2014ApJ...794..143B}
{Bardalez Gagliuffi} D.~C.,  et~al., 2014, \mn@doi [\apj]
  {10.1088/0004-637X/794/2/143}, \href
  {http://cdsads.u-strasbg.fr/abs/2014APJ...794..143B} {794, 143}

\bibitem[\protect\citeauthoryear{{Baron} et~al.,}{{Baron}
  et~al.}{2015}]{2015ApJ...802...37B}
{Baron} F.,  et~al., 2015, \mn@doi [\apj] {10.1088/0004-637X/802/1/37}, \href
  {http://adsabs.harvard.edu/abs/2015ApJ...802...37B} {802, 37}

\bibitem[\protect\citeauthoryear{{Bayo}, {Rodrigo}, {Barrado y Navascu{\'e}s},
  {Solano}, {Guti{\'e}rrez}, {Morales-Calder{\'o}n}  \& {Allard}}{{Bayo}
  et~al.}{2008}]{2008A&A...492..277B}
{Bayo} A.,  {Rodrigo} C.,  {Barrado y Navascu{\'e}s} D.,  {Solano} E.,
  {Guti{\'e}rrez} R.,  {Morales-Calder{\'o}n} M.,   {Allard} F.,  2008, \mn@doi
  [\aap] {10.1051/0004-6361:200810395}, \href
  {http://adsabs.harvard.edu/abs/2008A%26A...492..277B} {492, 277}

\bibitem[\protect\citeauthoryear{{Bertelli}, {Bressan}, {Chiosi}, {Fagotto}  \&
  {Nasi}}{{Bertelli} et~al.}{1994}]{1994A&AS..106..275B}
{Bertelli} G.,  {Bressan} A.,  {Chiosi} C.,  {Fagotto} F.,   {Nasi} E.,  1994,
  \aaps, \href {http://adsabs.harvard.edu/abs/1994A%26AS..106..275B} {106, 275}

\bibitem[\protect\citeauthoryear{{Bressan}, {Marigo}, {Girardi}, {Salasnich},
  {Dal Cero}, {Rubele}  \& {Nanni}}{{Bressan}
  et~al.}{2012}]{2012MNRAS.427..127B}
{Bressan} A.,  {Marigo} P.,  {Girardi} L.,  {Salasnich} B.,  {Dal Cero} C.,
  {Rubele} S.,   {Nanni} A.,  2012, \mn@doi [\mnras]
  {10.1111/j.1365-2966.2012.21948.x}, \href
  {https://ui.adsabs.harvard.edu/#abs/2012MNRAS.427..127B} {427, 127}

\bibitem[\protect\citeauthoryear{{Burgasser}, {Kirkpatrick}, {Reid}, {Brown},
  {Miskey}  \& {Gizis}}{{Burgasser} et~al.}{2003}]{2003ApJ...586..512B}
{Burgasser} A.~J.,  {Kirkpatrick} J.~D.,  {Reid} I.~N.,  {Brown} M.~E.,
  {Miskey} C.~L.,   {Gizis} J.~E.,  2003, \mn@doi [\apj] {10.1086/346263},
  \href {http://cdsads.u-strasbg.fr/abs/2003ApJ...586..512B} {586, 512}

\bibitem[\protect\citeauthoryear{{Burgasser}, {Geballe}, {Leggett},
  {Kirkpatrick}  \& {Golimowski}}{{Burgasser}
  et~al.}{2006}]{2006ApJ...637.1067B}
{Burgasser} A.~J.,  {Geballe} T.~R.,  {Leggett} S.~K.,  {Kirkpatrick} J.~D.,
  {Golimowski} D.~A.,  2006, \mn@doi [\apj] {10.1086/498563}, \href
  {http://cdsads.u-strasbg.fr/abs/2006ApJ...637.1067B} {637, 1067}

\bibitem[\protect\citeauthoryear{{Burgasser}, {Looper}, {Kirkpatrick}, {Cruz}
  \& {Swift}}{{Burgasser} et~al.}{2008}]{2008ApJ...674..451B}
{Burgasser} A.~J.,  {Looper} D.~L.,  {Kirkpatrick} J.~D.,  {Cruz} K.~L.,
  {Swift} B.~J.,  2008, \mn@doi [\apj] {10.1086/524726}, \href
  {http://cdsads.u-strasbg.fr/abs/2008ApJ...674..451B} {674, 451}

\bibitem[\protect\citeauthoryear{{Burgasser}, {Dhital}  \& {West}}{{Burgasser}
  et~al.}{2009}]{2009AJ....138.1563B}
{Burgasser} A.~J.,  {Dhital} S.,   {West} A.~A.,  2009, \mn@doi [\aj]
  {10.1088/0004-6256/138/6/1563}, \href
  {http://adsabs.harvard.edu/abs/2009AJ....138.1563B} {138, 1563}

\bibitem[\protect\citeauthoryear{{Burgasser}, {Cruz}, {Cushing}, {Gelino},
  {Looper}, {Faherty}, {Kirkpatrick}  \& {Reid}}{{Burgasser}
  et~al.}{2010}]{2010ApJ...710.1142B}
{Burgasser} A.~J.,  {Cruz} K.~L.,  {Cushing} M.,  {Gelino} C.~R.,  {Looper}
  D.~L.,  {Faherty} J.~K.,  {Kirkpatrick} J.~D.,   {Reid} I.~N.,  2010, \mn@doi
  [\apj] {10.1088/0004-637X/710/2/1142}, \href
  {http://cdsads.u-strasbg.fr/abs/2010ApJ...710.1142B} {710, 1142}

\bibitem[\protect\citeauthoryear{{Burgasser} et~al.,}{{Burgasser}
  et~al.}{2016}]{2016AAS...22743408B}
{Burgasser} A.~J.,  et~al., 2016, in American Astronomical Society Meeting
  Abstracts \#227. p. 434.08

\bibitem[\protect\citeauthoryear{{Burningham} et~al.,}{{Burningham}
  et~al.}{2013}]{2013MNRAS.433..457B}
{Burningham} B.,  et~al., 2013, \mn@doi [\mnras] {10.1093/mnras/stt740}, \href
  {http://adsabs.harvard.edu/abs/2013MNRAS.433..457B} {433, 457}

\bibitem[\protect\citeauthoryear{{Burningham}, {Marley}, {Line}, {Lupu},
  {Visscher}, {Morley}, {Saumon}  \& {Freedman}}{{Burningham}
  et~al.}{2017}]{2017MNRAS.470.1177B}
{Burningham} B.,  {Marley} M.~S.,  {Line} M.~R.,  {Lupu} R.,  {Visscher} C.,
  {Morley} C.~V.,  {Saumon} D.,   {Freedman} R.,  2017, \mn@doi [\mnras]
  {10.1093/mnras/stx1246}, \href
  {http://adsabs.harvard.edu/abs/2017MNRAS.470.1177B} {470, 1177}

\bibitem[\protect\citeauthoryear{{Burrows}, {Sudarsky}  \& {Hubeny}}{{Burrows}
  et~al.}{2006}]{2006ApJ...640.1063B}
{Burrows} A.,  {Sudarsky} D.,   {Hubeny} I.,  2006, \mn@doi [\apj]
  {10.1086/500293}, \href {http://cdsads.u-strasbg.fr/abs/2006ApJ...640.1063B}
  {640, 1063}

\bibitem[\protect\citeauthoryear{{Caballero}}{{Caballero}}{2009}]{2009A&A...507..251C}
{Caballero} J.~A.,  2009, \mn@doi [\aap] {10.1051/0004-6361/200912596}, \href
  {http://adsabs.harvard.edu/abs/2009A%26A...507..251C} {507, 251}

\bibitem[\protect\citeauthoryear{{Caballero}}{{Caballero}}{2012}]{2012Obs...132....1C}
{Caballero} J.~A.,  2012, The Observatory, \href
  {http://adsabs.harvard.edu/abs/2012Obs...132....1C} {132, 1}

\bibitem[\protect\citeauthoryear{{Caballero} \& {Montes}}{{Caballero} \&
  {Montes}}{2012}]{2012Obs...132..176C}
{Caballero} J.~A.,  {Montes} D.,  2012, The Observatory, \href
  {http://adsabs.harvard.edu/abs/2012Obs...132..176C} {132, 176}

\bibitem[\protect\citeauthoryear{{Caballero}, {Genebriera}, {Miret}, {Tobal}
  \& {Cairol}}{{Caballero} et~al.}{2012}]{2012Obs...132..252C}
{Caballero} J.~A.,  {Genebriera} J.,  {Miret} F.~X.,  {Tobal} T.,   {Cairol}
  J.,  2012, The Observatory, \href
  {http://adsabs.harvard.edu/abs/2012Obs...132..252C} {132, 252}

\bibitem[\protect\citeauthoryear{{Casagrande}, {Ram{\'{\i}}rez},
  {Mel{\'e}ndez}, {Bessell}  \& {Asplund}}{{Casagrande}
  et~al.}{2010}]{2010A&A...512A..54C}
{Casagrande} L.,  {Ram{\'{\i}}rez} I.,  {Mel{\'e}ndez} J.,  {Bessell} M.,
  {Asplund} M.,  2010, \mn@doi [\aap] {10.1051/0004-6361/200913204}, \href
  {http://adsabs.harvard.edu/abs/2010A%26A...512A..54C} {512, A54}

\bibitem[\protect\citeauthoryear{{Chambers} et~al.,}{{Chambers}
  et~al.}{2016}]{2016arXiv161205560C}
{Chambers} K.~C.,  et~al., 2016, preprint, \href
  {http://cdsads.u-strasbg.fr/abs/2016arXiv161205560C} {} (\mn@eprint {arXiv}
  {1612.05560})

\bibitem[\protect\citeauthoryear{{Cohen}, {Wheaton}  \& {Megeath}}{{Cohen}
  et~al.}{2003}]{2003AJ....126.1090C}
{Cohen} M.,  {Wheaton} W.~A.,   {Megeath} S.~T.,  2003, \mn@doi [\aj]
  {10.1086/376474}, \href {http://adsabs.harvard.edu/abs/2003AJ....126.1090C}
  {126, 1090}

\bibitem[\protect\citeauthoryear{{Cruz}, {N{\'u}{\~n}ez}, {Burgasser},
  {Abrahams}, {Rice}, {Reid}  \& {Looper}}{{Cruz}
  et~al.}{2018}]{2018AJ....155...34C}
{Cruz} K.~L.,  {N{\'u}{\~n}ez} A.,  {Burgasser} A.~J.,  {Abrahams} E.,  {Rice}
  E.~L.,  {Reid} I.~N.,   {Looper} D.,  2018, \mn@doi [\aj]
  {10.3847/1538-3881/aa9d8a}, \href
  {http://adsabs.harvard.edu/abs/2018AJ....155...34C} {155, 34}

\bibitem[\protect\citeauthoryear{{Cushing}, {Vacca}  \& {Rayner}}{{Cushing}
  et~al.}{2004}]{2004PASP..116..362C}
{Cushing} M.~C.,  {Vacca} W.~D.,   {Rayner} J.~T.,  2004, \mn@doi [\pasp]
  {10.1086/382907}, \href {http://adsabs.harvard.edu/abs/2004PASP..116..362C}
  {116, 362}

\bibitem[\protect\citeauthoryear{{Cushing} et~al.,}{{Cushing}
  et~al.}{2008}]{2008ApJ...678.1372C}
{Cushing} M.~C.,  et~al., 2008, \mn@doi [\apj] {10.1086/526489}, \href
  {https://ui.adsabs.harvard.edu/abs/2008ApJ...678.1372C} {678, 1372}

\bibitem[\protect\citeauthoryear{{Cutri} et~al.,}{{Cutri}
  et~al.}{2013}]{2013wise.rept....1C}
{Cutri} R.~M.,  et~al., 2013, Technical report, {Explanatory Supplement to the
  AllWISE Data Release Products}

\bibitem[\protect\citeauthoryear{{Day-Jones} et~al.,}{{Day-Jones}
  et~al.}{2011}]{2011ASPC..448..833D}
{Day-Jones} A.~C.,  et~al., 2011, in {Johns-Krull} C.,  {Browning} M.~K.,
  {West} A.~A.,  eds,  Astronomical Society of the Pacific Conference Series
  Vol. 448, 16th Cambridge Workshop on Cool Stars, Stellar Systems, and the
  Sun. p.~833

\bibitem[\protect\citeauthoryear{{Deacon} et~al.,}{{Deacon}
  et~al.}{2014}]{2014ApJ...792..119D}
{Deacon} N.~R.,  et~al., 2014, \mn@doi [\apj] {10.1088/0004-637X/792/2/119},
  \href {http://adsabs.harvard.edu/abs/2014ApJ...792..119D} {792, 119}

\bibitem[\protect\citeauthoryear{{Deka-Szymankiewicz}, {Niedzielski},
  {Adamczyk}, {Adam{\'o}w}, {Nowak}  \& {Wolszczan}}{{Deka-Szymankiewicz}
  et~al.}{2018}]{2018A&A...615A..31D}
{Deka-Szymankiewicz} B.,  {Niedzielski} A.,  {Adamczyk} M.,  {Adam{\'o}w} M.,
  {Nowak} G.,   {Wolszczan} A.,  2018, \mn@doi [\aap]
  {10.1051/0004-6361/201731696}, \href
  {https://ui.adsabs.harvard.edu/#abs/2018A&A...615A..31D} {615, A31}

\bibitem[\protect\citeauthoryear{{Delgado-Donate}, {Clarke}, {Bate}  \&
  {Hodgkin}}{{Delgado-Donate} et~al.}{2004}]{2004MNRAS.351..617D}
{Delgado-Donate} E.~J.,  {Clarke} C.~J.,  {Bate} M.~R.,   {Hodgkin} S.~T.,
  2004, \mn@doi [\mnras] {10.1111/j.1365-2966.2004.07803.x}, \href
  {http://adsabs.harvard.edu/abs/2004MNRAS.351..617D} {351, 617}

\bibitem[\protect\citeauthoryear{{Delgado Mena} et~al.,}{{Delgado Mena}
  et~al.}{2015}]{2015A&A...576A..69D}
{Delgado Mena} E.,  et~al., 2015, \mn@doi [\aap] {10.1051/0004-6361/201425433},
  \href {http://adsabs.harvard.edu/abs/2015A%26A...576A..69D} {576, A69}

\bibitem[\protect\citeauthoryear{{Demarque}, {Woo}, {Kim}  \& {Yi}}{{Demarque}
  et~al.}{2004}]{2004ApJS..155..667D}
{Demarque} P.,  {Woo} J.-H.,  {Kim} Y.-C.,   {Yi} S.~K.,  2004, \mn@doi [\apjs]
  {10.1086/424966}, \href {http://adsabs.harvard.edu/abs/2004ApJS..155..667D}
  {155, 667}

\bibitem[\protect\citeauthoryear{{Demarque}, {Guenther}, {Li}, {Mazumdar}  \&
  {Straka}}{{Demarque} et~al.}{2008}]{2008Ap&SS.316...31D}
{Demarque} P.,  {Guenther} D.~B.,  {Li} L.~H.,  {Mazumdar} A.,   {Straka}
  C.~W.,  2008, \mn@doi [\apss] {10.1007/s10509-007-9698-y}, \href
  {http://adsabs.harvard.edu/abs/2008Ap%26SS.316...31D} {316, 31}

\bibitem[\protect\citeauthoryear{{Dhital}, {West}, {Stassun}  \&
  {Bochanski}}{{Dhital} et~al.}{2010}]{2010AJ....139.2566D}
{Dhital} S.,  {West} A.~A.,  {Stassun} K.~G.,   {Bochanski} J.~J.,  2010,
  \mn@doi [\aj] {10.1088/0004-6256/139/6/2566}, \href
  {http://adsabs.harvard.edu/abs/2010AJ....139.2566D} {139, 2566}

\bibitem[\protect\citeauthoryear{{Dotter}, {Chaboyer}, {Jevremovi{\'c}},
  {Kostov}, {Baron}  \& {Ferguson}}{{Dotter}
  et~al.}{2008}]{2008ApJS..178...89D}
{Dotter} A.,  {Chaboyer} B.,  {Jevremovi{\'c}} D.,  {Kostov} V.,  {Baron} E.,
  {Ferguson} J.~W.,  2008, \mn@doi [\apjs] {10.1086/589654}, \href
  {http://adsabs.harvard.edu/abs/2008ApJS..178...89D} {178, 89}

\bibitem[\protect\citeauthoryear{{Eisenhardt} et~al.,}{{Eisenhardt}
  et~al.}{2019}]{2019arXiv190808902E}
{Eisenhardt} P. R.~M.,  et~al., 2019, arXiv e-prints, \href
  {https://ui.adsabs.harvard.edu/abs/2019arXiv190808902E} {p. arXiv:1908.08902}

\bibitem[\protect\citeauthoryear{{Evans} et~al.,}{{Evans}
  et~al.}{2018}]{2018A&A...616A...4E}
{Evans} D.~W.,  et~al., 2018, \mn@doi [\aap] {10.1051/0004-6361/201832756},
  \href {http://adsabs.harvard.edu/abs/2018A%26A...616A...4E} {616, A4}

\bibitem[\protect\citeauthoryear{{Faherty} et~al.,}{{Faherty}
  et~al.}{2016}]{2016ApJS..225...10F}
{Faherty} J.~K.,  et~al., 2016, \mn@doi [\apjs] {10.3847/0067-0049/225/1/10},
  \href {http://cdsads.u-strasbg.fr/abs/2016ApJS..225...10F} {225, 10}

\bibitem[\protect\citeauthoryear{{Fekel}}{{Fekel}}{1988}]{1988AJ.....95..215F}
{Fekel} F.~C.,  1988, \mn@doi [\aj] {10.1086/114630}, \href
  {http://adsabs.harvard.edu/abs/1988AJ.....95..215F} {95, 215}

\bibitem[\protect\citeauthoryear{{Filippazzo}, {Rice}, {Faherty}, {Cruz}, {Van
  Gordon}  \& {Looper}}{{Filippazzo} et~al.}{2015}]{2015ApJ...810..158F}
{Filippazzo} J.~C.,  {Rice} E.~L.,  {Faherty} J.,  {Cruz} K.~L.,  {Van Gordon}
  M.~M.,   {Looper} D.~L.,  2015, \mn@doi [\apj] {10.1088/0004-637X/810/2/158},
  \href {http://adsabs.harvard.edu/abs/2015ApJ...810..158F} {810, 158}

\bibitem[\protect\citeauthoryear{{Flower}}{{Flower}}{1996}]{1996ApJ...469..355F}
{Flower} P.~J.,  1996, \mn@doi [\apj] {10.1086/177785}, \href
  {http://adsabs.harvard.edu/abs/1996ApJ...469..355F} {469, 355}

\bibitem[\protect\citeauthoryear{{Gagn{\'e}} et~al.,}{{Gagn{\'e}}
  et~al.}{2018}]{2018ApJ...856...23G}
{Gagn{\'e}} J.,  et~al., 2018, \mn@doi [\apj] {10.3847/1538-4357/aaae09}, \href
  {http://adsabs.harvard.edu/abs/2018ApJ...856...23G} {856, 23}

\bibitem[\protect\citeauthoryear{{Gaia Collaboration} et~al.,}{{Gaia
  Collaboration} et~al.}{2016a}]{2016A&A...595A...1G}
{Gaia Collaboration} et~al., 2016a, \mn@doi [\aap]
  {10.1051/0004-6361/201629272}, \href
  {http://adsabs.harvard.edu/abs/2016A%26A...595A...1G} {595, A1}

\bibitem[\protect\citeauthoryear{{Gaia Collaboration} et~al.,}{{Gaia
  Collaboration} et~al.}{2016b}]{2016A&A...595A...2G}
{Gaia Collaboration} et~al., 2016b, \mn@doi [\aap]
  {10.1051/0004-6361/201629512}, \href
  {http://cdsads.u-strasbg.fr/abs/2016A%26A...595A...2G} {595, A2}

\bibitem[\protect\citeauthoryear{{Gaia Collaboration} et~al.,}{{Gaia
  Collaboration} et~al.}{2018}]{2018A&A...616A...1G}
{Gaia Collaboration} et~al., 2018, \mn@doi [\aap]
  {10.1051/0004-6361/201833051}, \href
  {http://adsabs.harvard.edu/abs/2018A%26A...616A...1G} {616, A1}

\bibitem[\protect\citeauthoryear{{G{\'a}lvez-Ortiz}, {Solano}, {Lodieu}  \&
  {Aberasturi}}{{G{\'a}lvez-Ortiz} et~al.}{2017}]{2017MNRAS.466.2983G}
{G{\'a}lvez-Ortiz} M.~C.,  {Solano} E.,  {Lodieu} N.,   {Aberasturi} M.,  2017,
  \mn@doi [\mnras] {10.1093/mnras/stw3097}, \href
  {http://adsabs.harvard.edu/abs/2017MNRAS.466.2983G} {466, 2983}

\bibitem[\protect\citeauthoryear{{Gossage}, {Conroy}, {Dotter}, {Choi},
  {Rosenfield}, {Cargile}  \& {Dolphin}}{{Gossage}
  et~al.}{2018}]{2018ApJ...863...67G}
{Gossage} S.,  {Conroy} C.,  {Dotter} A.,  {Choi} J.,  {Rosenfield} P.,
  {Cargile} P.,   {Dolphin} A.,  2018, \mn@doi [\apj]
  {10.3847/1538-4357/aad0a0}, \href
  {http://adsabs.harvard.edu/abs/2018ApJ...863...67G} {863, 67}

\bibitem[\protect\citeauthoryear{{Gustafsson}, {Edvardsson}, {Eriksson},
  {J{\o}rgensen}, {Nordlund}  \& {Plez}}{{Gustafsson}
  et~al.}{2008}]{2008A&A...486..951G}
{Gustafsson} B.,  {Edvardsson} B.,  {Eriksson} K.,  {J{\o}rgensen} U.~G.,
  {Nordlund} {\AA}.,   {Plez} B.,  2008, \mn@doi [\aap]
  {10.1051/0004-6361:200809724}, \href
  {http://adsabs.harvard.edu/abs/2008A%26A...486..951G} {486, 951}

\bibitem[\protect\citeauthoryear{{Harlan} \& {Taylor}}{{Harlan} \&
  {Taylor}}{1970}]{1970AJ.....75..165H}
{Harlan} E.~A.,  {Taylor} D.~C.,  1970, \mn@doi [\aj] {10.1086/110956}, \href
  {https://ui.adsabs.harvard.edu/abs/1970AJ.....75..165H} {75, 165}

\bibitem[\protect\citeauthoryear{Hastie \& Stuetzle}{Hastie \&
  Stuetzle}{1989}]{princurve}
Hastie T.,  Stuetzle W.,  1989, \mn@doi [Journal of the American Statistical
  Association] {10.1080/01621459.1989.10478797}, 84, 502

\bibitem[\protect\citeauthoryear{{Herter} et~al.,}{{Herter}
  et~al.}{2008}]{2008SPIE.7014E..0XH}
{Herter} T.~L.,  et~al., 2008, in Ground-based and Airborne Instrumentation for
  Astronomy II. Edited by McLean, Ian S.; Casali, Mark M. Proceedings of the
  SPIE, Volume 7014, article id. 70140X, 8 pp. (2008).. ,
  \mn@doi{10.1117/12.789660}

\bibitem[\protect\citeauthoryear{{H{\o}g} et~al.,}{{H{\o}g}
  et~al.}{1997}]{1997A&A...323L..57H}
{H{\o}g} E.,  et~al., 1997, \aap, \href
  {http://adsabs.harvard.edu/abs/1997A%26A...323L..57H} {323, L57}

\bibitem[\protect\citeauthoryear{{Jofr{\'e}}, {Petrucci}, {Saffe}, {Saker}, {de
  la Villarmois}, {Chavero}, {G{\'o}mez}  \& {Mauas}}{{Jofr{\'e}}
  et~al.}{2015}]{2015A&A...574A..50J}
{Jofr{\'e}} E.,  {Petrucci} R.,  {Saffe} C.,  {Saker} L.,  {de la Villarmois}
  E.~A.,  {Chavero} C.,  {G{\'o}mez} M.,   {Mauas} P.~J.~D.,  2015, \mn@doi
  [\aap] {10.1051/0004-6361/201424474}, \href
  {https://ui.adsabs.harvard.edu/#abs/2015A&A...574A..50J} {574, A50}

\bibitem[\protect\citeauthoryear{{Kirkpatrick} et~al.,}{{Kirkpatrick}
  et~al.}{2010}]{2010ApJS..190..100K}
{Kirkpatrick} J.~D.,  et~al., 2010, \mn@doi [\apjs]
  {10.1088/0067-0049/190/1/100}, \href
  {http://adsabs.harvard.edu/abs/2010ApJS..190..100K} {190, 100}

\bibitem[\protect\citeauthoryear{{Kouwenhoven}, {Goodwin}, {Parker}, {Davies},
  {Malmberg}  \& {Kroupa}}{{Kouwenhoven} et~al.}{2010}]{2010MNRAS.404.1835K}
{Kouwenhoven} M.~B.~N.,  {Goodwin} S.~P.,  {Parker} R.~J.,  {Davies} M.~B.,
  {Malmberg} D.,   {Kroupa} P.,  2010, \mn@doi [\mnras]
  {10.1111/j.1365-2966.2010.16399.x}, \href
  {http://adsabs.harvard.edu/abs/2010MNRAS.404.1835K} {404, 1835}

\bibitem[\protect\citeauthoryear{{Kouwenhoven}, {Goodwin}, {Davies}, {Parker},
  {Kroupa}  \& {Malmberg}}{{Kouwenhoven} et~al.}{2011}]{2011ASPC..451....9K}
{Kouwenhoven} M.~B.~N.,  {Goodwin} S.~P.,  {Davies} M.~B.,  {Parker} R.~J.,
  {Kroupa} P.,   {Malmberg} D.,  2011, in {Qain} S.,  {Leung} K.,  {Zhu} L.,
  {Kwok} S.,  eds,  Astronomical Society of the Pacific Conference Series Vol.
  451, 9th Pacific Rim Conference on Stellar Astrophysics. p.~9

\bibitem[\protect\citeauthoryear{{Kurucz}}{{Kurucz}}{1992}]{1992IAUS..149..225K}
{Kurucz} R.~L.,  1992, in {Barbuy} B.,  {Renzini} A.,  eds,  IAU Symposium Vol.
  149, The Stellar Populations of Galaxies. p.~225

\bibitem[\protect\citeauthoryear{{Lindegren} et~al.,}{{Lindegren}
  et~al.}{2016}]{2016A&A...595A...4L}
{Lindegren} L.,  et~al., 2016, \mn@doi [\aap] {10.1051/0004-6361/201628714},
  \href {https://ui.adsabs.harvard.edu/abs/2016A&A...595A...4L} {595, A4}

\bibitem[\protect\citeauthoryear{{Lindegren} et~al.,}{{Lindegren}
  et~al.}{2018}]{2018A&A...616A...2L}
{Lindegren} L.,  et~al., 2018, \mn@doi [\aap] {10.1051/0004-6361/201832727},
  \href {http://adsabs.harvard.edu/abs/2018A%26A...616A...2L} {616, A2}

\bibitem[\protect\citeauthoryear{{Line}, {Teske}, {Burningham}, {Fortney}  \&
  {Marley}}{{Line} et~al.}{2015}]{2015ApJ...807..183L}
{Line} M.~R.,  {Teske} J.,  {Burningham} B.,  {Fortney} J.~J.,   {Marley}
  M.~S.,  2015, \mn@doi [\apj] {10.1088/0004-637X/807/2/183}, \href
  {http://adsabs.harvard.edu/abs/2015ApJ...807..183L} {807, 183}

\bibitem[\protect\citeauthoryear{{Looper} et~al.,}{{Looper}
  et~al.}{2008}]{2008ApJ...686..528L}
{Looper} D.~L.,  et~al., 2008, \mn@doi [\apj] {10.1086/591025}, \href
  {http://cdsads.u-strasbg.fr/abs/2008ApJ...686..528L} {686, 528}

\bibitem[\protect\citeauthoryear{{Lucas}, {Roche}, {Allard}  \&
  {Hauschildt}}{{Lucas} et~al.}{2001}]{2001MNRAS.326..695L}
{Lucas} P.~W.,  {Roche} P.~F.,  {Allard} F.,   {Hauschildt} P.~H.,  2001,
  \mn@doi [\mnras] {10.1046/j.1365-8711.2001.04666.x}, \href
  {http://adsabs.harvard.edu/abs/2001MNRAS.326..695L} {326, 695}

\bibitem[\protect\citeauthoryear{{Luck}}{{Luck}}{2017}]{2017AJ....153...21L}
{Luck} R.~E.,  2017, \mn@doi [\aj] {10.3847/1538-3881/153/1/21}, \href
  {https://ui.adsabs.harvard.edu/#abs/2017AJ....153...21L} {153, 21}

\bibitem[\protect\citeauthoryear{{Maldonado}, {Villaver}  \&
  {Eiroa}}{{Maldonado} et~al.}{2013}]{2013A&A...554A..84M}
{Maldonado} J.,  {Villaver} E.,   {Eiroa} C.,  2013, \mn@doi [\aap]
  {10.1051/0004-6361/201321082}, \href
  {https://ui.adsabs.harvard.edu/#abs/2013A&A...554A..84M} {554, A84}

\bibitem[\protect\citeauthoryear{{Mamajek} \& {Hillenbrand}}{{Mamajek} \&
  {Hillenbrand}}{2008}]{2008ApJ...687.1264M}
{Mamajek} E.~E.,  {Hillenbrand} L.~A.,  2008, \mn@doi [\apj] {10.1086/591785},
  \href {http://adsabs.harvard.edu/abs/2008ApJ...687.1264M} {687, 1264}

\bibitem[\protect\citeauthoryear{{Marocco} et~al.,}{{Marocco}
  et~al.}{2014}]{2014MNRAS.439..372M}
{Marocco} F.,  et~al., 2014, \mn@doi [\mnras] {10.1093/mnras/stt2463}, \href
  {http://adsabs.harvard.edu/abs/2014MNRAS.439..372M} {439, 372}

\bibitem[\protect\citeauthoryear{{Marocco} et~al.,}{{Marocco}
  et~al.}{2017}]{2017MNRAS.470.4885M}
{Marocco} F.,  et~al., 2017, \mn@doi [\mnras] {10.1093/mnras/stx1500}, \href
  {http://adsabs.harvard.edu/abs/2017MNRAS.470.4885M} {470, 4885}

\bibitem[\protect\citeauthoryear{{Martin} et~al.,}{{Martin}
  et~al.}{2017}]{2017ApJ...838...73M}
{Martin} E.~C.,  et~al., 2017, \mn@doi [\apj] {10.3847/1538-4357/aa6338}, \href
  {http://adsabs.harvard.edu/abs/2017ApJ...838...73M} {838, 73}

\bibitem[\protect\citeauthoryear{{McCarthy}, {Sandiford}, {Boyd}  \&
  {Booth}}{{McCarthy} et~al.}{1993}]{1993PASP..105..881M}
{McCarthy} J.~K.,  {Sandiford} B.~A.,  {Boyd} D.,   {Booth} J.,  1993, \mn@doi
  [\pasp] {10.1086/133250}, \href
  {http://adsabs.harvard.edu/abs/1993PASP..105..881M} {105, 881}

\bibitem[\protect\citeauthoryear{{McLean}, {McGovern}, {Burgasser},
  {Kirkpatrick}, {Prato}  \& {Kim}}{{McLean}
  et~al.}{2003}]{2003ApJ...596..561M}
{McLean} I.~S.,  {McGovern} M.~R.,  {Burgasser} A.~J.,  {Kirkpatrick} J.~D.,
  {Prato} L.,   {Kim} S.~S.,  2003, \mn@doi [\apj] {10.1086/377636}, \href
  {http://cdsads.u-strasbg.fr/abs/2003ApJ...596..561M} {596, 561}

\bibitem[\protect\citeauthoryear{{Niedzielski}, {Deka-Szymankiewicz},
  {Adamczyk}, {Adam{\'o}w}, {Nowak}  \& {Wolszczan}}{{Niedzielski}
  et~al.}{2016}]{2016A&A...585A..73N}
{Niedzielski} A.,  {Deka-Szymankiewicz} B.,  {Adamczyk} M.,  {Adam{\'o}w} M.,
  {Nowak} G.,   {Wolszczan} A.,  2016, \mn@doi [\aap]
  {10.1051/0004-6361/201527362}, \href
  {http://adsabs.harvard.edu/abs/2016A%26A...585A..73N} {585, A73}

\bibitem[\protect\citeauthoryear{{Nordstr{\"o}m} et~al.,}{{Nordstr{\"o}m}
  et~al.}{2004}]{2004A&A...418..989N}
{Nordstr{\"o}m} B.,  et~al., 2004, \mn@doi [\aap] {10.1051/0004-6361:20035959},
  \href {https://ui.adsabs.harvard.edu/#abs/2004A&A...418..989N} {418, 989}

\bibitem[\protect\citeauthoryear{{Noyes}, {Hartmann}, {Baliunas}, {Duncan}  \&
  {Vaughan}}{{Noyes} et~al.}{1984}]{1984ApJ...279..763N}
{Noyes} R.~W.,  {Hartmann} L.~W.,  {Baliunas} S.~L.,  {Duncan} D.~K.,
  {Vaughan} A.~H.,  1984, \mn@doi [\apj] {10.1086/161945}, \href
  {https://ui.adsabs.harvard.edu/abs/1984ApJ...279..763N} {279, 763}

\bibitem[\protect\citeauthoryear{{Oh}, {Price-Whelan}, {Hogg}, {Morton}  \&
  {Spergel}}{{Oh} et~al.}{2017}]{2017AJ....153..257O}
{Oh} S.,  {Price-Whelan} A.~M.,  {Hogg} D.~W.,  {Morton} T.~D.,   {Spergel}
  D.~N.,  2017, \mn@doi [\aj] {10.3847/1538-3881/aa6ffd}, \href
  {https://ui.adsabs.harvard.edu/#abs/2017AJ....153..257O} {153}

\bibitem[\protect\citeauthoryear{{Pace}}{{Pace}}{2013}]{2013A&A...551L...8P}
{Pace} G.,  2013, \mn@doi [\aap] {10.1051/0004-6361/201220364}, \href
  {https://ui.adsabs.harvard.edu/abs/2013A&A...551L...8P} {551, L8}

\bibitem[\protect\citeauthoryear{{Perruchot} et~al.,}{{Perruchot}
  et~al.}{2008}]{2008SPIE.7014E..0JP}
{Perruchot} S.,  et~al., 2008, in Ground-based and Airborne Instrumentation for
  Astronomy II. p. 70140J, \mn@doi{10.1117/12.787379}

\bibitem[\protect\citeauthoryear{{Pietrinferni}, {Cassisi}, {Salaris}  \&
  {Castelli}}{{Pietrinferni} et~al.}{2004}]{2004ApJ...612..168P}
{Pietrinferni} A.,  {Cassisi} S.,  {Salaris} M.,   {Castelli} F.,  2004,
  \mn@doi [\apj] {10.1086/422498}, \href
  {http://adsabs.harvard.edu/abs/2004ApJ...612..168P} {612, 168}

\bibitem[\protect\citeauthoryear{{Pinfield}, {Jones}, {Lucas}, {Kendall},
  {Folkes}, {Day-Jones}, {Chappelle}  \& {Steele}}{{Pinfield}
  et~al.}{2006}]{2006MNRAS.368.1281P}
{Pinfield} D.~J.,  {Jones} H.~R.~A.,  {Lucas} P.~W.,  {Kendall} T.~R.,
  {Folkes} S.~L.,  {Day-Jones} A.~C.,  {Chappelle} R.~J.,   {Steele} I.~A.,
  2006, \mn@doi [\mnras] {10.1111/j.1365-2966.2006.10213.x}, \href
  {http://adsabs.harvard.edu/abs/2006MNRAS.368.1281P} {368, 1281}

\bibitem[\protect\citeauthoryear{{Ricker} et~al.,}{{Ricker}
  et~al.}{2015}]{2015JATIS...1a4003R}
{Ricker} G.~R.,  et~al., 2015, \mn@doi [Journal of Astronomical Telescopes,
  Instruments, and Systems] {10.1117/1.JATIS.1.1.014003}, \href
  {https://ui.adsabs.harvard.edu/abs/2015JATIS...1a4003R} {1, 014003}

\bibitem[\protect\citeauthoryear{{Saffe}}{{Saffe}}{2011}]{2011RMxAA..47....3S}
{Saffe} C.,  2011, \rmxaa, \href
  {http://adsabs.harvard.edu/abs/2011RMxAA..47....3S} {47, 3}

\bibitem[\protect\citeauthoryear{{Sartoretti} et~al.,}{{Sartoretti}
  et~al.}{2018}]{2018A&A...616A...6S}
{Sartoretti} P.,  et~al., 2018, \mn@doi [\aap] {10.1051/0004-6361/201832836},
  \href {https://ui.adsabs.harvard.edu/\#abs/2018A&A...616A...6S} {616, A6}

\bibitem[\protect\citeauthoryear{{Saumon} \& {Marley}}{{Saumon} \&
  {Marley}}{2008}]{2008ApJ...689.1327S}
{Saumon} D.,  {Marley} M.~S.,  2008, \mn@doi [\apj] {10.1086/592734}, \href
  {http://adsabs.harvard.edu/abs/2008ApJ...689.1327S} {689, 1327}

\bibitem[\protect\citeauthoryear{{Skrutskie} et~al.,}{{Skrutskie}
  et~al.}{2006}]{2006AJ....131.1163S}
{Skrutskie} M.~F.,  et~al., 2006, \mn@doi [\aj] {10.1086/498708}, \href
  {http://cdsads.u-strasbg.fr/abs/2006AJ....131.1163S} {131, 1163}

\bibitem[\protect\citeauthoryear{{Skumanich}}{{Skumanich}}{1972}]{1972ApJ...171..565S}
{Skumanich} A.,  1972, \mn@doi [\apj] {10.1086/151310}, \href
  {http://adsabs.harvard.edu/abs/1972ApJ...171..565S} {171, 565}

\bibitem[\protect\citeauthoryear{{Slesnick}, {Hillenbrand}  \&
  {Carpenter}}{{Slesnick} et~al.}{2004}]{2004ApJ...610.1045S}
{Slesnick} C.~L.,  {Hillenbrand} L.~A.,   {Carpenter} J.~M.,  2004, \mn@doi
  [\apj] {10.1086/421898}, \href
  {http://adsabs.harvard.edu/abs/2004ApJ...610.1045S} {610, 1045}

\bibitem[\protect\citeauthoryear{{Smart}, {Marocco}, {Caballero}, {Jones},
  {Barrado}, {Beam{\'{\i}}n}, {Pinfield}  \& {Sarro}}{{Smart}
  et~al.}{2017}]{2017MNRAS.469..401S}
{Smart} R.~L.,  {Marocco} F.,  {Caballero} J.~A.,  {Jones} H.~R.~A.,  {Barrado}
  D.,  {Beam{\'{\i}}n} J.~C.,  {Pinfield} D.~J.,   {Sarro} L.~M.,  2017,
  \mn@doi [\mnras] {10.1093/mnras/stx800}, \href
  {http://cdsads.u-strasbg.fr/abs/2017MNRAS.469..401S} {469, 401}

\bibitem[\protect\citeauthoryear{{Smart}, {Marocco}, {Sarro}, {Barrado},
  {Beam{\'\i}n}, {Caballero}  \& {Jones}}{{Smart}
  et~al.}{2019}]{2019MNRAS.485.4423S}
{Smart} R.~L.,  {Marocco} F.,  {Sarro} L.~M.,  {Barrado} D.,  {Beam{\'\i}n}
  J.~C.,  {Caballero} J.~A.,   {Jones} H.~R.~A.,  2019, \mn@doi [\mnras]
  {10.1093/mnras/stz678}, \href
  {https://ui.adsabs.harvard.edu/abs/2019MNRAS.485.4423S} {485, 4423}

\bibitem[\protect\citeauthoryear{{Smith}, {Lucas}, {Burningham}, {Jones},
  {Smart}, {Andrei}, {Catal{\'a}n}  \& {Pinfield}}{{Smith}
  et~al.}{2014}]{2014MNRAS.437.3603S}
{Smith} L.,  {Lucas} P.~W.,  {Burningham} B.,  {Jones} H.~R.~A.,  {Smart}
  R.~L.,  {Andrei} A.~H.,  {Catal{\'a}n} S.,   {Pinfield} D.~J.,  2014, \mn@doi
  [\mnras] {10.1093/mnras/stt2156}, \href
  {http://adsabs.harvard.edu/abs/2014MNRAS.437.3603S} {437, 3603}

\bibitem[\protect\citeauthoryear{{Smith} et~al.,}{{Smith}
  et~al.}{2015}]{2015MNRAS.454.4476S}
{Smith} L.~C.,  et~al., 2015, \mn@doi [\mnras] {10.1093/mnras/stv2290}, \href
  {http://adsabs.harvard.edu/abs/2015MNRAS.454.4476S} {454, 4476}

\bibitem[\protect\citeauthoryear{{Smith} et~al.,}{{Smith}
  et~al.}{2018}]{2018MNRAS.474.1826S}
{Smith} L.~C.,  et~al., 2018, \mn@doi [\mnras] {10.1093/mnras/stx2789}, \href
  {http://adsabs.harvard.edu/abs/2018MNRAS.474.1826S} {474, 1826}

\bibitem[\protect\citeauthoryear{{Sneden}}{{Sneden}}{1973}]{1973ApJ...184..839S}
{Sneden} C.,  1973, \mn@doi [\apj] {10.1086/152374}, \href
  {http://adsabs.harvard.edu/abs/1973ApJ...184..839S} {184, 839}

\bibitem[\protect\citeauthoryear{{Soubiran}, {Jasniewicz}, {Chemin}, {Crifo},
  {Udry}, {Hestroffer}  \& {Katz}}{{Soubiran}
  et~al.}{2013}]{2013A&A...552A..64S}
{Soubiran} C.,  {Jasniewicz} G.,  {Chemin} L.,  {Crifo} F.,  {Udry} S.,
  {Hestroffer} D.,   {Katz} D.,  2013, \mn@doi [\aap]
  {10.1051/0004-6361/201220927}, \href
  {https://ui.adsabs.harvard.edu/#abs/2013A&A...552A..64S} {552}

\bibitem[\protect\citeauthoryear{{Sousa}, {Santos}, {Israelian}, {Lovis},
  {Mayor}, {Silva}  \& {Udry}}{{Sousa} et~al.}{2011}]{2011A&A...526A..99S}
{Sousa} S.~G.,  {Santos} N.~C.,  {Israelian} G.,  {Lovis} C.,  {Mayor} M.,
  {Silva} P.~B.,   {Udry} S.,  2011, \mn@doi [\aap]
  {10.1051/0004-6361/201015646}, \href
  {http://adsabs.harvard.edu/abs/2011A%26A...526A..99S} {526, A99}

\bibitem[\protect\citeauthoryear{{Stassun}, {Corsaro}, {Pepper}  \&
  {Gaudi}}{{Stassun} et~al.}{2018}]{2018AJ....155...22S}
{Stassun} K.~G.,  {Corsaro} E.,  {Pepper} J.~A.,   {Gaudi} B.~S.,  2018,
  \mn@doi [\aj] {10.3847/1538-3881/aa998a}, \href
  {http://adsabs.harvard.edu/abs/2018AJ....155...22S} {155, 22}

\bibitem[\protect\citeauthoryear{{Stephens} et~al.,}{{Stephens}
  et~al.}{2009}]{2009ApJ...702..154S}
{Stephens} D.~C.,  et~al., 2009, \mn@doi [\apj] {10.1088/0004-637X/702/1/154},
  \href {http://adsabs.harvard.edu/abs/2009ApJ...702..154S} {702, 154}

\bibitem[\protect\citeauthoryear{{Sterzik} \& {Durisen}}{{Sterzik} \&
  {Durisen}}{2003}]{2003A&A...400.1031S}
{Sterzik} M.~F.,  {Durisen} R.~H.,  2003, \mn@doi [\aap]
  {10.1051/0004-6361:20030073}, \href
  {http://adsabs.harvard.edu/abs/2003A%26A...400.1031S} {400, 1031}

\bibitem[\protect\citeauthoryear{{Stevens}, {Stassun}  \& {Gaudi}}{{Stevens}
  et~al.}{2017}]{2017AJ....154..259S}
{Stevens} D.~J.,  {Stassun} K.~G.,   {Gaudi} B.~S.,  2017, \mn@doi [\aj]
  {10.3847/1538-3881/aa957b}, \href
  {http://adsabs.harvard.edu/abs/2017AJ....154..259S} {154, 259}

\bibitem[\protect\citeauthoryear{{Strassmeier}, {Washuettl}, {Granzer},
  {Scheck}  \& {Weber}}{{Strassmeier} et~al.}{2000}]{2000A&AS..142..275S}
{Strassmeier} K.,  {Washuettl} A.,  {Granzer} T.,  {Scheck} M.,   {Weber} M.,
  2000, \mn@doi [\aaps] {10.1051/aas:2000328}, \href
  {http://adsabs.harvard.edu/abs/2000A%26AS..142..275S} {142, 275}

\bibitem[\protect\citeauthoryear{{Takeda}, {Ohkubo}  \& {Sadakane}}{{Takeda}
  et~al.}{2002}]{2002PASJ...54..451T}
{Takeda} Y.,  {Ohkubo} M.,   {Sadakane} K.,  2002, \mn@doi [Publications of the
  Astronomical Society of Japan] {10.1093/pasj/54.3.451}, \href
  {https://ui.adsabs.harvard.edu/#abs/2002PASJ...54..451T} {54, 451}

\bibitem[\protect\citeauthoryear{{Takeda}, {Ohkubo}, {Sato}, {Kambe}  \&
  {Sadakane}}{{Takeda} et~al.}{2005}]{2005PASJ...57...27T}
{Takeda} Y.,  {Ohkubo} M.,  {Sato} B.,  {Kambe} E.,   {Sadakane} K.,  2005,
  \mn@doi [Publications of the Astronomical Society of Japan]
  {10.1093/pasj/57.1.27}, \href
  {https://ui.adsabs.harvard.edu/#abs/2005PASJ...57...27T} {57, 27}

\bibitem[\protect\citeauthoryear{{Takeda}, {Ford}, {Sills}, {Rasio}, {Fischer}
  \& {Valenti}}{{Takeda} et~al.}{2007}]{2007ApJS..168..297T}
{Takeda} G.,  {Ford} E.~B.,  {Sills} A.,  {Rasio} F.~A.,  {Fischer} D.~A.,
  {Valenti} J.~A.,  2007, \mn@doi [The Astrophysical Journal Supplement Series]
  {10.1086/509763}, \href
  {https://ui.adsabs.harvard.edu/#abs/2007ApJS..168..297T} {168, 297}

\bibitem[\protect\citeauthoryear{{Tsvetkov}, {Popov}  \& {Smirnov}}{{Tsvetkov}
  et~al.}{2008}]{2008AstL...34...17T}
{Tsvetkov} A.~S.,  {Popov} A.~V.,   {Smirnov} A.~A.,  2008, \mn@doi [Astronomy
  Letters] {10.1134/S1063773708010039}, \href
  {https://ui.adsabs.harvard.edu/abs/2008AstL...34...17T} {34, 17}

\bibitem[\protect\citeauthoryear{{Tull}}{{Tull}}{1998}]{1998SPIE.3355..387T}
{Tull} R.~G.,  1998, in {D'Odorico} S.,  ed.,  \procspie Vol. 3355, Optical
  Astronomical Instrumentation. pp 387--398, \mn@doi{10.1117/12.316774}

\bibitem[\protect\citeauthoryear{{Umbreit}, {Burkert}, {Henning}, {Mikkola}  \&
  {Spurzem}}{{Umbreit} et~al.}{2005}]{2005ApJ...623..940U}
{Umbreit} S.,  {Burkert} A.,  {Henning} T.,  {Mikkola} S.,   {Spurzem} R.,
  2005, \mn@doi [\apj] {10.1086/428602}, \href
  {http://adsabs.harvard.edu/abs/2005ApJ...623..940U} {623, 940}

\bibitem[\protect\citeauthoryear{{Upgren}, {Grossenbacher}, {Penhallow},
  {MacConnell}  \& {Frye}}{{Upgren} et~al.}{1972}]{1972AJ.....77..486U}
{Upgren} A.~R.,  {Grossenbacher} R.,  {Penhallow} W.~S.,  {MacConnell} D.~J.,
  {Frye} R.~L.,  1972, \mn@doi [\aj] {10.1086/111308}, \href
  {https://ui.adsabs.harvard.edu/abs/1972AJ.....77..486U} {77, 486}

\bibitem[\protect\citeauthoryear{{Valenti} \& {Fischer}}{{Valenti} \&
  {Fischer}}{2005}]{2005ApJS..159..141V}
{Valenti} J.~A.,  {Fischer} D.~A.,  2005, \mn@doi [The Astrophysical Journal
  Supplement Series] {10.1086/430500}, \href
  {https://ui.adsabs.harvard.edu/#abs/2005ApJS..159..141V} {159, 141}

\bibitem[\protect\citeauthoryear{{Valenti} \& {Piskunov}}{{Valenti} \&
  {Piskunov}}{1996}]{1996A&AS..118..595V}
{Valenti} J.~A.,  {Piskunov} N.,  1996, \aaps, \href
  {http://adsabs.harvard.edu/abs/1996A%26AS..118..595V} {118, 595}

\bibitem[\protect\citeauthoryear{{Vogt} et~al.,}{{Vogt}
  et~al.}{1994}]{1994SPIE.2198..362V}
{Vogt} S.~S.,  et~al., 1994, in {Crawford} D.~L.,  {Craine} E.~R.,  eds,
  \procspie Vol. 2198, Instrumentation in Astronomy VIII. p.~362,
  \mn@doi{10.1117/12.176725}

\bibitem[\protect\citeauthoryear{{Weinberg}, {Shapiro}  \&
  {Wasserman}}{{Weinberg} et~al.}{1987}]{1987ApJ...312..367W}
{Weinberg} M.~D.,  {Shapiro} S.~L.,   {Wasserman} I.,  1987, \mn@doi [\apj]
  {10.1086/164883}, \href {http://adsabs.harvard.edu/abs/1987ApJ...312..367W}
  {312, 367}

\bibitem[\protect\citeauthoryear{{Yoss}}{{Yoss}}{1961}]{1961ApJ...134..809Y}
{Yoss} K.~M.,  1961, \mn@doi [\apj] {10.1086/147209}, \href
  {https://ui.adsabs.harvard.edu/abs/1961ApJ...134..809Y} {134, 809}

\bibitem[\protect\citeauthoryear{{Zhang}, {Homeier}, {Pinfield}, {Lodieu},
  {Jones}, {Allard}  \& {Pavlenko}}{{Zhang} et~al.}{2017}]{2017MNRAS.468..261Z}
{Zhang} Z.~H.,  {Homeier} D.,  {Pinfield} D.~J.,  {Lodieu} N.,  {Jones}
  H.~R.~A.,  {Allard} F.,   {Pavlenko} Y.~V.,  2017, \mn@doi [\mnras]
  {10.1093/mnras/stx350}, \href
  {http://adsabs.harvard.edu/abs/2017MNRAS.468..261Z} {468, 261}

\bibitem[\protect\citeauthoryear{{da Silva} et~al.,}{{da Silva}
  et~al.}{2006}]{2006A&A...458..609D}
{da Silva} L.,  et~al., 2006, \mn@doi [\aap] {10.1051/0004-6361:20065105},
  \href {https://ui.adsabs.harvard.edu/#abs/2006A&A...458..609D} {458, 609}

\makeatother
\end{thebibliography}

\newpage
\appendix
\section{Observing log}
\label{app}

\begin{table*}
    \centering
    \begin{tabular}{l c c c c c}
        \hline
        Name & Night & Exp. Time & Standard & Standard & Standard Exp. Time \\
             & (UT)  & DIT (s) $\times$ NDIT & & $V$ mag & DIT (s) $\times$ NDIT \\
        \hline
        HD~164507\,B & 2018-04-28 & 240 $\times$ 8 & HD165029 & 6.42 & 10 $\times$ 4 \\
        V478~Lyr\,C & 2018-04-29 & 300 $\times$ 8 & HD192538 & 6.46 & 10 $\times$ 4\\
        CD-28 8692\,B & 2018-04-28 & 240 $\times$ 8 & HD98949 & 7.52 & 10 $\times$ 4 \\ 
        2MASS~J18392740+4424510 & 2018-04-29 & 300 $\times$ 8 & HD192538 & 6.46 & 10 $\times$ 4 \\
        2MASS~J01390902+8110003 & 2018-10-16 & 180 $\times$ 4 & HD8424 & 6.36 & 5 $\times$ 4 \\
        2MASS~J01385969+8110084 & 2018-10-16 & 240 $\times$ 8 & HD8424 & 6.36 & 5 $\times$ 4 \\
        2MASS~J23253550+4608163 & 2018-10-16 & 120 $\times$ 4 & HD219290 & 6.31 & 5 $\times$ 4 \\
        2MASS~J23253519+4608098 & 2018-10-16 & 300 $\times$ 8 & HD219290 & 6.31 & 5 $\times$ 4 \\
        LT~UMa\,B & 2019-04-16 & 300 $\times$ 8 & HD91311 & 6.53 & 30 $\times$ 4 \\
        \hline
    \end{tabular}
    \caption{Log for the Palomar TripleSpec observations.}
    \label{tab:obslog}
\end{table*}

\bsp	
\label{lastpage}
\end{document}